\documentclass[epj]{svjour}
\usepackage{graphics}
\DeclareMathAccent{\pol}{\mathord}{letters}{"7E}
\usepackage{graphicx} 
\usepackage{amssymb,amsmath} 

\usepackage{epstopdf}
\graphicspath{{figures/}} 
\usepackage{url}   
\usepackage{nicefrac}  
\begin{document}
\title{The Structure of the Nucleon: Elastic Electromagnetic Form Factors}
\author{V.~Punjabi\inst{1},  C.F.~Perdrisat\inst{2},  M.K.~Jones\inst{3},   E.J.~Brash\inst{3,4}, and C.E.~Carlson\inst{2}
}                     
\offprints{punjabi@jlab.org (V. Punjabi)}          
\institute{Norfolk State University, Norfolk, VA 23504, USA \and  The College of William \& Mary, Williamsburg, VA 23187, USA \and Thomas Jefferson National Accelerator Facility, Newport News, VA 23606, USA \and Christopher Newport University, Newport News, VA 23606, USA}
\date{Received: \today / Revised version: \today}
%
\abstract{
Precise proton and neutron form factor measurements at Jefferson Lab,
using spin observables, have recently made a significant contribution to the
unraveling of the internal structure of the nucleon. Accurate experimental measurements of the
nucleon form factors are a test-bed for understanding how the nucleon's static properties
and dynamical behavior emerge from QCD, the theory of the strong interactions between quarks. 
There has been enormous theoretical progress, since the publication of the Jefferson Lab proton form
factor ratio data, aiming at reevaluating the picture of the nucleon. We will
review the experimental and theoretical developments in this field and discuss the outlook for the future.
} 
\titlerunning{The Structure of the Nucleon}
\maketitle
\tableofcontents
\section{Introduction} 

One of the fundamental goals of nuclear physics is to understand the structure and behavior of strongly interacting matter 
in terms of its basic
constituents, quarks and gluons. Quantum chromodynamics (QCD) is the theory of the strong interaction, responsible for binding 
quarks through the exchange of gluons to form  hadrons (baryons and mesons). The electromagnetic form factors are among the 
most basic quantities containing information 
about the internal structure of the proton and neutron, together known as nucleons. The challenge of understanding the nucleon's
structure and dynamics has occupied a central place in nuclear physics. High energy electron scattering provides one of the most powerful 
tools to investigate the structure of nucleons.

Early electron scattering experiments with nuclei were motivated by a need to verify predictions of the then current models
of the electromagnetic interaction of electrons with nuclei, and in particular with the proton and neutron; Rosenbluth
predicted that high energy electrons would be scattered dominantly by the magnetic moment of the proton \cite{rosenbluth}. 
Available accelerators in the early fifties had energies smaller than 50 MeV, and provided information on the nuclear
radius of elements from Be to Pb. The first clear evidence that the proton has a structure was obtained at the
High Energy Physics Laboratory (HEPL) at Stanford 
in the period form 1953 to 1956, under the leadership of Robert Hofstadter \cite{hofs55}. A proton charge radius of 0.77 fm was 
extracted by Chambers and Hofstadter \cite{Chambers} from the electron-proton data obtained using the electron beams with  
energies up to 550 MeV at the HEPL, confirming that the proton has a finite size. In his review 
paper Hofstadter \cite{hofstadter56} discussed in detail the extraction of proton charge and 
magnetization radii between 0.72 and 0.80 fm using different models.
Almost sixty years after the work of Hofstadter \cite{hofstadter56}, the question of whether the proton radius is 0.8775~(51)~fm, the CODATA \cite{Mohr:2012tt} value from $ep$ elastic scattering, or
is 0.84087~(39)~fm \cite{pohl:2013}, the muonic
hydrogen value, which is a difference of seven standard deviations, is being discussed intensely.

A similar change in accepted concepts occurred when the data,  
which was obtained at the Thomas Jefferson National 
Accelerator Facility (TJNAF) or Jefferson Lab (JLab)
for the proton's  electric to magnetic form factor ratio, $G_{Ep}/G_{Mp}$ from
double polarization experiments and 
completed in 2000 at a four momentum transfer squared, $Q^2$, of up to 
5.6 GeV$^2$ \cite{jones,punjabi05B,gayou:2002,Puckett:2011}, differed drastically from the form factor results 
obtained with the cross sections data using 
the Rosenbluth separation method \cite{hand63,janssens,price,litt,berger,bartel,bork,simon,walker,andivahisA,christy,qattan05}. 
The standard form factor database up to 1990's had been entirely defined by cross section measurements, and
suggested that, for $Q^2 \lesssim 6$ GeV$^2$, the ratio $\mu_p G_{Ep}/G_{Mp} \approx 1$, where $\mu_p$ is the magnetic moment of the proton. 
The double polarization experiments at JLab demonstrated that  $\mu_p G_{Ep}/G_{Mp}$ decreased approximately 
linearly with $Q^2$ for $Q^2 > 0.5$~GeV$^2$, dropping to a value of 0.35
at $Q^2$ =  5.6~GeV$^2$ which was the highest $Q^2$ investigated  at that time. 

In the last several years, the field of nucleon structure has been investigated extensively. Two recent experiments at JLab have increased 
the $Q^2$ range of the $G_{En}$ and $ \mu_{p}G_{Ep}/G_{Mp}$ data. These data have triggered much activity in the determination of the 
flavor separated form factors of the dressed up- and down quarks in the nucleon. Experiments at Mainz, JLab and  MIT-Bates have resulted in a much better coverage of the low $Q^2$ range for $G_{Ep}$,  contributing to the intensive discussion of a possible 
disagreement between electron elastic scattering and muonic hydrogen data used to determine the proton radius. 
Several experiments have measured the e$^{+}$/e$^{-}$ cross section ratio for the proton with a level of precision that tests recent calculations which included two photon exchange contributions to the cross sections. A direct measurement of the two photon contribution to the double polarization 
observables was undertaken at JLab. The results confirm the expectation that two photon exchange 
affects the proton form factor ratio at less than the percent level. Combining these various experiments and theoretical calculations, the possible role of two photon exchange in bridging the gap in the extraction of the proton form factors from cross section and double polarization observables can be investigated.

\subsection{History of Elastic Electron Scattering on the Nucleon}

Elastic electron proton scattering has evolved since the history making series of experiments with electron beams of the
HEPL at Stanford in 1950s. 
Under the leadership of R. Hofstadter, a series of crucial results were obtained from cross section measurements \cite{hofs53}. 
Several fundamental pieces of information were established following these experiments, including the approximate $1/Q^8$
decrease of the cross section with $Q^2$, establishing the approximate shape
of the charge distribution, and a first value for the proton radius. Theoretical
work evolved in parallel with these experimental ``firsts'', leading to the description of the elastic electron scattering
in terms of the lowest order process,  the exchange of a single virtual
photon with negative invariant mass squared; this lowest order contribution, also called the Born term, was expected to be dominant
because of the smallness of the electromagnetic coupling constant $\alpha_{EM}$.  Fundamental expressions for the hadronic current
and the definition of two invariant form factors, $F_1$ and $F_2$, later named the Dirac and Pauli form factors, of the 
Born term, were issues of this period. In 1957 Yennie, Levy and Ravenhall \cite{yennie57} derived 
an expression for the ${ep}$ cross section in terms of these two form factors, 
$F_1$ and $F_2$, following Rosenbluth's work \cite{rosenbluth}, as given in Eq. \ref{eq:csF2F1}.

The possibility of  measuring either the polarization transferred to the recoil proton, or the asymmetry if the target proton or neutron 
is polarized, with longitudinally polarized electrons, was discussed in a paper by Akhiezer {\it et al.} \cite{akhiezer1957} in 1957. It was to be more 
than 30 years before such experiments, which require a polarized electron beam, could be performed with good accuracy. 
Further papers on double polarization experiments followed, including Scofield \cite{Scofield:1959zz}, Akhiezer and Rekalo \cite{akh1B,akh2B}, Dombey \cite{dombey}, and
Arnold, Carlson and Gross \cite{arnold}. 

The construction of the Continuous Electron Beam Accelerator Facility (CEBAF) at JLab, in Virginia, led to 
an intensive program of nucleon form factor measurements, first for the proton and then for the neutron, and a significant breakthrough 
in our understanding of the proton structure. 
In the Born  approximation the transferred polarization has only two non-zero components, both in the 
reaction plane defined by the beam and scattered electron, one along the recoil proton momentum, and the other perpendicular to it.
For the proton, polarization transfer has been used most often at JLab \cite{jones,punjabi05B,gayou:2002,Puckett:2011}; it requires 
a re-scattering of the proton to measure its polarization. 
For the neutron, target asymmetry has now been used successfully \cite{zhu,warren,Riordan:2010} for the determination of 
the $G_{En}/G_{Mn}$ ratio; this requires a polarized target of 
either ${}^2\pol{\rm{H}}$  or ${}^3\pol{\rm{He}}$, which limits the maximum electron current that can be
tolerated without significant depolarization of the target.

Among the earliest electron-proton scattering polarization experiments is a search in 1963 at the Orsay linear accelerator, for 
a one-photon/two-photon interference 
effect with an un-polarized electron beam and an un-polarized target. They searched for normal and transverse 
polarization components. The normal polarization component  
was found to be 0.040 $\pm$ 0.027; the transverse polarization component was 0.000 $\pm$ 0.028 at $Q^2$ of 0.61 GeV$^2$\cite{bizot}.

A similar single-spin experiment in 1970 with an un-polarized 15--18 GeV electron beam at the Stanford linear accelerator,
and a polarized proton target with polarization perpendicular to the reaction plane to characterize the interference of
the two-photon exchange with the single photon exchange (Born) process, produced asymmetries of order 1 to 2 \% in the
range of $Q^2$ 0.38 to 0.98 GeV$^2$ \cite{Powell:1970}. 

On the neutron side, the pioneering experiment of Madey {\it et al.} performed the first
recoil polarization measurement of $G_{En}$ at a $Q^2$-value of 0.255 GeV$^2$ in 1994 \cite{eden:1994} at the MIT BATES
Linear Accelerator; and the first double polarization measurement of the proton form factor ratio, $G_{Ep}/G_{Mp}$, was also carried out at the same lab 
in 1994-1995 by measuring the two polarization transfer components P$_{\ell}$ and P$_t$, at $Q^2$-values of 
0.38 to 0.50 GeV$^2$ \cite{milbrathA,milbrathB,barkhuff}. 

Also the mid nineties saw a number of double polarization experiments at Nationaal Instituut voor 
Kernfysica en Hoge Energie Fysica (NIKHEF) \cite{passchier} and Mainz Microtron (MAMI) \cite{herberg,ostrick} to determine 
the neutron electric form factor up to 
$Q^2\sim $1 GeV$^2$. All experiments used polarized electron beams and a polarized target, either ${}^2\pol{\rm{H}}$ or ${}^3\pol{\rm{He}}$. 

In this review, we will focus on the space-like nucleon form factors, as in the past 15 years they have 
been studied more extensively both experimentally 
and theoretically, compared to the time-like nucleon form factors \cite{Mirazita}. Also the strangeness form factors are not 
discussed in this review;  see Ref.~\cite{Armstrong} for a review of the field of
parity violating electron scattering and strangeness form factors.

This review is organized as follows. Section~\ref{sec:formal} describes the formalism of elastic electron scattering on the nucleon. 
In subsection~\ref{subsec:dpff}, the use of elastic differential cross section data 
to extract the two electromagnetic form factors of proton and neutron by the 
Rosenbluth, or longitudinal and transverse (LT)-separation technique is reviewed. Subsection~\ref{subsec:formalpol} discusses how the form factors are measured in 
the double polarization experiments.  The two photon exchange formalism is explained in subsection~\ref{subsec:twophoton}.

Section~\ref{sec:expstatus} is devoted to discuss the experimental status. Subsection~\ref{subsec:xsection} describes the experiments which 
extracted the electric and magnetic form factors 
for the proton and the neutron from measurements of cross sections.
Subsection~\ref{subsec:poltransfer} reviews the status of the double polarization experiments and discuss the results
for the proton and the neutron, obtained from the recoil polarization  method and beam-target asymmetry measurements. The role of two photon exchange 
contributions in the elastic $ep$ reaction in resolving the discrepancy between the proton's electric form factor extracted by recoil polarization versus 
the Rosenbluth separation technique are discussed in subsection~\ref{subsec:discrepancies}.  Subsection~\ref{subsec:protonradius} reviews the present 
status of the proton charge radius. Subsection~\ref{flavor} discusses the present status of flavor separation of 
nucleon form factors. 

Section~\ref{sec:theory} deals with the theoretical interpretations of the electromagnetic nucleon form factors. 
Subsection~\ref{subsec:models} reviews the models of the nucleon form factors. These models include conformal fits to the form factors,  
vector meson dominance, dispersion analysis, constituent quark models, pion cloud models, transverse densities, and correspondences with 
higher dimensional theories. Subsection~\ref{subsec:dse} describes the Dyson-Schwinger equations and diquark models. Subsection~\ref{subsec:dis} discusses 
links between deep-inelastic scattering and nucleon form factors
which includes perturbative QCD inspired models and 
generalized parton distribution (GPD) models. Subsection~\ref{subsec:lattice} describes lattice QCD calculations of nucleon form factors.

Section~\ref{sec:conclusion} summarizes the current issues and challenges in the area of electromagnetic form factors. This 
section closes 
with a discussion of future experiments at Jefferson Lab which will measure the proton and neutron form factors to $Q^2$~=10~GeV$^2$ or greater.
\section{Formalism of Elastic Electron Nucleon Scattering}
\label{sec:formal}
The lowest order approximation for electron nucleon scattering is the single virtual photon exchange process,
or Born term. The Born approximation is expected to provide a good lowest order description of elastic $eN$ scattering (with $N=p,n$) because
 of the weak electro-magnetic coupling of
the photon with the charge and the magnetic moment of the nucleon. The amplitude for the process is the 
product of the four-component leptonic and hadronic currents, $\ell_{\mu}$ and ${\mathcal J}_{\mu}$, and
can be written as:
\begin{eqnarray}
i{\mathcal M}&=&\frac{-i}{q_{\mu}^2}\ell_{\mu}{\mathcal J}^{\mu} \nonumber \\
&=&\frac{-ig_{\mu\nu}}{q_{\mu}^2}\left[ie\bar{u}(k')\gamma^{\nu}u(k)\right]\left[-ie\bar{v}(p')\Gamma^{\mu}(p',p)v(p)\right],\,
\end{eqnarray}
where $k,k',p,p'$ are the the four-momenta of the incident and scattered, electron and proton, respectively,
$\Gamma^{\mu}$ contains all information of the nucleon structure, and $g_{\mu\nu}$ 
is the metric tensor. To insure relativistic invariance and the correct parity property of the amplitude 
$\mathcal M$, $\Gamma^{\mu}$ can only contain p, p' and $\gamma^{\nu}$, besides numbers, 
masses and $Q^2$, defined as, $Q^2=-(\vec{q}^{~2}-\omega^2)=-q_{\mu}^{ 2}$, is the negative of the square of the 
invariant mass, $q_{\mu}$, of the virtual photon exchanged in the one-photon 
approximation of $e{\it N}$ scattering.

The most general form for the hadronic current for the spin $\frac{1}{2}$-nucleon, satisfying relativistic invariance 
and current conservation, and including an internal structure is:\\
\begin{equation}
{\mathcal J}^{\mu}=ie\overline{\nu}(p')\left[\gamma^{\mu}{{F_1(Q^2)}}+
\frac{i\sigma^{\mu\nu}q_{\nu}}{2M}\kappa_{j}{{F_2(Q^2)}}\right]\nu(p), \\
\label{eq:Jhadron}
\end{equation}
\noindent where $M$ is the nucleon mass; $\kappa_j,\ \mbox{with}\ j=p,n$ is the anomalous magnetic moment, in units of the nuclear magneton,
 $\mu_N=e\hbar/(2M_p)$. 
The Dirac and Pauli form factors, $F_1(Q^2)$ and $F_2(Q^2)$ are the only structure functions allowed 
in the Born term by relativistic invariance. As is now the most frequently used notation, $\kappa_jF_2\ \mbox{with}\ j=p,n$ will be written as $F_{2p}$ 
and $F_{2n}$, respectively.  
In the static limit, $Q^2=0$, $F_{1p}=1$, $F_{2p}=\kappa_p=1.7928$ and $F_{1n}=0$ and $F_{2n}=\kappa_n=-1.9130$, 
for the proton and neutron, respectively.

\subsection{Cross Section Experiments}
\label{subsec:dpff}
The Lab frame differential cross section for detection of the electron in elastic $ep$ or $en$ scattering is then:
\begin{eqnarray}
\frac{d\sigma}{d\Omega_e} &=& \left(\frac{d\sigma}{d\Omega}\right)_{Mott}\frac{E_e}{E_{beam}} \Big(F_1^2(Q^2) \nonumber \\
& + & \mbox{} \tau\Big[F_2^2(Q^2) + 2[F_1(Q^2)+F_2(Q^2)]^2\tan^2\frac{\theta_e}{2}\Big]\Big),
\label{eq:csF2F1} 
\end{eqnarray}
with $\tau=Q^2/4M^2$. The Mott cross section is:
\begin{equation}
\left(\frac{d\sigma}{d\Omega}\right)_{Mott}=\frac{\alpha^2\cos^2\frac{\theta}{2}}{4E_{beam}^2\sin^4\frac{\theta}{2}}.
\label{eq:csmott}
\end{equation}

The incident electron (beam) and scattered electron energies are labeled $E_{beam}$ and $E_e$, respectively. The fraction $E_{e}/E_{beam}$
in Eq.~(\ref{eq:csF2F1}) is the recoil correction to the Mott cross section.

Experimental cross section data are most easily analyzed in terms of another set of form factors, the Sachs form 
factors $G_{E}$ and $G_{M}$ \cite{walecka,ernst}. The relations between $G_{E}$ and $G_{M}$ and $F_1$ and $F_2$ for proton and neutron are:
\begin{eqnarray}
	G_{E({p,n})}& = & F_{1({p,n})}-\tau F_{2({p,n})} \nonumber \\
	G_{M({p,n})}& = & F_{1({p,n})}+F_{2({p,n})}.
	\label{eq:gepgmp}
\end{eqnarray}
The scattering cross section Eq.~(\ref{eq:csF2F1}) can then be written in a simpler form, without an interference term, leading to a 
separation method for $G_{E}^2$ and $G_{M}^2$ known as Rosenbluth (or Longitudinal-Transverse) technique, 
as will be seen below. Now the cross section is:
\begin{eqnarray}
\frac{d\sigma}{d\Omega_e} &=& \left(\frac{d\sigma}{d\Omega}\right)_{Mott}\frac{E_e}{E_{beam}}\frac{1}{1+\tau} \left( G_{E}^2 + 
\frac{\tau}{\epsilon}G_{M}^2\right),
\label{eq:csgegm}
\end{eqnarray}
\noindent 
where $\epsilon$ is the polarization of the virtual photon defined as:

\begin{equation}
\epsilon=\frac{1}{1+2(1+\tau)\tan^2\frac{\theta_e}{2}}.
\label{eq:epsilon}
\end{equation} 
\noindent

The Rosenbluth separation technique takes advantage 
of the linear dependence in $\epsilon$, in the reduced 
cross section $\sigma_{red}$, based on Eq.~(\ref{eq:csgegm}), as follows:
\begin{eqnarray}
\sigma_{red} &=& \frac{\epsilon(1+\tau)}{\tau}\frac{E_{beam}}{E_e}\left(\frac{d\sigma}{d\Omega}\right)_{e}) 
/ \left(\frac{d\sigma}{d\Omega}\right)_{Mott} \nonumber \\
&=& G_{M}^2+\frac{\epsilon}{\tau} G_{E}^2,
\label{eq:redcs}
\end{eqnarray}
\noindent showing that $\sigma_{red}$ is expected to have a linear dependence on $\epsilon$, with the slope proportional 
to $G_{E}^2$ and the intercept equal to $G_{M}^2$.

\subsection{Double Polarization Experiments}
\label{subsec:formalpol}
In 1968 and 1974 Akhiezer and Rekalo \cite{akh1B,akh2B} discussed the interest of measuring 
an interference term of the form  $G_{E}G_{M}$ by observing the transverse 
component of the recoiling proton polarization in $\pol{e} p~\rightarrow~e \pol {p}$  
at large $Q^2$, to obtain $G_E$ in the presence of a dominating $G_M$.
In a review paper Dombey \cite{dombey} emphasized the virtues of measurements with a polarized lepton beam 
on a polarized target to obtain polarization observables. 
Also later in 1981 Arnold, Carlson and Gross \cite{arnold} discussed in detail, that the best way 
to measure the neutron and proton form factors would be to use the $^2{\rm H}(\pol{e},e' \pol{n})p$ 
and $^1{\rm H}(\pol{e},e' \pol{p})$ reactions, respectively.

Indeed, both the recoil polarization and target asymmetry measurement methods have
been used successfully to measure the proton and neutron form factors to high 
four momentum transfer, $Q^2$, at JLab. The same methods have been used also at MIT-Bates, MAMI, and NIKHEF, to make precise 
proton and neutron form factor measurements at lower $Q^2$. Both methods are discussed below, with 
benefits and drawbacks of using polarized target and/or focal plane polarimeter.

\subsubsection{Recoil Polarization Method}

\label{subsubsec:poltrans}

\begin{figure}
\begin{center}
\resizebox{\columnwidth}{!}{%
\includegraphics{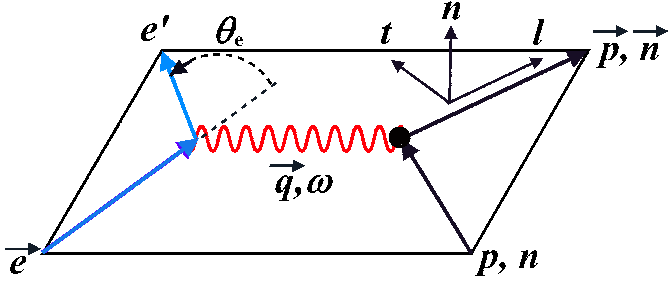}
}
\caption{Illustration of the kinematics and polarization of the recoil nucleon for 
$\pol {e} p\rightarrow e' \pol {p}$ and $\pol {e} n\rightarrow e' \pol {n}$.}
\label{fig:nlt}
\end{center}
\end{figure}

With a longitudinally polarized electron beam and an unpolarized target, the 
polarization of the incoming electron is transferred to the nucleon (proton or neutron) via exchange of a single virtual photon as 
shown in Fig. \ref{fig:nlt}. 
For elastic $ep$ 
scattering, in the single photon exchange approximation,  with a longitudinally polarized electron beam, the only non-zero
polarization transfer components are the longitudinal and transverse, $P_{\ell}$ and $P_t$. The normal
polarization transfer  component, $P_{n}$,  is zero.
 For single photon exchange, the 
transferred polarization components can be written in terms of the Sachs form factors as:

\begin{eqnarray}
I_o P_n & = & 0   \nonumber \\ 
I_oP_\ell & = & hP_e\frac{(E_{beam}+E_{e})}{M}\sqrt{\tau(1+\tau)}\tan^2\frac{\theta_e}{2}  G_M^2 \nonumber \\ 
I_oP_t & = & -hP_e2 \sqrt{\tau(1+\tau)}\tan\frac{\theta_e}{2} G_EG_M \nonumber \\
 I_o & = & G_E^2+\frac{\tau}{\epsilon}G_M^2 
 \label{eq:plpt}
\end{eqnarray}

\noindent
where $h = \pm 1$ are the beam helicity states, $P_e$  is the magnitude of polarization, and 
$\theta_{e}$ is the electron scattering angle.

The ratio of $G_E$ to $G_M$ is then directly obtained from the ratio of the two polarization
components $P_t$ and $P_{\ell}$ as:
\begin{equation}
\frac{G_{E}}{G_{M}}=-\frac{P_{t}}{P_{\ell}}\frac{(E_{beam}+E_{e})} 
{2M}\tan \frac{\theta_{e}}{2}.
\label{eq:ratio}
\end{equation}

The double-polarization approach to obtain 
the ratio at the large momentum transfer by measuring two polarization 
components simultaneously was first proposed at JLab in 1989~\cite{perpun}; it is based on 
a combination of spin precession in a magnetic spectrometer and using a proton polarimeter.
The major advantage of the method, compared to cross section measurements, 
is that in the Born approximation, for each $Q^2$, a single measurement of the azimuthal angular
distribution of the proton scattered in a secondary target gives both the longitudinal, $P_{\ell}$, and transverse, $P_t$, polarization.
Thus the ratio of electric to magnetic form factors of the proton is obtained directly from a simultaneous 
measurement of the two recoil polarization components. The knowledge of the beam polarization and of the analyzing power of the 
polarimeter is not needed to extract the ratio, $G_{E}/G_{M}$, strongly decreasing the 
systematic uncertainties. The kinematic factors in Eq.~(\ref{eq:ratio}) are
typically known to a precision far greater than the statistical precision of the recoil polarization components.

\subsubsection{Asymmetry with Polarized Targets}
\label{subsubsec:Asymmetry}

It was discussed by Dombey \cite{dombey} in a review paper in 1969 that the nucleon form factors 
can be extracted from the scattering of longitudinally polarized electrons off
a polarized nucleon target. 
In the one photon exchange approximation, the elastic electron nucleon
scattering cross section can be written as a sum of two parts: $\Sigma$, which corresponds to
the unpolarized elastic differential cross section given by Eq.~(\ref{eq:csgegm}), and a polarized part, $\Delta$, which is
non-zero only if the electron beam is longitudinal polarized \cite{donnelly,raskin};

\begin{equation} 
\sigma_{h} = \Sigma + h P_e  \Delta. \\
\label{eq:asymm}
\end{equation}

\begin{figure}
\begin{center}
\resizebox{0.45\textwidth}{!}{%
\includegraphics{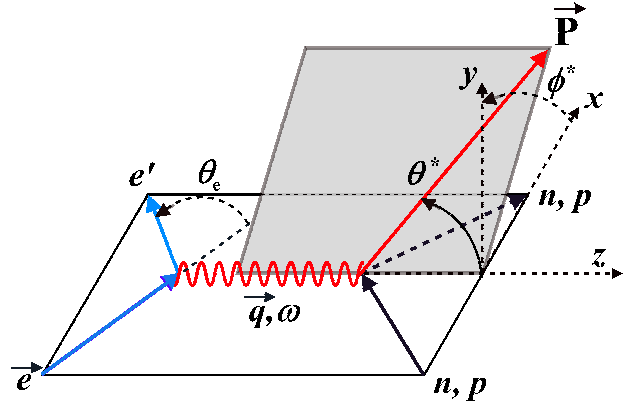}
}
\caption{Illustration of the kinematics and orientation of the target polarization $\pol{P}$, for the reaction $\pol{e} \pol{n} \rightarrow e' n$
and $\pol{e} \pol{p} \rightarrow e' p$. }
\label{fig:epkin_asym}
\end{center}
\end{figure}

The polarized part of the cross section, $\Delta$, with two terms related to the directions of the target polarization, 
$\vec{P}(\theta^{\ast} \phi^{\ast})$, is given by \cite{donnelly,raskin}: 
\begin{eqnarray}
\Delta & = & -2 \left(\frac{d\sigma}{d\Omega}\right)_{Mott} \frac{E_e}{E_{beam}}\tan\frac{\theta_e}{2} \sqrt{\frac{\tau}{1+\tau}} \nonumber \\
& &\Big(\sin\theta^{\ast}\cos\phi^{\ast}G_E G_M   \nonumber \\
&+& \sqrt{\tau\big[1+(1+\tau)\tan^2\frac{\theta_e}{2}\big]} \cos\theta^{\ast}G_M^2 \Big)  
\label{eq:delta}
\end{eqnarray}
\noindent
where $\theta^{\ast}$ and $\phi^{\ast}$ are the polar and 
azimuthal laboratory angles of the target polarization vector with $\vec q$ in the $\vec z$
direction and $\vec y$ normal to the electron scattering plane, as shown 
in Figure \ref{fig:epkin_asym}.

The physical asymmetry $A$ is then defined as 
\begin{equation}
A=\frac{\sigma_{+} -\sigma_{-}}{\sigma_{+} + \sigma_{-}}=\frac{\Delta}{\Sigma},
\label{eq:asy}
\end{equation}
\noindent
where $\sigma_+$ and $\sigma_-$ are the cross sections for the two beam helicity states. 

For a longitudinally polarized beam and polarized target, the measured asymmetry, $A_{meas}$, is 
related to the physical asymmetry, $A$, by 
\begin{equation}
A_{meas}=h P_{e}P_{target}A,
\label{eq:asy1}
\end{equation}
\noindent
where $P_{e}$ and $P_{target}$ are electron beam and target polarization, respectively, and 
$A$ can be obtained using Eqs.~(\ref{eq:csgegm}) and (\ref{eq:delta}),  
\begin{eqnarray}
A &=&-\frac{2\sqrt{\tau(1+\tau)}\tan\frac{\theta_e}{2}}{G_E^2+\frac{\tau}{\epsilon}G_M^2}
\Big[ \sin\theta^{\ast}\cos\phi^{\ast}G_E G_M \nonumber \\
&+& \sqrt{\tau\big[1+(1+\tau)\tan^2\frac{\theta_e}{2}\big]} \cos\theta^{\ast}G_M^2 \Big].
\label{eq:asy2}
\end{eqnarray}

From Eq.~(\ref{eq:asy2}), it is apparent that to extract $G_{E}$, the target polarization 
in the laboratory frame must be perpendicular with respect to the momentum transfer vector ${\vec q}$
and within the reaction plane, with $\theta^{\ast}= \pi/2$ and $\phi^{\ast}= 0^o$ or $180^o$. 
For these conditions, the physical asymmetry $A$ in Eq.~(\ref{eq:asy2}) simplifies to:
\begin{equation}
A_{perp}=\frac{-2\sqrt{\tau(1+\tau)}\tan\frac{\theta_e}{2} \frac{G_E}{G_M}}{(\frac{G_E}{G_M})^2+\frac{\tau}{\epsilon}}.
\label{eq:asy3}
\end{equation}
\noindent
As $(G_E/G_M)^2$ is quite small, $A_{perp}$ is approximately proportional to $G_E/G_M$. In practice, the second term in 
Eq.~(\ref{eq:asy2}) is not strictly zero due to the finite acceptance of the detectors, but these effects are small and depend
on kinematics only in first order and can be corrected for, so the ratio $G_E/G_M$ is not affected directly. One can also note that
$A_{perp} = P_t/(hP_e)$.

The discussion above is only applicable to a free electron-nucleon scattering. 
For a quasi-elastic electron scattering from a nuclear targets, like $^2$H or $^3$He, corrections are required for several nuclear effects. 

\subsection{Two-photon exchange}
\label{subsec:twophoton}
In the one-photon exchange process, the form factors depend only on $Q^2$ but not on other kinematic variables. 
A deviation in the form factors from constant when varying the kinematics (i.e. $E_{beam}$ and the scattered electron angle, 
keeping $Q^2$ constant) would indicate the presence of a mechanism beyond the Born approximation. 

In the general case, 
elastic $eN$ scattering can be described by three
complex amplitudes \cite{guichon,afanbrod,kivel:2012}:  $\tilde{G}_M$, $\tilde{G}_E$,
and $\tilde{F}_3$, the first two 
chosen as generalizations of the Sachs 
electric and magnetic form factors, $G_E$ and $G_M$, and the 
last one, $\tilde{F}_3$, vanishing in case of Born approximation.
The reduced cross section, $\sigma _{red}$ can be written as:
\begin{eqnarray}
\sigma_{red}&=& G_M^2 + \frac{\varepsilon}{\tau} G_E^2 + 2 G_M\Re\left(\delta \tilde{G}_M + \frac{\varepsilon}{M^2} \tilde F_3\right) 
\nonumber \\
&+& 2 \frac{\varepsilon}{\tau} G_E \Re \left(\delta\tilde G_E + \frac{\nu}{M^2}\tilde F_3\right),
\label{eq:siggen}
\end{eqnarray}
where
\begin{eqnarray}
\Re\tilde{G_M}(Q^2,\varepsilon)&=&G_M(Q^2)+\Re\delta \tilde{G_M}(Q^2,\varepsilon)  \\ 
\label{eq:regm}
\Re\tilde{G_E}(Q^2,\varepsilon)&=&G_E(Q^2)+\Re\delta \tilde{G_E}(Q^2, \varepsilon)~.
\label{eq:rege} 
\end{eqnarray}
and 
\begin{eqnarray}
\frac{\nu}{M^2}=\frac{s-u}{4M^2}=\sqrt{\tau(1-\tau)}\frac{1+\varepsilon}{1-\varepsilon}.
\label{eq:nu2gamma}
\end{eqnarray}
The polarization transfer components can be written as:
\begin{multline}
P_n =  \sqrt{\frac{2\varepsilon(1+\varepsilon)}{\tau}} \frac{h P_e}{\sigma _{red}} \Big[ -G_M \Im(\delta\tilde{G}_E  +  \frac{\nu}{M^2}\tilde F_3) \\
+   G_E \Im (\delta \tilde{G}_M +\frac{2\varepsilon}{1+\varepsilon}\frac{\nu}{M^2}\tilde{F}_3) \Big]
\label{eq:pngen}
\end{multline}
\begin{multline}
	P_t=-\sqrt{\frac{2\varepsilon(1-\varepsilon)}{\tau}} \frac{h P_e}{\sigma _{red}} \Big[ G_E G_M \\  
 + G_E\Re(\delta \tilde{G}_M) + G_M\Re(\delta \tilde{G}_E+\frac{\nu}{M^2}\tilde F_3) \Big]
\label{eq:ptgen}
\end{multline}
\begin{multline}
	P_{\ell}=\sqrt{(1-\varepsilon ^2)}\frac{h P_e}{\sigma _{red}} 
\Big[ G_M^2 \\ +  2 G_M \Re(\delta \tilde G_M+\frac{\varepsilon}{1+\varepsilon}\frac{\nu}{M^2}\tilde F_3)\Big]
\label{eq:plgen}
\end{multline}
\noindent
While the Sachs form factors depend only on $Q^2$, in the general case the amplitudes depend also on $\varepsilon$. 
The  reduced cross section
and the polarization transfer components, $P_t$ and $P_{\ell}$ are sensitive only to the real part
of the two-photon amplitudes. The  normal polarization transfer component, $P_n$, is sensitive to the imaginary parts of the two-photon 
amplitudes. In the Born approximation, only the first term remains from $\sigma_{red}$, $P_t$ and $P_{\ell}$ 
in Eqs.~(\ref{eq:siggen}), (\ref{eq:ptgen}) and (\ref{eq:plgen}) while $P_n$ is zero.

\section{Experimental Status}
\label{sec:expstatus}
The structure of the nucleons has been investigated experimentally with rigor  
over last 70 years using elastic electron scattering.
The two Sachs form factors, $G_{E}$ and $G_{M}$, required 
to describe the nucleon charge- and magnetization distribution have been 
traditionally obtained by cross section measurements. In the static limit,
the proton, $G_{Mp}$, and neutron, $G_{Mn}$, magnetic form factors are equal to the proton, $\mu_p$, and neutron,  $\mu_n$, magnetic moments while
the proton, $G_{Ep}$, and neutron, $G_{En}$, electric form factors are equal to unity and zero, respectively.  The earliest experiments at low $Q^2$ found:
\begin{equation}
G_{Ep} \approx \frac{G_{Mp}}{\mu_p} \approx \frac{G_{Mn}}{\mu_n} ,
\end{equation} 
 The $Q^2$ dependence of these form factors can be approximately characterized by a dipole form factor:  
\begin{equation}
G_D = \Big(1+\frac{Q^2}{0.71}\Big)^{-2},
\end{equation} 
 The data for $G_{Mp}$ have shown good consistency between different experiments
up to 30 GeV$^2$, however, the determination of $G_{Ep}$ 
at $Q^2$ greater than 2~GeV$^2$ has suffered from 
large error bars. The neutron electric form factor, $G_{En}$, is small and difficult to extract from cross section experiments. 
New experimental methods using spin observables were needed which pushed the development of polarized targets and new accelerators with high duty factor
and  polarized electron beams.

The recent generation of electron accelerators with high polarization and high current electron beams, at MIT-Bates, 
MAMI and JLab, have made it possible to investigate the internal structure of the nucleon with 
precision. In particular, the new series of experiments that measured spin observables, like
beam-target asymmetry and recoil polarization, have
allowed experiments to obtain the proton and neutron electromagnetic form factors accurately to large $Q^2$.
In this section we describe the proton and neutron form factors obtained from cross sections and double polarization experiments.
\subsection{Cross Section Experiments}
\label{subsec:xsection}

The electric and magnetic form factors of the nucleon can be extracted from measurements of cross sections at a constant $Q^{2}$ and different beam energies using Eq.~(\ref{eq:redcs}).  This is known as the Rosenbluth separation technique. The strict linearity of the reduced cross section is based on the dominance of one-photon exchange in the elastic electron-nucleon scattering reaction. In principle, only two $\epsilon$ points are needed to determine the slope, $G^2_{E}/\tau$, and intercept, $G^2_{M}$,  from Eq.~(\ref{eq:redcs}). Usually experiments have measured more than two $\epsilon$ points for a given $Q^2$. From a practical experimental viewpoint, more $\epsilon$ points allow better understanding and checks of systematic errors. From a theoretical viewpoint, the linearity of the $\epsilon$ dependence of the 
reduced cross section can be investigated with more $\epsilon$ points. One clear sign of a two-photon exchange contribution to the cross section would be 
a non-linearity in the $\epsilon$-dependence of the reduced cross section. Unfortunately, the two-photon exchange contribution can have a linear $\epsilon$-dependence which cannot be experimentally separated out in the cross section measurement and must be calculated  theoretically. The form factor data presented in the following sections were not corrected for hard two-photon exchange contributions when they
were  extracted from the elastic scattering measurements.

\subsubsection{Proton Form Factors}
\label{subsubsec:gmp}

\begin{figure}[b]
 	\begin{center}
 		\resizebox{\columnwidth}{!}{%
 			\includegraphics[angle=0]{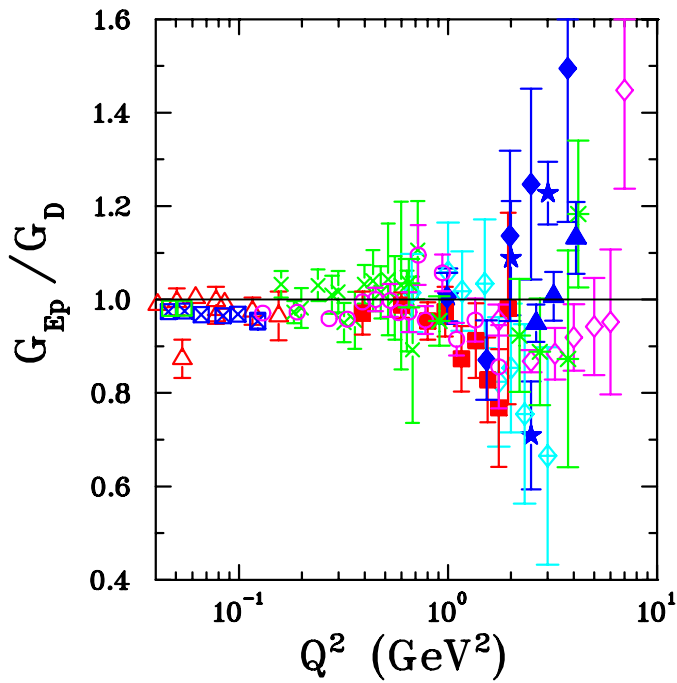}
 		}
 		\caption{$G_{Ep}$/$G_{D}$ extracted from cross section measurements versus $Q^2$. The data from before 1980 are:  open 
triangle (red) \cite{hand63}, multiplication sign (green)\cite{janssens}, open circle (magenta) \cite{price}, filled diamond (blue) \cite{litt}, filled 
square (red) \cite{berger}, crossed diamond (cyan) \cite{bartel},  crossed square (blue) \cite{bork} and open square (green) \cite{simon}.  The SLAC data from the 
1990's are filled star (blue) \cite{walker}  and open diamond (magenta) \cite{andivahisA}. The JLab data 
are asterisk (green) \cite{christy} and  filled triangle (blue) \cite{qattan05}.}
 		\label{fig:gepgd}
 	\end{center}
 \end{figure}

\begin{figure}[tbh]
 	\begin{center}
 		\resizebox{\columnwidth}{!}{%
 			\includegraphics[angle=0]{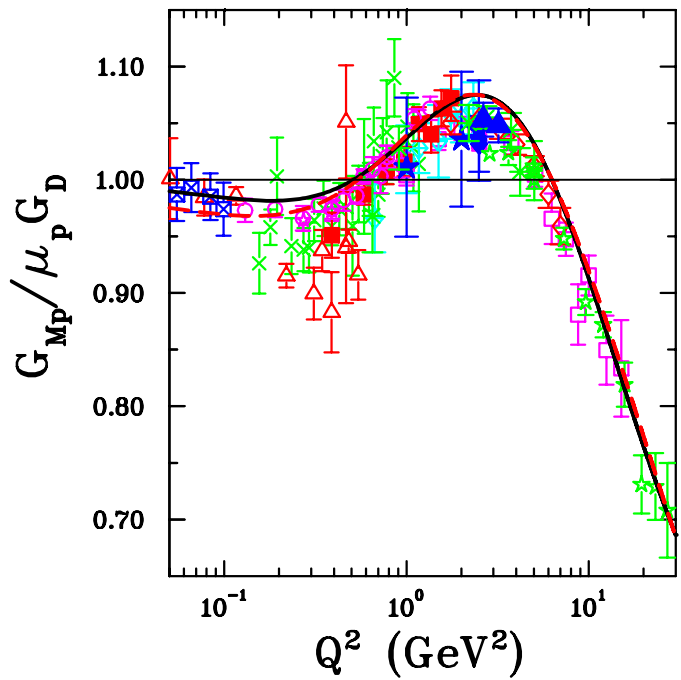}
 		}
 		\caption{$G_{Mp}$/$\mu_p$$G_{D}$ extracted from cross section measurements versus $Q^2$. The symbols are the same as in Fig.~\ref{fig:gepgd}. Additional 
data points at the highest $Q^2$, open square (magenta) \cite{kirk:1972xm} and open  star (green) \cite{sill}, were extracted from cross sections 
assuming $\mu_p$$G_{Ep}$/$G_{Mp}~=~1$. The solid (dashed) line is a fit by Ref.~\cite{kelly04} (Ref.~\cite{brash}) described in the text. 
 		}
 		\label{fig:gmpgd}
 	\end{center}
 \end{figure}

Extraction of the proton form factors from the cross section data is complicated by the strong dependence of the Mott cross section on the scattering angle. In addition, as can be seen in Eq.~(\ref{eq:redcs}), the relative contribution of the two form factors to the reduced cross section changes with $Q^2$. 
This difficulty in measuring both form factors within the same experiment with small errors bars was a problem 
from the first experiments using the Rosenbluth separation technique. 
 Ref.~\cite{hand63} contains a tabulation of measurements of $G_{Mp}$ and $G_{Ep}$ from the early 1960's which are plotted as open triangles in  Figs.~\ref{fig:gepgd} and \ref{fig:gmpgd}. From Ref.~\cite{hand63}, we have selected to plot only data
  which have relative error bars of less than 10\%. One can see that $G_{Ep}$ is measured with this precision only to  $Q^2 < 0.2$~GeV$^2$ while
  the $G_{Mp}$ data points are plotted only above $Q^2 = 0.2$~GeV$^2$.
  
  Throughout the 1960's, experiments increased their precision. Measurements were done at the Stanford Mark III accelerator of cross sections to 2\% statistical precision at $Q^2$ between 0.18 to 0.8~GeV$^2$ \cite{janssens}.
  Their $G_{Ep}$ and $G_{Mp}$ values are plotted as multiplication sign symbol in
  Figs.~\ref{fig:gepgd} and \ref{fig:gmpgd}. In 1971, results on $G_{Ep}$ and $G_{Mp}$ were published which combined cross section measurements  at the Cambridge Electron Accelerator with previous cross section measurements to cover a range of $Q^2$ between 0.13 to 1.75~GeV$^2$. The results are plotted in 
 Figs.~\ref{fig:gepgd} and \ref{fig:gmpgd} as open circles \cite{price}. There is good agreement with the $G_{Ep}$ and $G_{Mp}$  of Ref.~\cite{janssens} in the region of overlap, while $G_{Ep}/G_D$ for $Q^2 > 1$~GeV$^2$ starts to drop-off below unity. In 1970, a SLAC experiment measured $G_{Ep}$ and $G_{Mp}$ at $Q^2$ between 1 to 3.8~GeV$^2$ and found $G_{Ep}/G_D > 1$, which is opposite to the trend of Ref.~\cite{price}. These data are plotted as filled diamonds in Figs.~\ref{fig:gepgd} and \ref{fig:gmpgd}. In this same time period, $G_{Ep}$ and $G_{Mp}$ measurements were done at Bonn \cite{berger} for $Q^2$ between 0.34 to 1.94~GeV$^2$ and at DESY \cite{bartel} for $Q^2$ between 0.67 to 3.0~GeV$^2$. The data from Ref.~\cite{berger} and \cite{bartel} are plotted in Figs.~\ref{fig:gepgd} and \ref{fig:gmpgd} as
 filled square and crossed diamond, respectively. Both of these data sets agree with the downward trend in $G_{Ep}/G_D$ for $Q^2 > 1$~GeV$^2$ seen in Ref.\cite{price}, which disagrees with the rise in $G_{Ep}/G_D$ observed in Ref.~\cite{litt}.
 
 In the 1970's, a series of experiments at Mainz sought to measure  the $G_{Ep}$ and $G_{Mp}$ form factors at $Q^2$ below 0.1~GeV$^2$ with greater precision. The first
 \cite{bork} measured cross sections at $Q^2$ between 0.014 to 0.12~GeV$^2$. The second \cite{simon} did measurements up to $Q^2$ of 0.055~GeV$^2$ with an emphasis on extracting the charge radius of the proton. The $G_{Ep}$ and $G_{Mp}$ form factors of Ref.~\cite{bork} and \cite{simon} are plotted as crossed square and open square in Figs.~\ref{fig:gepgd} and \ref{fig:gmpgd}. These experiments demonstrate the precision that can be obtained in the measurement of $G_{Ep}$ at extremely low $Q^2$. In Sec.~\ref{subsec:protonradius}, results from these experiments for the proton charge radius are shown in Fig.~\ref{Rp_vs_t}.

In the late 1960's and early 1970's, experiments at SLAC  \cite{kirk:1972xm} pushed the limits of the $ep$ cross section measurements to $Q^2$~=~25~GeV$^2$.
In the 1990's, an experiment \cite{sill} at SLAC measured the $ep$ cross section at forward angles for $Q^2$ from 3 to 30~GeV$^2$ with improved statistical precision. Both experiments extracted $G_{Mp}$ under the assumption that $\mu_p$$G_{Ep}$/$G_{Mp}$~=~1. The data from Ref.~\cite{kirk:1972xm} (open squares) and \cite{sill} (open stars) are plotted in Fig.~\ref{fig:gmpgd}. The data are consistent with each other and show a drop-off in  $G_{Mp}$/$\mu_p$$G_{D}$ above $Q^2~=~7$~GeV$^2$.
 
 In the 1990's, at SLAC, two experiments were done which extended the precision and upper range of $Q^2$ for measurement of $G_{Ep}$ by the Rosenbluth separation technique. Ref.~\cite{walker} measured at $Q^2$ of 1.0, 2.0
 and 3.0~GeV$^2$. Ref.~\cite{andivahisA} measured at $Q^2$ between 1.75 and 9~GeV$^2$ and this experiment pushed the measurement of $G_{Ep}$ to the maximum $Q^2$ that has been done at this time. In Figs.~\ref{fig:gepgd} and \ref{fig:gmpgd}, the data of Ref.\cite{walker} and \cite{andivahisA} are plotted as filled stars and open diamonds, respectively. The $G_{Mp}$ data of both experiments agree with each other and previous experiments, while the $G_{Ep}/G_D$ are very different at $Q^2$~=~3~GeV$^2$. The $G_{Ep}/G_D$ of Ref.~\cite{andivahisA}
 agrees well with the early measurements of Ref.~\cite{price} at $Q^2$~=~1.75~GeV$^2$ and  have a flat $Q^2$ dependence with a slight rise  
 for $Q^2 > 1.75~$GeV$^2$. A global reanalysis of cross section experiments was done by Ref.~\cite{arring03} in 2003. This reanalysis found that the cross section measurements from different experiments were consistent with each other in extraction of  both $G_E$ and $G_M$, though it excluded the small angle data of Ref.~\cite{walker}
 from the global analysis.
 
 The $G_{Ep}/G_{Mp}$ data measured by the recoil polarization method will be discussed in the upcoming Sec.~\ref{subsec:poltransfer}, but the effect on the extraction of $G_{Mp}$ will be briefly discussed here.  In 2002, an extraction of $G_{Mp}$ was done from the cross section data of the previous experiments using the constraint that 
 \begin{equation}
 G_{Ep}/G_{Mp} = 1.0 - 0.13\times(Q^2 - 0.04), \nonumber
 \end{equation} 
 which originates from a fit to $G_{Ep}/G_{Mp}$ from the recoil polarization experiments \cite{brash}.
 The fit to the extracted $G_{Mp}$ is shown in Fig.~\ref{fig:gmpgd} as a dashed line and the fit is a few percent larger than $G_{Mp}$ from the standard Rosenbluth method. Also shown in Fig.~\ref{fig:gmpgd} as a solid line is a fit by Ref.~\cite{kelly04} to the $G_{Mp}$ extracted by Ref.~\cite{brash} with additional low $Q^2$ $G_{Mp}$ values from Ref.~\cite{hohler}. Since the time of these fits, the effects of two-photon exchange on the extraction of $G_{Ep}$  and $G_{Mp}$ have been calculated by theorists and this topic will be discussed in Sec.~\ref{subsec:discrepancies}.


 Taking advantage of the high duty  factor of modern accelerators, cross-sections can be 
 measured to high precision over a range of $\epsilon$ in a relatively short time period. This was recently done with spectacular precision by an experiment at MAMI. Cross sections were measured at 1422 kinematic settings covering a $Q^2$ range from 0.004 to 1.0~GeV$^2$ with average point-to-point systematic error of 0.37\%  \cite{Bernauer:2010wm}. Data were taken with the three separate spectrometers of MAMI at 6 different beam energies. With this large data set, the authors extracted $G_{Ep}$ and $G_{Mp}$ by fits to their cross section data rather than the traditional Rosenbluth separation technique.  The group has published a long paper \cite{Bernauer:2013tpr} on the same data set and included other world data in their fits. The sensitivity to different functional forms for the fits was investigated by using many different spline and polynomial forms. The fits also included 31 normalization parameters for possible systematic effects with cross sections measured with the different spectrometers and in different run periods.  The fits which had   a reduced $\chi^2 < 1.16$ had a maximum difference in their cross sections of 0.7\%.
 
 At JLab,  $G_{Ep}$ and $G_{Mp}$ were measured using the Rosenbluth separation method at $Q^2$ between 0.4 to 5.5~GeV$^2$ as part of an experiment to measure inelastic cross sections on a range of nuclei \cite{christy}. The results are plotted in Figs.~\ref{fig:gepgd} and \ref{fig:gmpgd} as asterisk and agree with previous measurements.

 Instead of detecting the elastically scattered electron, $(e,e^{\prime})$, an experiment 
 which detected the elastically scattered proton to
identify elastic reactions,  $(e,p)$, was run at JLab in 2002 \cite{qattan05}. 
The same experimental approach of extracting the form factors by measuring elastic cross sections
at fixed $Q^2$ and different $\epsilon$ by varying the beam energy was used. The experimental method takes advantage of the fact
that the proton momentum is constant for all $\epsilon$ at a fixed $Q^2$. 
In addition for  $(e,p)$, the detected proton rate and the radiation corrections 
have a smaller dependence on $\epsilon$ compared to  $(e,e^{\prime})$ experiments. 
All this combines to  reduce the $\epsilon$ dependent systematic error compared to $(e,e^{\prime})$ experiments.
The form factors were measured at $Q^2$ = 2.64, 3.10 and 4.60 GeV$^2$.
In Fig.~\ref{fig:gepgd} and \ref{fig:gmpgd}, the measurements of $G_{Ep}$ and $G_{Mp}$ from the $(e,p)$ reaction are plotted. The agreement between the form
factors extracted by the different experiments is excellent. Experiments which detect either
the scattered electrons or the scattered protons have different systematics, so the agreement between
the two techniques indicates that the experimental systematic errors are understood.

\begin{figure}[tbh]
	\begin{center}
		\resizebox{\columnwidth}{!}{%
			\includegraphics{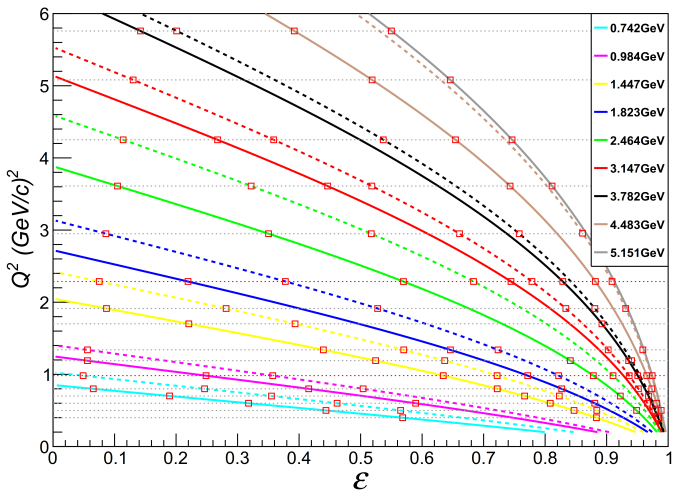}
		}
		\caption{Kinematic coverage of the JLab experiment E05-017. Each kinematic point of the experiment is plotted for $Q^2$ versus $\epsilon$. The solid and dashed lines are constant energy and the dotted lines are constant $Q^2$.}
		\label{fig:super-rosen}
	\end{center}
\end{figure}

With the success of the first JLab $(e,p)$ experiment, a subsequent experiment, E05-017,  was run at JLab in Hall C in 2007 \cite{e05017}.
The experiment measured cross-sections at a total of 102 kinematic settings covering  a wide $Q^2$ range from 0.4 to 5.76~GeV$^2$
with at least three $\epsilon$ points per $Q^2$. The emphasis was  to measure at each $Q^2$ as wide an $\epsilon$ range as possible. Fig.~\ref{fig:super-rosen} plots the $Q^2$ versus $\epsilon$ for all kinematic points  of E05-017. To obtain multiple $\epsilon$ at the each $Q^2$, 17 different beam energies were needed for the experiment. This number of beam energies in a relatively short time period demonstrate the amazing capabilities in the operation of the Continuous Electron Beam Accelerator at JLab.
At $Q^2$~=~1~GeV$^2$, thirteen $\epsilon$ points were measured ranging from $\epsilon$~=~0.05 to 0.98, with eight of the points
above $\epsilon$ = 0.8. Similarly for $Q^2$~=~2.3~GeV$^2$, ten $\epsilon$ points were measured ranging from $\epsilon$~=~0.07 to 0.92, 
with five of the points above $\epsilon$ = 0.7.  The wide range of  $\epsilon$ at a fixed $Q^2$ allows a check of the non-linearity 
in the $\epsilon$  dependence of the cross-section which would be a sign of two-photon exchange contributions effecting the cross-sections.
The effects from two-photon exchange contributions could have a dramatic  $\epsilon$  dependence near  $\epsilon$~=~1.

\subsubsection{Neutron Form Factors}
\label{subsubsec:gmn}

The neutron has zero charge and therefore  $G_{En}$  has to be zero at $Q^2=0$. The
slope of  $G_{En}$  at $Q^2=0$ is related to the mean-square radius of the neutron  (see Eq.~(\ref{eq:sloperms}) for the proton). Also, the mean-square radius of the neutron, 
$<\,r_{en}{^2}\,>$, can be expressed in terms of the neutron-electron scattering length. The
neutron-electron scattering length, $b_{ne}$, can be determined from
total transmission cross sections for epithermal and thermal neutrons scattering 
on the atomic electrons in noble gas targets \cite{Krohn:1973re} and
lead and  bismuth targets \cite{Aleksandrov:1986mw,Koester:1995nx,Kopecky:1997rw}.  Using results from the total
transmission experiments, the Particle Data Group published a recommended value of  $<\,r_{en}{^2}\,>\,=\,- 0.1161 \pm 0.0022~\mbox{fm}^2 $ or $dG_{En}/dQ^2 = 0.01935 \pm 0.00037~\mbox{fm}^2$ \cite{Agashe:2014kda}. 

\begin{figure}
	\begin{center}
		\resizebox{\columnwidth}{!}{%
			\includegraphics[angle=0]{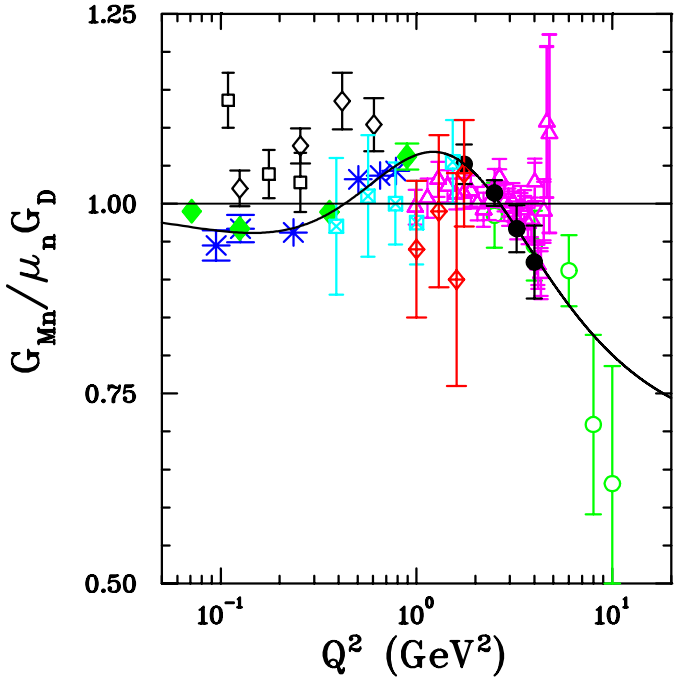}
		}
		\caption{$G_{Mn}/\mu_n$$G_D$ versus $Q^2$. The symbols corresponds to the data from 
			open circle (green) \cite{rockB}, filled circle (black) \cite{lung}, open square (black) \cite{marko}, 
			open diamond (black) \cite{bruins}, asterisk (blue) \cite{anklin} and \cite{Anklin:1998ae}, 
			filled diamond (green) \cite{kubon}, open triangle (magenta) \cite{lachniet:2008}, square with cross inside (cyan) \cite{bartel} and 
diamond with cross inside (red) \cite{Arnold:1988us}. The solid line is a fit by Kelly \cite{kelly04} which was done in 2004 before the Hall B 
measurement \cite{lachniet:2008} and  excluded the data of Ref. \cite{marko} and \cite{bruins}.}
		\label{fig:gmn_nopol}
	\end{center}
\end{figure}

With no free neutron target, cross section experiments have to make measurements on a deuteron target to extract the neutron form factors.  Experiments have measured cross sections for quasi-elastic single arm $d(e,e')$ and coincidence $d(e,e'p)n$ reactions.  Primarily, the quasi-elastic $ed$ reaction is a measurement of $G_{Mn}$ with limited sensitivity to $G_{En}$, since $G_{En}$ is near zero and much smaller than the proton contribution to the cross section.
Single arm quasi-elastic $ed$ scattering by Ref.~\cite{bartel,hanson,hughes} can be used to extract $G_{Mn}$ but this requires theoretical knowledge of the large final state interactions at low $Q^2$, which leads to a sizable theoretical  uncertainty.  Coincidence cross section measurements in the $d(e,e'p)n$ reaction were done by \cite{budnitz,dunning}. Detection of the neutron in coincidence reduces the theoretical uncertainty and the proton contribution in the extraction, but the uncertainty on the knowledge of the neutron detection efficiency becomes important. In the 1990's, $G_{Mn}$  was measured at MIT-Bates at $Q^2$ of 0.11, 0.18 and 0.26~GeV$^2$ using the $d(e,e'n)p$ reaction and the data are plotted in Fig.~\ref{fig:gmn_nopol} as open squares.  This experiment measured the neutron detection efficiency using the $^2H(\gamma,pn)$ reaction.

Extracting $G_{Mn}$ from the ratio of 
cross sections of the quasi-elastic $d(e,e'n)p$ to $d(e,e'p)n$ reactions is the least sensitive method to uncertainties in the calculation of the deuteron wave function, final state interactions and meson exchange contributions. In the early 1970's, at DESY, pioneering experiments measuring the ratio of quasi-free cross sections for the $d(e,e'n)$ to $d(e,e'p)$ reactions were performed  by Ref.~\cite{bartel} and \cite{stein}. These experiments extracted $G_{Mn}$ at $Q^2$ = 0.4, 0.57, 0.78, 1.0 and 1.5~GeV$^2$  and the $G_{Mn}$ data are plotted in Fig.~\ref{fig:gmn_nopol} as squares with cross.

 In 1995, at Bonn, the ratio of quasi-elastic $d(e,e'n)p$ to $d(e,e'p)n$ cross sections was used to extract $G_{Mn}$. 
 $G_{Mn}$  was measured at $Q^2$ of 0.13, 0.25,
0.42 and 0.61~GeV$^2$ \cite{bruins} and is plotted in Fig.~\ref{fig:gmn_nopol} with
empty diamond. The neutron detection efficiency was measured {\it in situ} using the $^{1}H(\gamma,\pi^{-})n$ reaction. 
The photons were produced by bremsstrahlung in the hydrogen target.

 A series of measurements of the ratio of  $d(e,e'n)p$ to $d(e,e'p)n$ cross sections were made in which the neutron detector efficiency was measured by taking the neutron detector to Paul Scherrer Institute (PSI) and  using the neutron beam line. A LED system was used to monitor gain and baseline shifts at PSI and during the experiments. The first experiment was done at NIKHEF and  $G_{Mn}$  was measured at $Q^2$ of 0.61 and 0.70~GeV$^2$ \cite{anklin}. The next experiment was done at MAMI and measurements of $G_{Mn}$ were done at $Q^2$ of 0.24, 0.50, 0.65 and 0.78~GeV$^2$ \cite{Anklin:1998ae}. The results from these two experiments are plotted as an asterisk in Fig.~\ref{fig:gmn_nopol}. These were followed by more measurements of $G_{Mn}$ at MAMI which extend the $Q^2$ range. The latter experiment was done at $Q^2$ values of 0.071, 0.125, 0.36 and
0.89~GeV$^2$ with statistical errors at the 1.5\% level and are plotted  in Fig.~\ref{fig:gmn_nopol} as filled diamonds.   Efficiency measurements at PSI were done before and after each experiment and the measurements were consistent. Within this series of experiments, the $G_{Mn}$ at matching $Q^2$ are in excellent agreement, but are smaller than $G_{Mn}$ measured at Bonn \cite{bruins} and MIT-Bates \cite{marko}. Ref.~\cite{jourdan} suggested that the Bonn experiment had miscalculated their neutron efficiency, because a 
contribution from pion electroproduction was not taken into account when determining the neutron efficiency which led to an overestimate of $G_{Mn}$. In their reply \cite{bruins1}, Bruins {\it et al} responded that in the peaking approximation the contribution from pion electroproduction to their kinematics is negligible. In calculating the contribution from electroproduction, each paper uses a different data set to extrapolate to the measured kinematic region,
so, as stated in Ref.~\cite{bruins1}, the only way to conclusively settle the disagreement is to measure pion electroproduction and photoproduction in the kinematics of the experiment.

At the present time, the highest $Q^2$ measurement of $G_{Mn}$ was done at SLAC in the early 1980's
by measuring quasi-free $ed$ cross sections \cite{rockB}. This experiment measured  $G_{Mn}$ from $Q^2$ of 2.5 to 10~GeV$^2$. Another SLAC experiment \cite{Arnold:1988us} measured $G_{Mn}$ from $Q^2$ of 1.0 to 1.75~GeV$^2$.  In the 1990's, at SLAC, a Rosenbluth separation experiment was done for the quasi-free $ed$ reaction at $Q^2$ = 1.75, 2.5, 3.25 and 4.0~GeV$^2$ and both $G_{Mn}$ and $G_{En}$ were extracted \cite{lung}. The $G_{En}$ values were consistent with zero with large error bars.   All experiments have consistent $G_{Mn}$ values in the region of overlapping kinematics and their  $G_{Mn}$ values are plotted in Fig.~\ref{fig:gmn_nopol}. 

The experiment at JLab in Hall~B using CLAS measured $G_{Mn}$ in fine $Q^2$ bins from $Q^2$ between 1.0 to 4.8~GeV$^2$ \cite{lachniet:2008}.
The data are plotted as open triangles in Fig.~\ref{fig:gmn_nopol}). A unique feature of this experiment was a dual cell design with liquid hydrogen and deuterium cells separated by 4.7~cm. This allowed measurement of the neutron detection efficiency by the $H(e,e^{\prime}\pi^{+})n$ reaction to be done simultaneously with cross section measurements. The Hall B data  overlaps nicely with the SLAC measurements. Though at $Q^2$ near 4.8~GeV$^2$, the Hall B data set suggest a less rapid $Q^2$ fall-off, then measured in the SLAC high $Q^2$ data. 

Experiments have used elastic $ed$ cross sections to determine the neutron form factors. The scattering by an electron from
the spin 1 deuteron requires 3 form factors in the hadronic current operator, 
for the charge, quadrupole and 
magnetic distributions, $G_C$, $G_Q$ and $G_{Md}$, respectively. In the original
impulse approximation form of the cross section 
developed by Gourdin \cite{gourdin}, the elastic $ed$ cross section is: 
\begin{equation}
\frac{d\sigma}{d\Omega}=\frac{d\sigma}{d\Omega}_{Mott}\left(A(Q^2)+B(Q^2)\tan^2(\frac{\theta_e}{2})\right),
\label{eq:edxn}
\end{equation}
\noindent
where 
\begin{eqnarray}
A(Q^2) & = & G_C^2(Q^2)+\frac{8}{9}\eta^{2}G_Q^{2}(Q^2)+
\frac{2}{3}\eta(1+\eta)G_{Md}^{2}, \nonumber \\
 B(Q^2) & = & \frac{3}{4}\eta(1+\eta)^2G_{Md}^2(Q^2), 
\end{eqnarray}
with $\eta=Q^2/4M_D^2$. The charge, quadrupole and magnetic form factors 
can be written in terms of the isoscalar electric and magnetic form factors as follows:
\begin{eqnarray}
	G_C&=&G_{E}^{S}C_E,~~\mbox{  }~~G_Q=G_{E}^{S}C_Q ,\nonumber \\ G_{Md}&=&\frac{M_D}{M_p}(
	G_{M}^{S}C_S+\frac{1}{2}G_{E}^{S}C_L), 
\end{eqnarray}
\noindent
where the coefficients $C_E$, $C_Q$, $C_L$ and $C_S$
are Fourier transforms of specific combinations of the S- and D-state 
deuteron wave functions, $u(r)$ and $w(r)$~\cite{gourdin}. 
The isoscalar ($G_i^S$) and isovector ($G_i^V$)
magnetic ($i = M$) and electric ($i = E$) form factors are defined as~:
\begin{equation}
G_{i}^S = G_{i p} + G_{i n} \mbox{  and  } G_{i}^V = G_{i p} - G_{i n}.
\end{equation}
Theoretical knowledge of the deuteron wave function is needed to extract the form factors which is a major systematic uncertainty. Both the magnetic and electric form factors can be extracted from the elastic $ed$ cross sections, though the $Q^2$ range is limited by the theoretical uncertainties.  In the 1960's, elastic $ed$ cross section measurements were done which determined $G_{Mn}$ and $G_{En}$ for $Q^2 < 0.12$~GeV$^2$ \cite{benaksasB,grossetete}.

\begin{figure}[tb]
	\begin{center}
		\resizebox{\columnwidth}{!}{%
			\includegraphics{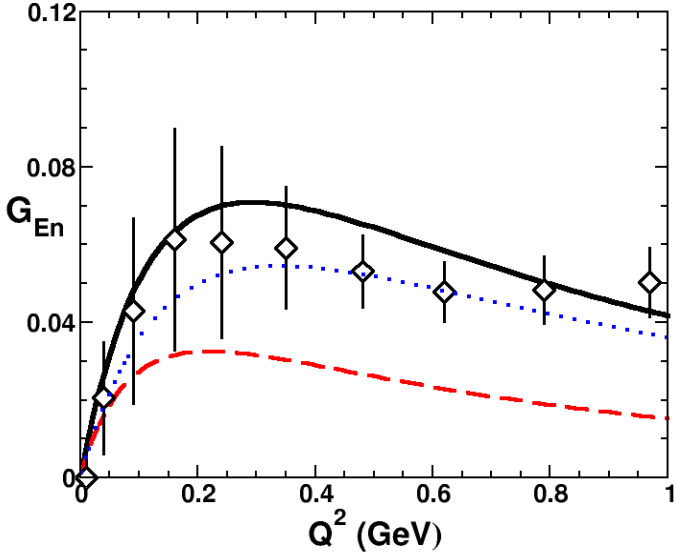}
		}
		\caption{$G_{En}$ extracted from the $ed$ elastic reaction.  $G_{En}$ using the fit form of Eq.~(\ref{eq:platchkov}) from Ref.~\cite{platchkov} fitted to $A(Q^2)$ data with either  the Nijmegen solid line (black) or a Reid soft core dashed (red) $NN$ potential in the theoretical calculation of the deuteron wave function. 
The dotted line (blue) is Eq.~(\ref{eq:galster}). The diamond points are from Ref.~\cite{schiavil}.}
		\label{fig:gen_platchov}
	\end{center}
\end{figure}

The 1971 DESY experiment of Galster {\it et al.} \cite{galster} measured elastic 
$ed$ cross sections for forward scattered electrons for $Q^2$ up to 0.6 GeV$^2$. At these kinematics, the cross section in Eq.~(\ref{eq:edxn}) is dominated by the $A(Q^2)$ term and $G_{Md}$  contributes less than 5\% to the $A(Q^2)$ term. The $A(Q^2)$ data was fitted using different deuteron wave functions and by using 
\begin{equation}
G_{Ep} = \frac{G_{Mp}}{\mu_p} = \frac{G_{Mn}}{\mu_n} = G_D,
\end{equation} 
with different parametrization of $G_{En}$. The lowest $\chi^2$ for a fit was obtained using the Feshbach-Lomon 
\cite{feshbach} deuteron wave function and the following fitting function:
\begin{equation}
G_{En}(Q^2)=-\frac{\mu_n \tau}{1+5.6\tau}G_{D}(Q^2).
\label{eq:galster}
\end{equation}
This fit is plotted in Fig.~\ref{fig:gen_platchov} as a dotted line.

The most recent experiment to measure the elastic $ed$ cross section to 
determine $G_{En}$ is that of Platchkov {\it et al.} \cite{platchkov}. 
These data extend to $Q^2$ of 0.7 
GeV$^2$, with significantly smaller statistical uncertainties than all previous 
experiments. 
The form factor  $A(Q^2)$ is very sensitive to the deuteron
wave function, and therefore to the $NN$ interaction. Furthermore, the shape of 
$A(Q^2)$ cannot be explained by the impulse approximation alone. Corrections for  meson exchange
currents (MEC) and a small contribution from relativistic effects were found 
to significantly improve the 
agreement between calculations and the measured shape of $A(Q^2)$. 
When fitting the $A(Q^2)$ data,  a modified form of the Galster fit,
\begin{equation}
G_{En}(Q^2)=-\frac{a\mu_n \tau G_{D}}{1+b\tau},
\label{eq:platchkov}
\end{equation}
was used. Several $NN$ potentials which including meson exchange currents as well as relativistic corrections were used to calculate the deuteron wave function. 
 In Fig.~\ref{fig:gen_platchov}, the fits of $G_{En}$ extracted from fitting $a$ and $b$ in Eq.~(\ref{eq:platchkov}) to the  measured $A(Q^2)$ are plotted when using the Nijmegen (black solid line) or  a Reid soft core (red dashed) $NN$ potential to calculate the deuteron wave function. Both fits to $A(Q^2)$ had similar $\chi^2$ and the spread between the line gives a sense of the theoretical uncertainty in extracting $G_{En}$ from the elastic $ed$ cross section. In 2001, an extraction of $G_{En}$ was performed using the entire elastic $ed$ cross section and polarization data \cite{schiavil} and the results are plotted in Fig.~\ref{fig:gen_platchov} as diamonds with the error bars showing the theoretical uncertainty. These data show the limit of using the $ed$ elastic reaction to determine $G_{En}$ and the need to use the quasi-elastic $ed$ polarization observables to extract $G_{En}$/$G_{Mn}$ which will be discussed in Sec.~\ref{NeutronFFpol}.

\subsection{Double Polarization Experiments}
\label{subsec:poltransfer}

Both the recoil polarization method, and the asymmetry measurement using polarized target, have been used to measure 
the proton and the neutron form factors. Here we first describe the proton form factor results; the neutron form 
factor results will be discussed in the next subsection.
 
\subsubsection{Proton Form Factors}
\label{ProtonFFpol}
The earliest polarization experiments, measuring the polarization of the recoil proton \cite{bizot}, or
measuring the asymmetry using a polarized proton target \cite{Powell:1970} with unpolarized electron beams,
were done to search for two photon effects.  

The first experiment with polarized electron beam and polarized target was done at the Stanford 
Linear Accelerator Center (SLAC) in 1970's \cite{alguard}. 
This experiment measured the beam-target asymmetry $A=\frac{\sigma_{+} -\sigma_{-}}{\sigma_{+} + \sigma_{-}}$
at $Q^2$ = 0.765 GeV$^2$. The experiment showed that the results and the theoretical values were in 
good agreement if the signs of $G_{Ep}$ and $G_{Mp}$ are the same. 

The recoil polarization method was used for the first time in an experiment at the MIT-Bates laboratory to measure
the proton form factor ratio $G_{Ep}/G_{Mp}$. This experiment determined 
$G_{Ep}/G_{Mp}$ for a free proton \cite{milbrathA,milbrathB}, as well as for a bound proton in a deuterium 
target \cite{barkhuff}, at $Q^2$-values of 0.38 and 0.5 GeV$^2$. The success of this experiment highlighted the fact that 
the recoil polarization transfer technique would be of great interest for future measurements of $G_{E}$ and 
$G_{M}$ at higher $Q^2$ values, for both the proton and the neutron. 

Next, using the same method of measuring the recoil polarization in $^1{\rm H}(\pol{e},e' \pol{p})$ reaction, 
the ratio $G_{Ep}/G_{Mp}$  was measured at MAMI at $Q^2$-values of 0.373, 0.401 and 0.441 Gev$^2$ \cite{pospischil}. The 
ratio results were found to be 
in agreement with those of Milbrath {\it et al.} \cite{milbrathA,milbrathB} as well as Rosenbluth measurements. 

In the late 1990's and 2000's measurements using the recoil polarization method 
were made at JLab in Hall A and Hall C \cite{gayou:2001,strauch,mac,hu,Paolone:2010} at low $Q^2$ values, as calibration 
measurements for other polarization experiments. Two new high precision ratio measurements at low $Q^2$ were made in Hall A at JLab; the first 
in 2006 measured the ratio in the range of $Q^2$ 
from 0.2 to 0.5 GeV$^2$ \cite{Ron:2011}, the second in 2008 measured it at $Q^2$ of 0.3 to 0.7 GeV$^2$ \cite{Zhan:2011}. 

The proton form factor ratio has also been obtained by measuring the beam-target asymmetry 
in the $^1\pol{\rm H}(\pol{e},e'p)$ reaction at a $Q^2$ of 1.51 GeV$^2$ in a Hall C  experiment at JLab in 
elastic $ep$ scattering \cite{Jones:2006}. This is 
the highest $Q^2$ at which the $G_{Ep}/G_{Mp}$ ratio has been obtained from a beam-target asymmetry measurement. The same method was 
used  by the BLAST group at MIT-Bates \cite{crawford}; this experiment measured the ratio at $Q^2$ values of 0.2 to 0.6 GeV$^2$ with 
high precision.  
\begin{figure}
\begin{center}
\resizebox{0.485\textwidth}{!}{%
\includegraphics[angle=0]{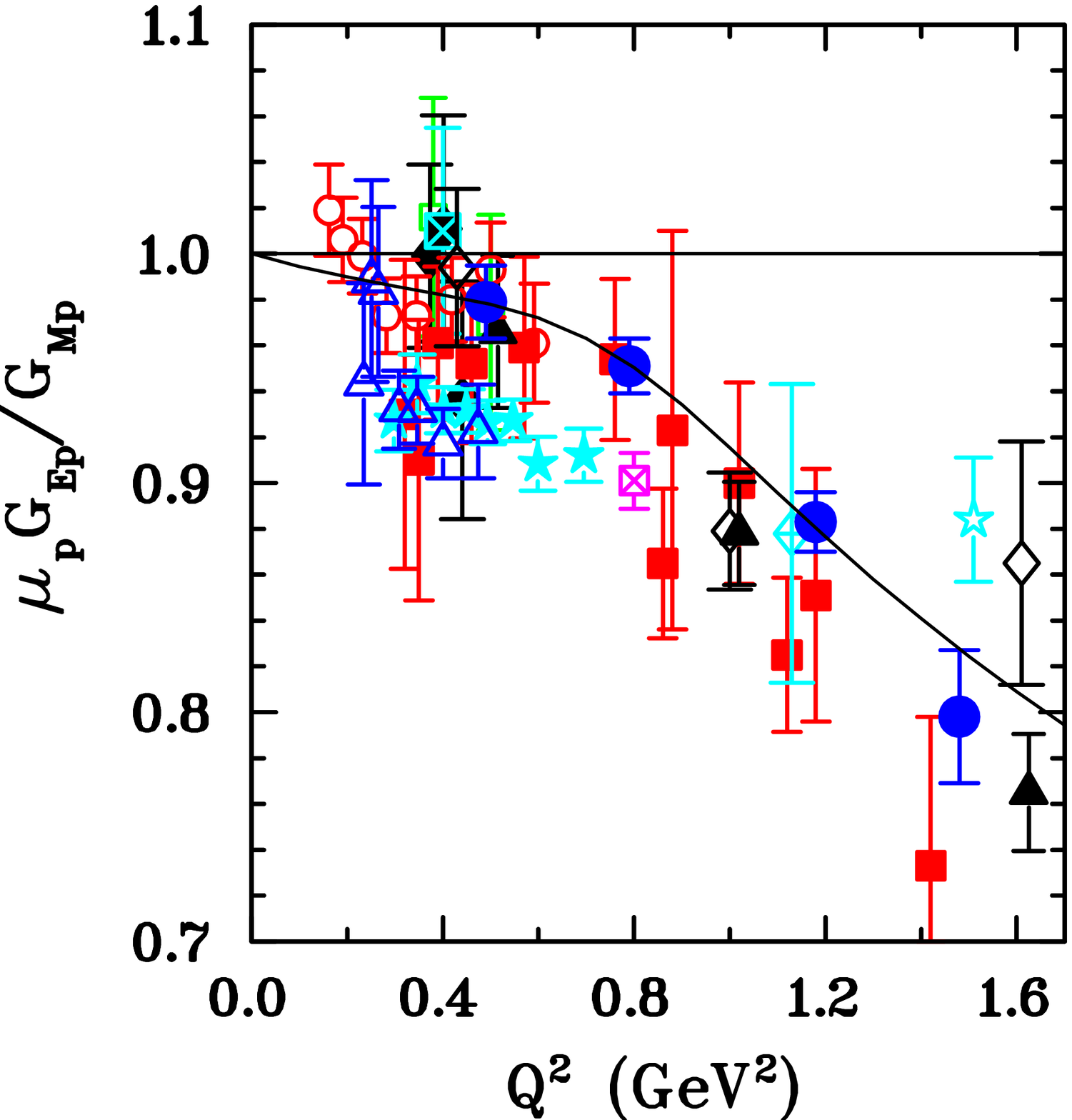}
}
\caption{Ratio $\mu_p G_{Ep}/G_{Mp}$  for $Q^2$ smaller than 1.7 GeV$^2$ from number of different experiments from several laboratories; 
Open square (green) \cite{milbrathA,milbrathB}, filled diamond (black) \cite{pospischil}, filled circle (blue) \cite{jones,punjabi05B}, filled square \cite{gayou:2001}, 
filled triangle (black) \cite{strauch}, crossed diamond (cyan) \cite{mac}, open diamond (black) \cite{hu}, crossed square (magenta) \cite{Paolone:2010}, 
open triangle (blue) \cite{Ron:2011}, filled star (cyan) \cite{Zhan:2011}, open star (cyan) \cite{Jones:2006}, open circle (red) \cite{crawford}. The 
curve is a 7 parameter 
fit given in Eq. \ref{eq:gepfit}, with ratio $\mu_p G_{Ep}/G_{Mp}$ constrained 
to 1 at $Q^2$=0; the fit is of the Kelly type  \cite{kelly04}, polynomial over polynomial, with $1/Q^2$ behavior at large $Q^2$.}
\label{fig:gepgd_low_pol}
\end{center}
\end{figure}

Figure \ref{fig:gepgd_low_pol} shows all the low $Q^2$ data obtained from recoil polarization experiments 
\cite{milbrathA,jones,punjabi05B,pospischil,gayou:2001,strauch,mac,hu,Paolone:2010,Ron:2011,Zhan:2011} 
and beam-target asymmetry measurements \cite{Jones:2006,crawford} obtained at  MIT-Bates, MAMI, and JLab. 
As can be seen from figure \ref{fig:gepgd_low_pol}, data from different experiments are in general agreement. 
The slow decrease of the data starts at $Q^2 \approx$ 0.5 GeV$^2$ and continues to 1.7 GeV$^2$. 

A real break-through was made towards the understanding of the internal structure of the proton, when 
two JLab Hall A and one Hall C experiments obtained the elastic electromagnetic 
form factor ratio of the proton, $G_{E}^{p}/G_{M}^{p}$ at $Q^2$'s larger than 1 GeV$^2$, from the measured
recoil proton polarization components $P_t$ and $P_{\ell}$, using the recoil polarization method. 
The first of these experiments measured the proton form factor ratios for $Q^{2}$ 
from 0.5 to 3.5 GeV$^{2}$ in 1998 \cite{jones,punjabi05B}, the second from 4.0, 4.8 and 5.6 GeV$^2$ 
in 2000 \cite{gayou:2002,Puckett:2011} 
and the third in 2007-8 up to 8.4 GeV$^2$ \cite{Puckett:2010}. 

In the first JLab experiment GEp(1), elastic $ep$ events were selected by detecting the scattered electrons and 
the recoiling protons in coincidence, using the two identical high-resolution
spectrometers (HRS) of Hall A \cite{nimhallA}. One of the HRS was equipped with a focal plane polarimeter (FPP) 
to detect the polarization of the recoil protons. The FPP consisted of two front detectors to track incident protons, followed by a graphite 
analyzer and two rear detectors to track scattered particles. The polarization of the recoiling 
proton was obtained from the asymmetry of the azimuthal distribution of the proton after re-scattering in the graphite analyzer
of the polarimeter.

In the second JLab experiment, GEp(2), the ratio, $G_{Ep}/G_{Mp}$ was measured at $Q^{2}$ = 4.0, 4.8 and 5.6 GeV$^{2}$ with
an overlap point at $Q^{2}$ = 3.5 GeV$^{2}$ ~\cite{gayou:2002,Puckett:2011}. 
Several changes were made compared to the first experiment, to extend the measurement to
higher $Q^2$. First, to increase the coefficient-of-merit (COM) of the focal plane 
polarimeter (FPP), a CH$_{2}$ analyzer was used instead of the graphite; hydrogen has 
much higher analyzing power \cite{spinka,dmiller} than carbon \cite{cheung}; and, to 
increase the fraction of events with the proton interacting in the analyzer, the
thickness of the analyzer was increased from 50 cm of graphite to 100~cm of CH$_{2}$.
Second, to achieve complete solid angle matching with the 
HRS detecting the proton and determining its polarization, a large frontal area lead-glass calorimeter was constructed
 and replaced the second HRS used in GEp(1).
At the largest $Q^{2}$ of GEp(2) of 5.6 GeV$^2$, the solid angle of the electromagnetic calorimeter was 6 times that of the HRS.  

\begin{figure}
\begin{center}
\includegraphics[width = 85 mm]{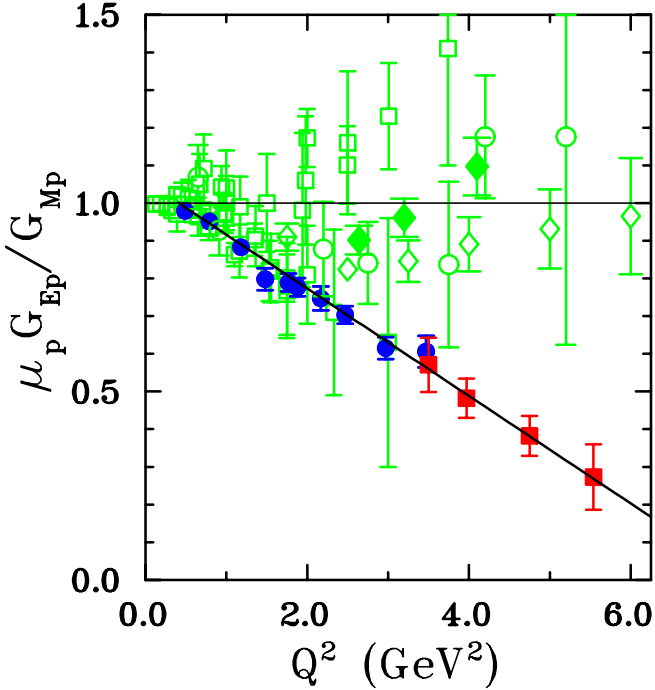}
\caption{The ratio $\mu_p G_{Ep}/G_{Mp}$ from the two JLab experiments 
filled circle (blue) \cite{jones,punjabi05B} and filled square (red) \cite{gayou:2002}, compared to Rosenbluth separation results in (green), 
open diamond \cite{andivahisA}, open circle \cite{christy}, filled diamond \cite{qattan05}, and 
open square \cite{price,litt,berger,bartel,walker}. The fit to the data is as in Gayou {\it et al.} \cite{gayou:2002}.} 
\label{fig:gepgmp_pol_cs}
\end{center}
\end{figure} 

The results from the first two JLab experiments \cite{jones,punjabi05B,gayou:2002,Puckett:2011}, are plotted 
in Fig.~\ref{fig:gepgmp_pol_cs} as the ratio $\mu_{p}G_{Ep}/G_{Mp}$ versus $Q^2$, where they are 
compared with Rosenbluth separation data \cite{price,litt,berger,bartel,walker,andivahisA,christy,qattan05}.
As can be seen from this figure, for the polarization data at the larger $Q^2$'s the statistical
uncertainties are small, unlike those of the  cross sections data, underlining
the difficulties in obtaining $G_{Ep}$ by the Rosenbluth separation method at larger $Q^2$'s; the Rosenbluth data also show a large scatter 
among the results from different experiments. 

The $\mu_{p}G_{Ep}/G_{Mp}$ ratio results from JLab \cite{jones,punjabi05B,gayou:2002} showed conclusively for the first
time, a clear deviation of this ratio from unity, starting at $Q^2\simeq 1$~GeV$^2$; older data from \cite{berger,price,bartel}
showed such a decreasing ratio, but with much larger statistical and systematic uncertainties, as seen in Fig. \ref{fig:gepgmp_pol_cs}. 
The most important feature of the JLab data is the sharp decrease of the ratio
$\mu_p G_{Ep}/G_{Mp}$ from 1, starting at $Q^2$ $\approx$ 1 GeV$^2$ to a value of
$\sim 0.35$ at $Q^2$= 5.6 GeV$^2$, indicating that $G_{Ep}$ falls 
faster with increasing $Q^2$ than $G_{Mp}$,  thus clearly highlighting a definite difference between the spatial distributions of charge and 
magnetization at short distances. This was the first definite experimental indication
that the $Q^2$ dependence of $G_{Ep}$ and $G_{Mp}$ are different. These results were very surprising at the time (1998-2002), as 
they appeared to contradict the previously accepted belief that the ratio $\mu_{p}G_{Ep}/G_{Mp}$ remains close to 1, a consensus 
based on the Rosenbluth separation results up to 6 GeV$^2$, as illustrated in Fig.~\ref{fig:gepgmp_pol_cs}.  

As discussed above, the two methods available to determine the proton form factors $G_{Ep}$ and $G_{Mp}$,
the Rosenbluth separation and polarization transfer,  give definitively 
different results; the difference cannot be bridged by either simple re-normalization 
of the Rosenbluth data \cite{arring03}, or by variation of the polarization data within the 
quoted statistical and systematic uncertainties. This discrepancy has been known
for sometime now, and has been the subject of extensive discussion and investigation. 
A possible explanation is the contribution from the hard two-photon exchange process, which affects the polarization transfer components at the 
level of only a few percent, but has drastic effects on the  Rosenbluth 
separation results. This will be discussed in detail in section \ref{subsec:discrepancies}.

Following the unexpected results from the two first polarization transfer experiments in Hall A at JLab, GEp(1)
and GEp(2), a third experiment in Hall C, GEp(3), was carried out to extend the $Q^2$-range to $\approx$ 9 GeV$^2$. 
Two new detectors were built to carry out this experiment: a large solid-angle electromagnetic 
calorimeter and a double focal plane polarimeter (FPP). The recoil protons were detected in the 
high momentum spectrometer (HMS) equipped with two new FPPs in series. The scattered electrons were
detected in a new lead glass calorimeter (BigCal) built for this purpose out of
1744 glass bars, 4x4 cm$^2$ each, and a length of 20$X_0$, with a total frontal area of 2.6 m$^2$ 
which provided complete kinematical matching to the HMS solid angle. 
This experiment was completed in the spring of 2008 and measured the form 
factor ratio at $Q^2$ of 5.2, 6.7 and 8.5 GeV$^2$. 

\begin{figure}
\begin{center}
\includegraphics[width=\columnwidth,angle=0]{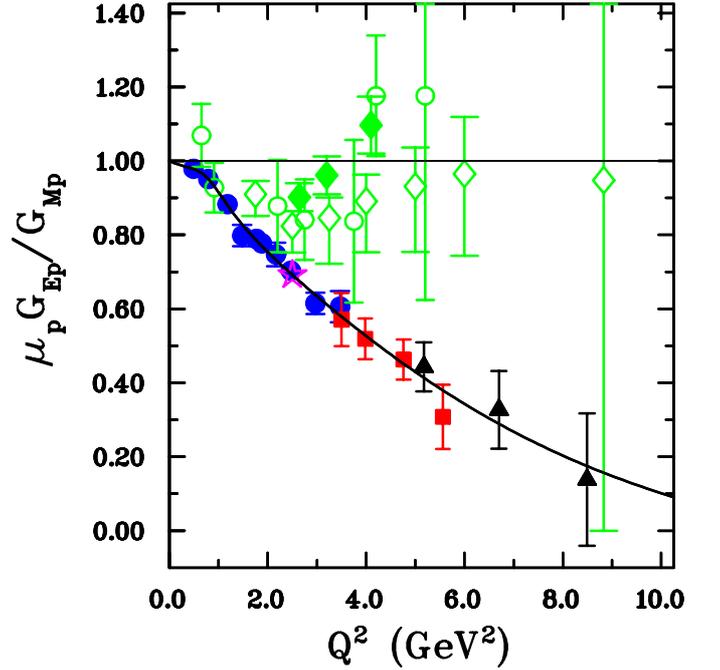}
\caption{All data for the ratio $\mu_p G_{Ep}/G_{Mp}$ obtained from the three large $Q^2$ recoil polarization experiments at 
JLab (filled circle (blue) \cite{punjabi05B}, filled star (magenta) \cite{Meziane:2010}, filled square (red) \cite{Puckett:2011} and 
filled triangle (black) \cite{Puckett:2010}) compared to Rosenbluth
separation data (green), open diamond \cite{andivahisA}, open circle \cite{christy}, filled diamond \cite{qattan05}. The curve is the same as in 
Figure \ref{fig:gepgd_low_pol}, a 7 parameter fit given in Eq. \ref{eq:gepfit}.} 
\label{fig:gepgmp_large_qsqr_pol}
\end{center}
\end{figure}

Figure \ref{fig:gepgmp_large_qsqr_pol} shows the results from the three JLab 
experiments \cite{jones,gayou:2002,Puckett:2011,punjabi05B,Puckett:2010}, as the 
ratio $\mu_{p}G_{Ep}/G_{Mp}$ versus $Q^2$. The uncertainties shown for the recoil polarization data are statistical 
only. 

The striking feature of the results of the GEp(3) experiment is the continued, strong and almost linear
decrease of the ratio with increasing $Q^2$, albeit with some indication of a slowdown at the highest $Q^2$. 
The GEp(3) overlap point at $5.2$ GeV$^2$ is in good agreement with the two 
surrounding points from the GEp(2) data ~\cite{gayou:2002,Puckett:2011}. The GEp(3) experiment used a completely different 
apparatus in a $Q^2$ range where direct comparison with the Hall A recoil polarization 
results from the GEp(2) experiment is possible. This comparison provides an important confirmation of the reproducibility 
of the results obtained with the recoil polarization technique.
Additionally, the results of the high-statistics survey of the $\epsilon$-dependence 
of $G_{Ep}/G_{Mp}$ at $Q^2=2.5$ GeV$^2$, obtained from the GEp($2 \gamma$) experiment \cite{Meziane:2010}, which ran at 
the same time as the GEp(3) experiment is shown as a magenta star in  
Fig.~\ref{fig:gepgmp_large_qsqr_pol}, and is in excellent agreement with the results from the GEp(1) experiment 
in Hall A ~\cite{jones,punjabi05B} at $Q^2=2.47$ GeV$^2$.

The results of the three JLab GEp experiments are the most precise measurements to date of the proton form factor 
ratio in this range of $Q^2$, hence they represent a very significant advancement of the experimental knowledge of
the structure of the nucleon. The proton electromagnetic form factor results from Jefferson Lab 
at high values of the four-momentum transfer $Q^2$ have had a big impact on progress in hadronic physics; these results have 
required a significant rethinking of nucleon structure which will be discussed in the theory section.

\subsubsection{Neutron Form Factors}
\label{NeutronFFpol}

The early measurements of the form factors of the neutron are discussed in section 
\ref{subsubsec:gmn}; in this section only double polarization measurements are discussed.
The recoil polarization and beam-target asymmetry, both techniques that have been used 
to measure $G_{Ep}$ and $G_{Mp}$, also have been used to measure $G_{En}$ and $G_{Mn}$. However, as there 
are no free neutron targets, measurements of $G_{En}$ and $G_{Mn}$ are more difficult than $G_{Ep}$ and $G_{Mp}$.
To make these measurements, complex light targets like $^2H$ and $^3He$ must be used in quasi elastic scattering. 
First, the recoil polarization experiments, and next the beam-target asymmetry experiments to extract $G_{En}$, will be described. 

The use of the recoil polarization technique to measure the neutron charge form factor was made first at the MIT-Bates laboratory
in the late 80's using the exclusive $^2{\rm H}(\pol{e},e'\pol{n})p$ reaction \cite{eden}. The advantage of using a 
deuteron target is that theoretical calculations predict the extracted neutron form factor results to be 
insensitive to effects like, final state interaction (FSI), meson exchange currents (MEC), isobar 
configurations (IC), and to the choice of the deuteron wave function \cite{arenhovelA,rekalo2,laget}. 
In this experiment, the neutron form factor  $G_{En}$ was obtained from the measured transverse 
polarization component $P'_t$ of the recoiling neutron, 
and known beam polarization, $P_e$, at a $Q^2$ of 0.255 GeV$^2$. 
The relation between the polarization transfer coefficient $P_{t}$,  
the beam polarization, $P_e$, and the measured neutron polarization component, $P'_t$, is 
$P'_t=P_e P_{t}$, the polarization transfer coefficient $P_{t}$ given by Eq.~(\ref{eq:plpt}), is for a free neutron.
This early experiment  demonstrated the feasibility of extracting 
$G_{En}$ from the quasi-elastic $^2{\rm H}(\pol{e},e'\pol{n})p$ reaction with the recoil polarization
technique, with the possibility of extension to larger $Q^2$ values.

The recoil polarization transfer method was next used at MAMI \cite{herberg,ostrick} using the same 
reaction $^2{\rm H}(\pol{e},e'\pol{n})p$ to determine $G_{En}$, at a $Q^2$
of 0.15 and 0.34 GeV$^2$. However, in this experiment the
recoil neutron polarization components $P_t$ and $P_l$ were measured simultaneously, using a dipole with vertical B-field to 
precess the neutron polarization in the reaction plane; the ratio $P_t/P_l$, is related directly
to $G_E/G_M$ as shown in Eq.~(\ref{eq:ratio}), again for a free neutron.  
As discussed earlier for the proton, the measurement of the ratio $P_t/P_l$, has some  
advantage over the measurement of $P_t$ only; in the ratio the 
electron beam polarization and the polarimeter analyzing power cancel; as a result the systematic
uncertainty is small. Also the model dependence for a bound neutron, which occurs via the
dependence of the neutron wave function on the nuclear
binding, cancels in these polarization observables in leading
order for the extraction of the form factor. 

Next, the electric form factor $G_{En}$ was obtained at $Q^2$ = 0.3, 0.6 and 0.8 GeV$^2$ 
from the measured ratio of polarization transfer components, $P_t/P_l$, in another experiment at MAMI \cite{glazier:2004};
Glazier {\it et al.} concluded that the results from this experiment were in good agreement with all 
other $G_{En}$ results from double-polarization measurements.


The first double polarization experiment to measure $G_{En}$
at JLab by Madey {\it et al.} \cite{madey,plaster}, obtained the neutron form factor ratios
$G_E/G_M$ at $Q^2$ values of 0.45, 1.13 and 1.45 GeV$^2$ using the same method of measuring the
recoil neutron polarization components $P_t$ and $P_l$ simultaneously. 
The neutron charge form factor $G_{En}$ was calculated from the measured ratio, using the best-fit values of $G_{Mn}$.
This was the first experiment that determined $G_{En}$ with small statistical and systematic 
uncertainty to relatively high $Q^2$. 
Madey {\it et al.} concluded that a successful theoretical models must be able to
predict both neutron and proton electromagnetic form factors simultaneously. The neutron electric form factor
is more sensitive to small components of the nucleon wave function, and differences between model predictions
for $G_{En}$ tend to differ with increasing $Q^2$;  hence the new data from this experiment to larger $Q^2$ provided a challenging test 
for theoretical model calculations.  


\begin{figure}
\begin{center}
\resizebox{0.45\textwidth}{!}{%
\includegraphics[angle=0]{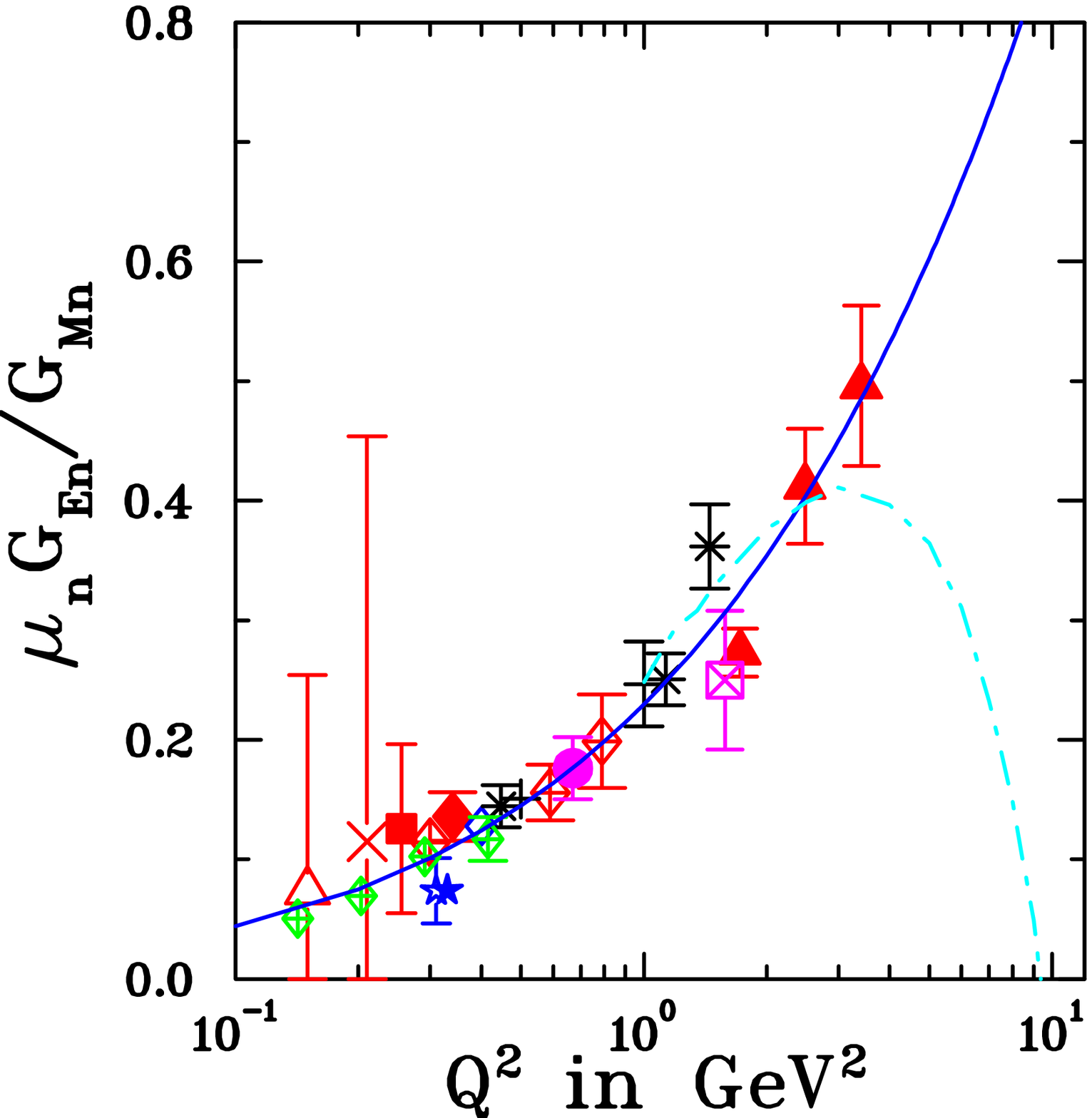}
}
\caption{The complete data base for $G_{En}/G_{Mn}$ from double polarization experiments; recoil polarization with deuterium 
target \cite{eden:1994,herberg,ostrick,glazier:2004,madey,plaster}, asymmetry with polarized 
deuterium target: \cite{passchier,zhu,warren,geis:2008}, and asymmetry with polarized $^3He$ 
target: \cite{cejones,thompson,meyerhoff:1994,becker,rohe,bermuth,golak:2001,Riordan:2010,schlimme:2013}. 
The symbols for the data are filled square (red) \cite{eden}, filled diamond (red) \cite{ostrick}, open triangle (red) \cite{herberg}, 
multiplication sign (red) \cite{passchier}, crossed diamond (red) \cite{glazier:2004}, open star (blue) \cite{meyerhoff:1994},
filled star (blue) \cite{becker}, open diamond (blue) \cite{golak:2001}, filled circle (magenta) \cite{bermuth}, asterisk (black) \cite{madey},
plus sign (black) \cite{warren}, crossed diamond (green) \cite{geis:2008}, filled triangle (red) \cite{Riordan:2010}, cross diamond (red) \cite{schlimme:2013}. The solid (blue)
line is a fit using Eq.~(\ref{eq:genfit}). $G_{En}/G_{Mn}$ from  the DSE model of Ref.~\cite{cloet:2008} are plotted as a dash-dotted (cyan) line.}
\label{fig:gen_pol}
\end{center}
\end{figure} 

The first measurement of $G_{En}$ using the  beam-target asymmetry was made at NIKHEF 
at a $Q^2$ of 0.21 GeV$^2$ using the $^2\pol {\rm H}(\pol {e},e'n)p$ reaction \cite{passchier}. 
The experiment used a polarized electron beam from the storage ring and a vector polarized 
deuterium gas target, internal to the storage ring; $G_{En}$ was extracted from the measured sideways 
spin-correlation parameter in quasi-free scattering. Passchier {\it et al.} concluded that their result puts
strong constraints on $G_{En}$ up to $Q^2$ = 0.7 GeV$^2$ when combined with the 
measured $G_{En}$ slope from Kopecky {\it et al.} \cite{Kopecky} at $Q^2$=0 GeV$^2$ and the elastic electron-deuteron 
scattering data from Platchkov {\it et al.} \cite{platchkov}. 

The neutron electric form factor at $Q^2$ = 0.5 and 1.0 GeV$^2$ was extracted from 
measurements of the beam-target asymmetry using the $^2\pol{\rm H}(\pol {e},e'n)p$ reaction in 
quasi elastic kinematics, at JLab in Hall C \cite{zhu,warren}; the 
polarized electrons were scattered off a solid polarized deuterated ammonia (${ND_3}$) target
in which the deuteron polarization was perpendicular to the momentum transfer. 
This was the first experiment to obtain $G_{En}$ at a relatively large $Q^2$ using a polarized target. 

There was a measurement of $G_{En}/G_{Mn}$ at the MIT-Bates lab in mid 2000's, using
a longitudinally polarized electron beam and a vector-polarized $^2H$ target internal to the
storage ring over a range of $Q^2$ between 0.10 and 0.55 GeV$^2$ \cite{geis:2008}; in this experiment the quasi-elastically scattered
electrons were detected in coincidence with recoil neutrons in the BLAST detector. They used the parametrization of 
Friedrich and Walcher \cite{Friedrich} for $G_{Mn}$ to calculate $G_{En}$.  Geis {\it et al.} \cite{geis:2008} concluded that their data are 
in excellent agreement with VMD based models of Lomon \cite{lomon} and Belushkin \cite{Belushkin:2006qa}, and also 
agree with the meson-cloud calculation
of Miller \cite{miller02b}. 

All the experiments described above used either a polarized or an unpolarized deuterium target. 

In 1984 Blankleider and Woloshyn \cite{blankleider}, proposed that to 
measure $G_{En}$ or $G_{Mn}$, a polarized $^3$He target could be used. 
Their argument was that the ground state of $^3$He  is dominated by the 
spatially symmetric S-state in which the  two proton spins point in opposite directions, hence the spin of 
the nucleus is largely carried by the neutron. Therefore, polarized $^3$He target effectively serve as a 
polarized neutron target; and in the quasi-elastic scattering region the spin-dependent properties 
are dominated by the neutron in the $^3$He target. 

The first two experiments that used a polarized $^3$He target and
measured the asymmetry with polarized electrons in spin-dependent quasi-elastic scattering were done 
at MIT-Bates Laboratory \cite{cejones,thompson}; these experiments extracted the value of $G_{En}$
at a $Q^2$=0.16 and 0.2 GeV$^2$, using the model of Blankleider and Woloshyn \cite{blankleider}. 
However, Thompson {\it et al.} \cite{thompson} pointed out  
that significant corrections were necessary at $Q^2$=0.2 GeV$^2$, for spin-dependent 
quasi elastic scattering on polarized $^3$He according to the calculation of Laget \cite{laget};
hence no useful information on $G_{En}$ could be extracted from these measurements; but Thompson {\it et al.} \cite{thompson}
concluded that at higher $Q^2$ values the relative contribution of the 
polarized protons becomes significantly less and a precise measurements of $G_{En}$ using polarized 
$^3$He target will become possible. 

The neutron electric form factor $G_{En}$ was  obtained in the early 1990's in several 
experiments at MAMI; these experiments measured the beam-target asymmetry 
in the exclusive quasi-elastic scattering of polarized electrons from polarized $^3$He in the 
$^3\pol{\rm He}(\pol{e},e'n)pp$ reaction \cite{meyerhoff:1994,becker,rohe,bermuth}. 
The first of these experiments at MAMI obtained  $G_{En}$ at $Q^2$ = 0.31 GeV$^2$ \cite{meyerhoff:1994}; and the next 
experiment measured $G_{En}$ at $Q^2$ of 0.35 GeV$^2$ \cite{becker} and 0.67 GeV$^2 $ \cite{rohe,bermuth} using the 
same experimental setup. The value of $G_{En}$ at $Q^2$ of 0.35 GeV$^2$ \cite{becker} was later 
corrected by Golak {\it et al.} \cite{golak:2001}, based 
on Faddeev solutions and with some MEC corrections. The size of these corrections is expected to 
decrease with $Q^2$, although the corrections become increasingly difficult to calculate with increasing $Q^2$.

The GEn(1) experiment in Hall A at JLab measured the ratio $G_{En}/G_{Mn}$ in 2006 
at a $Q^2$ = 1.72, 2.48, and 3.41 GeV$^2$ using the reaction $^3\pol{\rm He}(\pol{e},e'n)pp$ in quasi-elastic 
kinematics \cite{Riordan:2010}. Longitudinally polarized electrons were scattered off a
polarized target in which the nuclear polarization was oriented perpendicular to the momentum transfer. The scattered electrons 
were detected in a magnetic spectrometer in coincidence with knocked out neutrons, that were detected in a large hadron detector. The ratio
$G_{En}/G_{Mn}$ was obtained from the measured beam-target asymmetry. Riordan {\it et al.} \cite{Riordan:2010} concluded that this 
experiment more than doubled the $Q^2$
range over which $G_{En}$ is known, and this fact greatly sharpens the mapping of the nucleon's constituents, and provides a
new benchmark for comparison with theory.

There is yet another recent experiment at MAMI, that measured the charge form factor of the neutron at a $Q^2$ of 1.58 GeV$^2$
using the polarized $^3$He target and longitudinally polarized electron beam \cite{schlimme:2013}.  To reduce systematic errors, 
data were taken for four different target polarization orientations. The data of this experiment are in very good agreement with the data of 
Riordan {\it et al.} \cite{Riordan:2010}.

Figure~\ref{fig:gen_pol} shows the ratio $G_{En}/G_{Mn}$  versus $Q^2$ obtained from all double polarization experiments in the last two decades.
The results from each experiment are shown as different symbols as explained in the figure caption. The solid curve is a fit to the data as given by 
Eq.~(\ref{eq:genfit}).

\begin{figure}
	\begin{center}
		\resizebox{\columnwidth}{!}{%
			\includegraphics[angle=0]{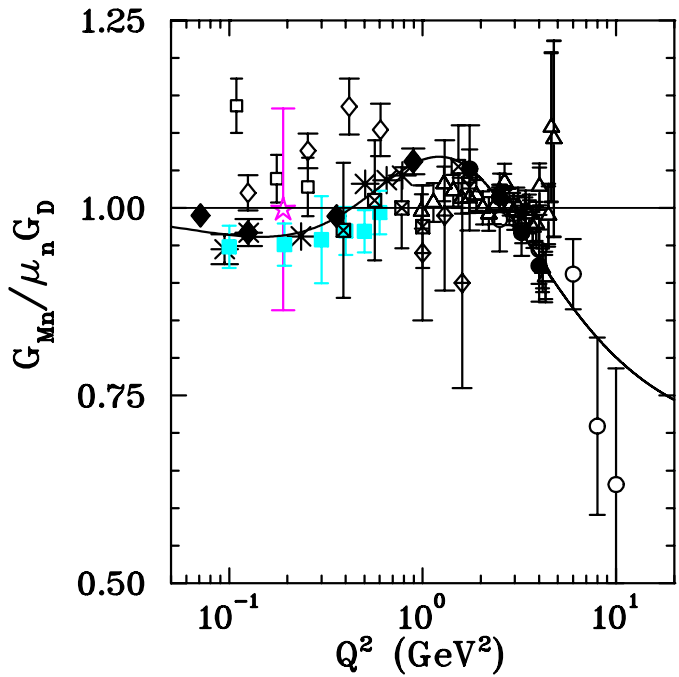}
		}
		\caption{$G_{Mn}/\mu_n$$G_D$ versus $Q^2$. The data from polarization experiments: the open star (magenta) \cite{gao1}, and the 
                        filled square (cyan) \cite{anderson}. The data from 
			cross sections (black) are same as in Figure \ref{fig:gmn_nopol}.The solid line is a fit by Kelly \cite{kelly04}.}
		\label{fig:gmn_pol}
	\end{center}
\end{figure}

The first experiment to obtain the magnetic form factor of the neutron, $G_{Mn}$, from polarization observables
was done at the MIT-Bates laboratory. This experiment obtained
$G_{Mn}$ from the measured beam-target asymmetry in inclusive quasi-elastic scattering 
of polarized electrons from polarized $^3$He target at $Q^2$ of 0.19 GeV$^2$ \cite{gao1}; the uncertainty 
on $G_{Mn}$ was dominated by the statistics, with a relatively small contribution from model dependence of the analysis. 
The second experiment was done at JLab in Hall A; this experiment extracted $G_{Mn}$ for $Q^2$ values between 0.1 and 0.6 GeV$^2$,
by measuring the transverse asymmetry in the $^3\pol{\rm He}(\pol{e},e')$ reaction 
in quasi-free kinematics \cite{xu00,xu02,anderson}. The values of $G_{Mn}$ were obtained with a full Faddeev calculation 
at $Q^2$ of 0.1 and 0.2 GeV$^2$, and in the plane wave impulse approximation (PWIA) at $Q^2$ of 0.3 to 0.6 GeV$^2$. 
It was argued by the authors of this paper that the PWIA extraction of $G_{Mn}$ is reasonably reliable in  the $Q^2$ range of 0.3 to 
0.6 GeV$^2$; however, a more precise extraction of $G_{Mn}$ requires fully relativistic three-body 
calculations. The $G_{Mn}$ data from both double polarization experiments \cite{gao1,xu00,xu02,anderson}
are shown in Fig. \ref{fig:gmn_pol}, together with the data from earlier unpolarized measurements 
\cite{rockB,lung,marko,bruins,anklin,Anklin:1998ae,kubon,lachniet:2008,bartel,Arnold:1988us}. There is some scatter in the data, but they are
close to $G_{Mn}/\mu_n$$G_D ~\approx~ 1.0$, except two open circles data point at a larger $Q^2$ values.

\subsection{Discrepancies between Cross Section and Polarization Results}
\label{subsec:discrepancies}

As discussed in Sec.~\ref{ProtonFFpol} and illustrated in Figs. \ref{fig:gepgmp_pol_cs} and \ref{fig:gepgmp_large_qsqr_pol}, there is a clear discrepancy between the extracted values of the proton form factor ratio, $G_{Ep}/G_{Mp}$, from polarization transfer and double polarization experiments, and those obtained from cross section measurements using the Rosenbluth separation technique.  As also mentioned previously, one possible explanation for this discrepancy is related to the hard two-photon exchange process; more properly, it is the interference between single-photon and two-photon exchange processes that has the potential to complicate the extraction of the form factors in Rosenbluth separation experiments. 

Two groups simultaneously suggested that the difference between cross section and double polarization results might be due to
previously neglected two-hard-photon exchange processes; these were Guichon and Vanderhaeghen \cite{guichon}, 
and Blunden $et~al.$ \cite{blundenA}.  In general, cross section data
 require large radiative  corrections, whereas
double-polarization ratios do not.   This is because radiative corrections affect the longitudinal and transverse polarization observables 
similarly, and thus the residual correction for double polarization is at the few percent level.  A number
 of calculations of the two-hard-photon contribution have been published over the last decade.  A partial list
 of calculations of the contribution of the two-hard-photon process to the cross section includes
 \cite{afanbrod,arring04,Kondratyuk:2005kk,bystritskiy:2007,vanderhaeghen00,carlson:2007twophoton}.
 
The signature of two-photon exchange processes would be an observed $\epsilon$-dependence; in the case of cross section experiments the effect is an non-linearity in the reduced cross section as a function of $\epsilon$, whereas in the polarization transfer experiments, one might expect an $\epsilon$-dependence in the ratio of the measured proton polarization components.  There have been several experimental attempts over the last decade to search for these dependencies.

\subsubsection{Results from the $G_{Ep}(2\gamma)$ Experiment}
\label{subsubsec:gep2g}

The $G_{Ep}(2\gamma)$ experiment in Hall C at JLab recently measured the
ratio $-\sqrt{(1+\epsilon)/2\epsilon}(P_t/P_\ell$), which strictly equals $G_{Ep}/G_{Mp}$ in the
Born approximation, at a central value $Q^2$=2.49 GeV$^2$,
and for three values of $\epsilon$: 0.152, 0.635 and 0.785, with 
very small error bars \cite{Meziane:2010}. $P_t$ and $P_{\ell}$ are the transverse and longitudinal 
components of the polarization
transferred to the proton, respectively.  Simultaneously, values of $P_{\ell}/P_{\ell,Born}$ were 
obtained at the two larger $\epsilon$ values, using the lowest $\epsilon$ data point to
determine the analyzing power of the polarimeter for the common central proton momentum of 2.06~GeV/c.  For these data, 
radiative corrections were calculated using the model independent calculation of \cite{afanasev:radcor}.  The calculation includes
the vacuum polarization correction, the electron vertex correction, and internal and external bremsstrahlung corrections. In general, polarization observables are rather insensitive to radiative corrections, as they are by definition a ratio of a polarized cross section to an unpolarized one.  Moreover, the bremsstrahlung correction can be reduced drastically by applying missing mass or
inelasticity cut, as was done in this experiment through elastic event selection.  

\begin{figure}
\centering
\includegraphics[angle=90,width=0.45\textwidth]{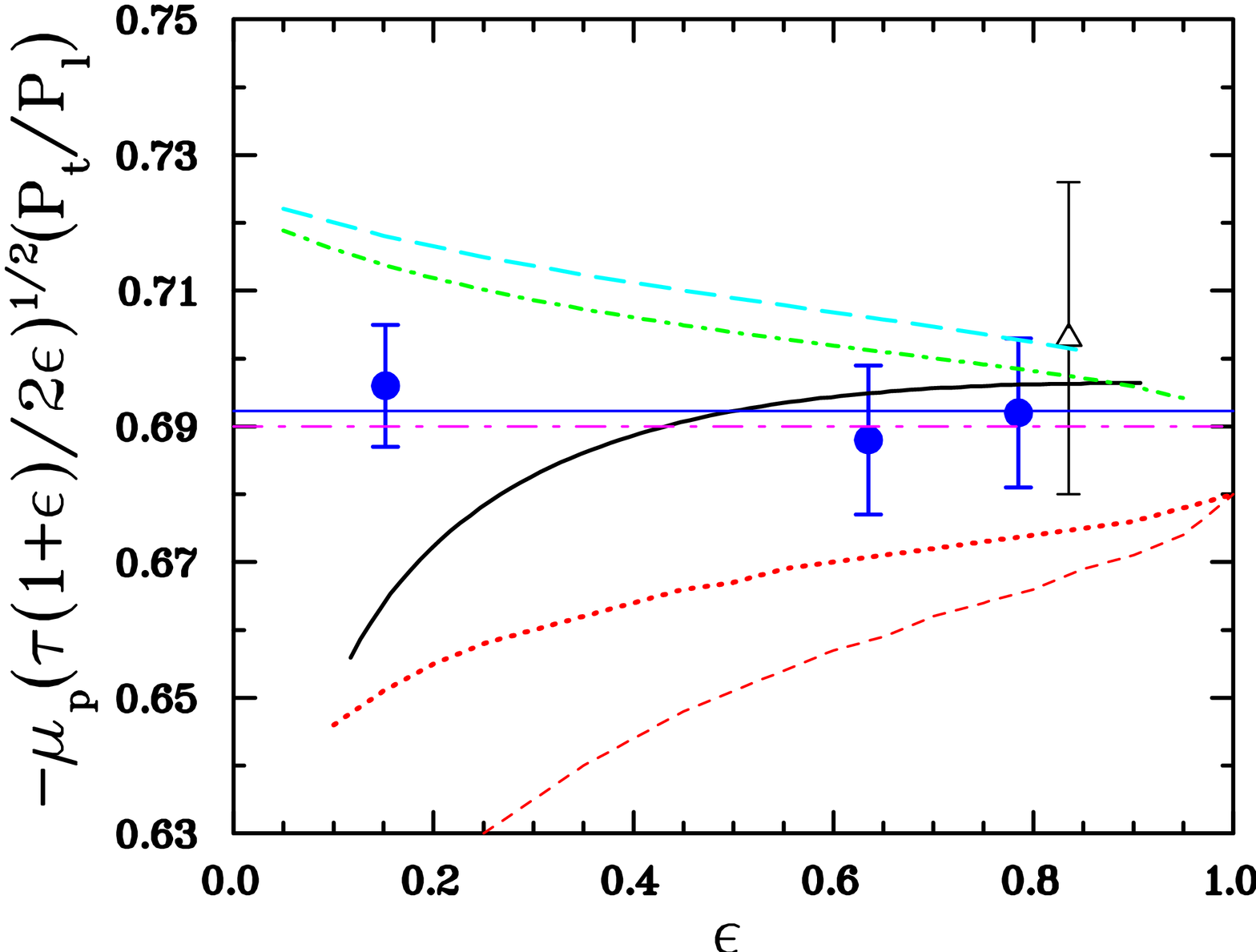}
\includegraphics[angle=90,width=0.45\textwidth]{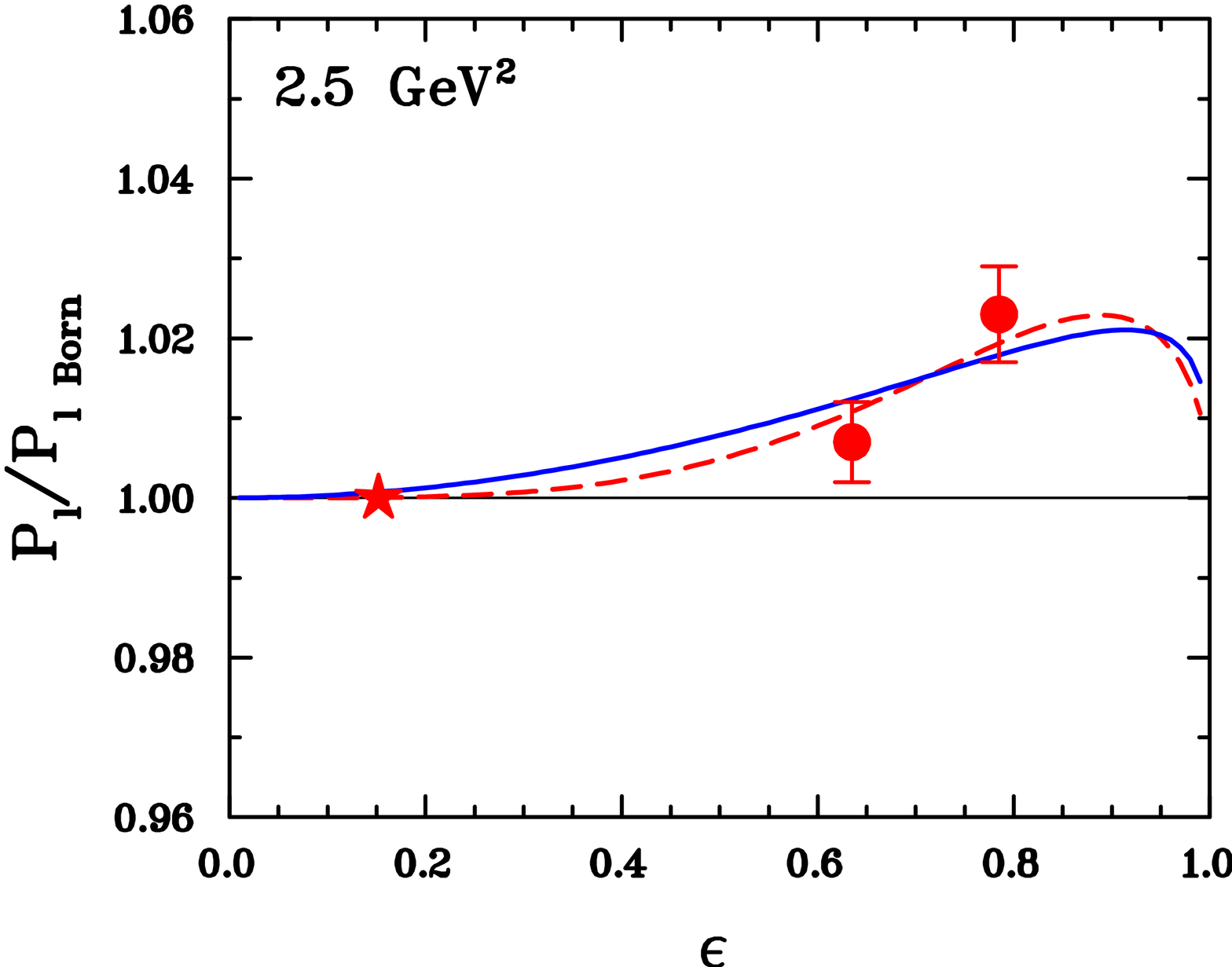}
\caption{\label{gep2gammafig} Results of the $G_{Ep}(2\gamma)$ experiment at JLab \cite{Meziane:2010}. The ratio plotted in the top panel 
is equal to $\mu_pG_{Ep}/G_{Mp}$ in the Born (one-photon-exchange) approximation with the data of Ref.~\cite{Meziane:2010} plotted as filled circles (blue) and
the data point of Ref.~\cite{punjabi05B} plotted as empty triangle (black).
The theoretical curves are:  long dot-dashed (magenta)~\cite{bystrit}, short dot-dashed (green) ~\cite{blundenB}, long 
dashed (cyan) ~\cite{Borisyuk:2014ssa,Borisyuk:2013hja}, solid (black)~\cite{afanbrod}, dotted (red) ~\cite{Guttmann:2010au} with 
wave functions from~\cite{braun06}, and short-dashed (red)~\cite{Guttmann:2010au} with wave functions from~\cite{Chernyak:1987nu}.
The bottom panel shows $P_{\ell}/P_{\ell,Born}$ for the two higher $\epsilon$ points of the experiment plotted as filled circles (red). The lowest 
$\epsilon$ point was used to determine the polarimeter analyzing power. The curves are fits from~\cite{Guttmann:2010au}; 
the dashed (red) curve corresponds to "Fit1" and the solid (blue) curve corresponds to "Fit2" as specified in this reference.} 
\end{figure}

The results of these two measurements are shown in Fig.~\ref{gep2gammafig}. The measured ratio 
$-\sqrt{(1+\epsilon)/2\epsilon}(P_t/P_\ell)$
appears to have no $\epsilon$ dependence within the small statistical and systematic uncertainties of the experiment.  In contrast, the ratio $P_{\ell}/P_{\ell,Born}$, displayed in the bottom panel of Fig.~\ref{gep2gammafig}, shows a systematic deviation
 from unity at the largest
$\epsilon$ value of up to 4.5 standard deviations. Such a behavior can be explained (see the curve in lower panel
 of Fig.~\ref{gep2gammafig}) within the context of recent work described in \cite{Guttmann:2010au} which shows that the corrections to the three form factors required in the presence of the 
interference of the one- and two-photon terms do not cancel one another as $\epsilon\rightarrow 1$. 

\subsubsection{Results from $e^+/e^-$ Scattering Experiments}
\label{subsubsec:hallb}

The most direct way to characterize a hard two-photon contribution to the elastic {\it ep} cross section is to
compare $e^+p$ and $e^-p$ scattering.  There are recent results from two experiments (VEPP-3 in Novosibirsk and in Hall B at JLab) which attempt to determine the two-hard-photon
 contribution via measurements of the ratio, $R$, of the elastic $e^+ p$ and $e^- p$ scattering cross sections. In addition, the OLYMPUS experiment at DESY (which also will measure this ratio) is currently in the data analysis phase, and results are expected to be published soon.
 
In general, the lepton-proton elastic scattering cross section is proportional to the square of the sum of the Born amplitude and all higher-order QED correction amplitudes. The ratio of $e^\pm p$ elastic scattering cross sections can be written~\cite{Adikaram:2014ykv} as:
\begin{eqnarray}
R = \frac{\sigma(e^+p)}{\sigma(e^-p)} \approx
\frac{1+\delta_{even}-\delta_{2\gamma}-\delta_{brem}}
     {1+\delta_{even}+\delta_{2\gamma}+\delta_{brem}} \nonumber \\
\approx  1 - 2 ( \delta_{2\gamma} + \delta_{brem})/(1+\delta_{even}) ~,
\label{eq:R}
\end{eqnarray}
where $\delta_{even}$ is the total charge-even (relative to lowest order) radiative correction factor, and $\delta_{2\gamma}$ and 
$\delta_{brem}$ are the fractional two-photon-exchange and lepton-proton bremsstrahlung interference contributions, respectively.    
After calculating and correcting for the  charge-odd $\delta_{brem}$ term,  the corrected cross section
 ratio is:
\begin{equation}\label{eq:R2g}
R' \approx 1 - \frac{2 \delta_{2\gamma}}{(1+\delta_{even})}.
\end{equation}
The hard two-photon contribution to the $ep$ scattering cross section is:
\begin{equation}
\delta_{2\gamma} = \frac{2Re{\bigl(\mathcal{M}_{1\gamma}^{\dagger} \mathcal{M}_{2\gamma}^{hard}\bigr)}}{|\mathcal{M}_{1\gamma}|^2}, \label{eq2}
\end{equation}
where $\mathcal{M}_{1\gamma}$ and $ \mathcal{M}_{2\gamma}^{hard}$ are the single- and two-photon hard scattering amplitudes, respectively.
For the purposes of comparing to theoretical predictions, the results are sometimes presented as the ratio $R_{2\gamma} = (1 - \delta_{2\gamma}) / (1 + \delta_{2\gamma})$.

\begin{figure}
\centering
\includegraphics[width=0.45\textwidth,clip]{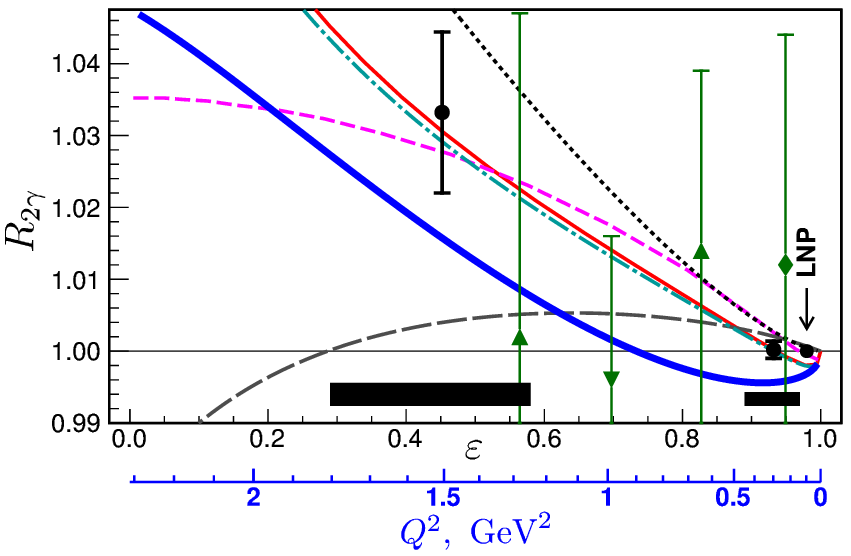}
\includegraphics[width=0.45\textwidth,clip]{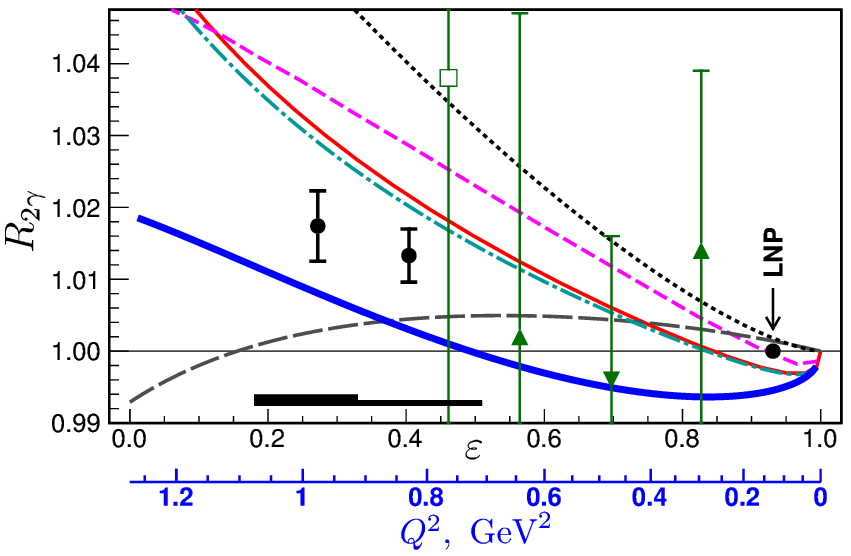}
\caption{\label{vepp3}Experimental data together with theoretical predictions for the ratio $R_{2\gamma}$ as a function of~$\epsilon$ or~$Q^2$. The 
top and bottom panels correspond respectively to \mbox{Run-I} and \mbox{Run-II}. The data points are: open squares (green) ~\cite{Browman:1965},
downward triangles (green)~\cite{Anderson:1966},
diamonds (green) ~\cite{Bartel:1967},
upward triangles (green) ~\cite{Anderson:1968}, and
circles (black)~\cite{Rachek:2014fam}.
The theoretical curves are from 
\cite{Borisyuk:2008es} dash-dotted (green),
\cite{blundenB} thin solid (red),
\cite{Bernauer:2013tpr} thick solid (blue),
\cite{TomasiGustafsson:2009pw} long-dashed (black),
\cite{arrsick} short-dashed (magenta), and
\cite{Qattan:2011ke} dotted (black).
}
\end{figure}

\begin{figure} 
\begin{center}    
\includegraphics[width=0.49\textwidth,clip=true]{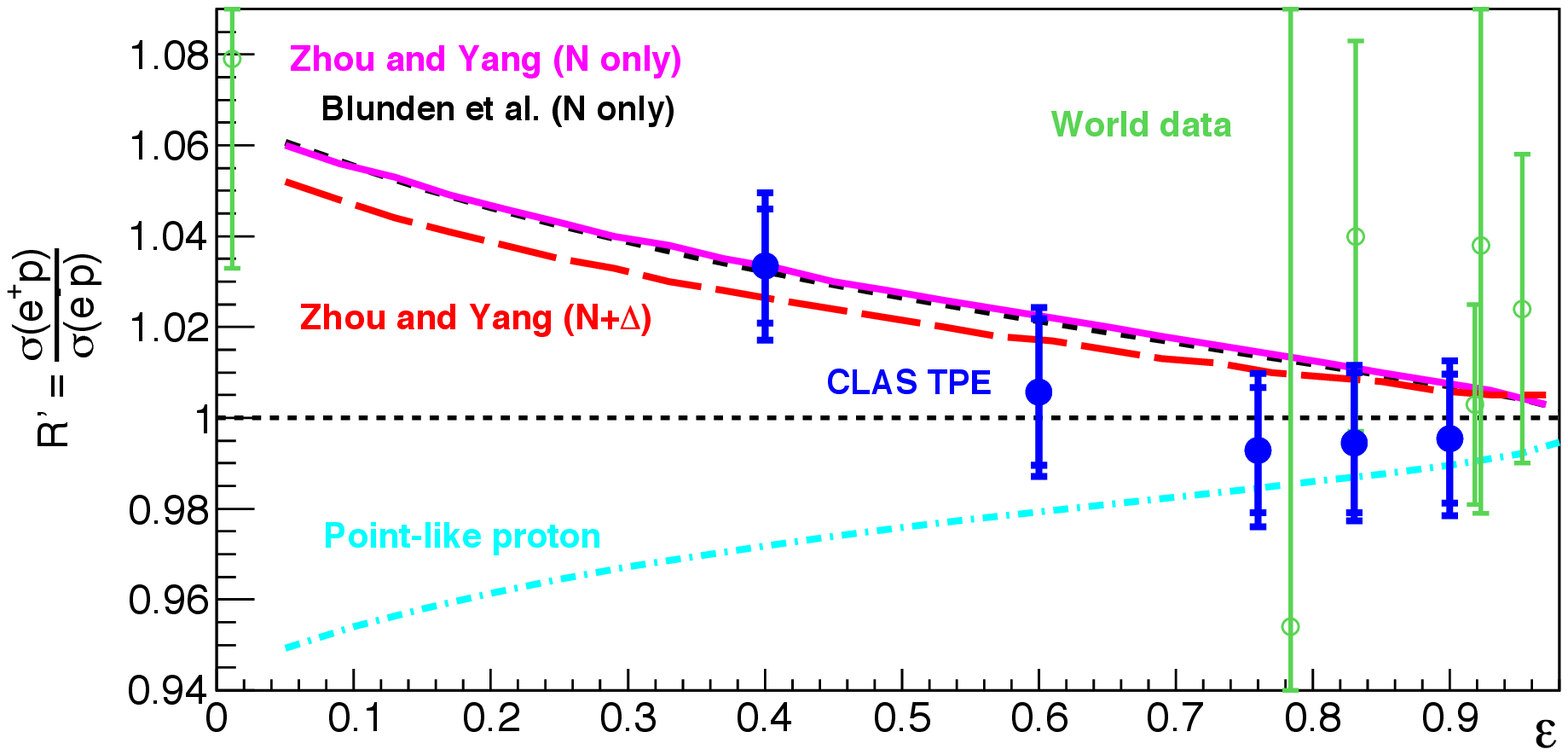}\\
\includegraphics[width=0.49\textwidth,clip=true]{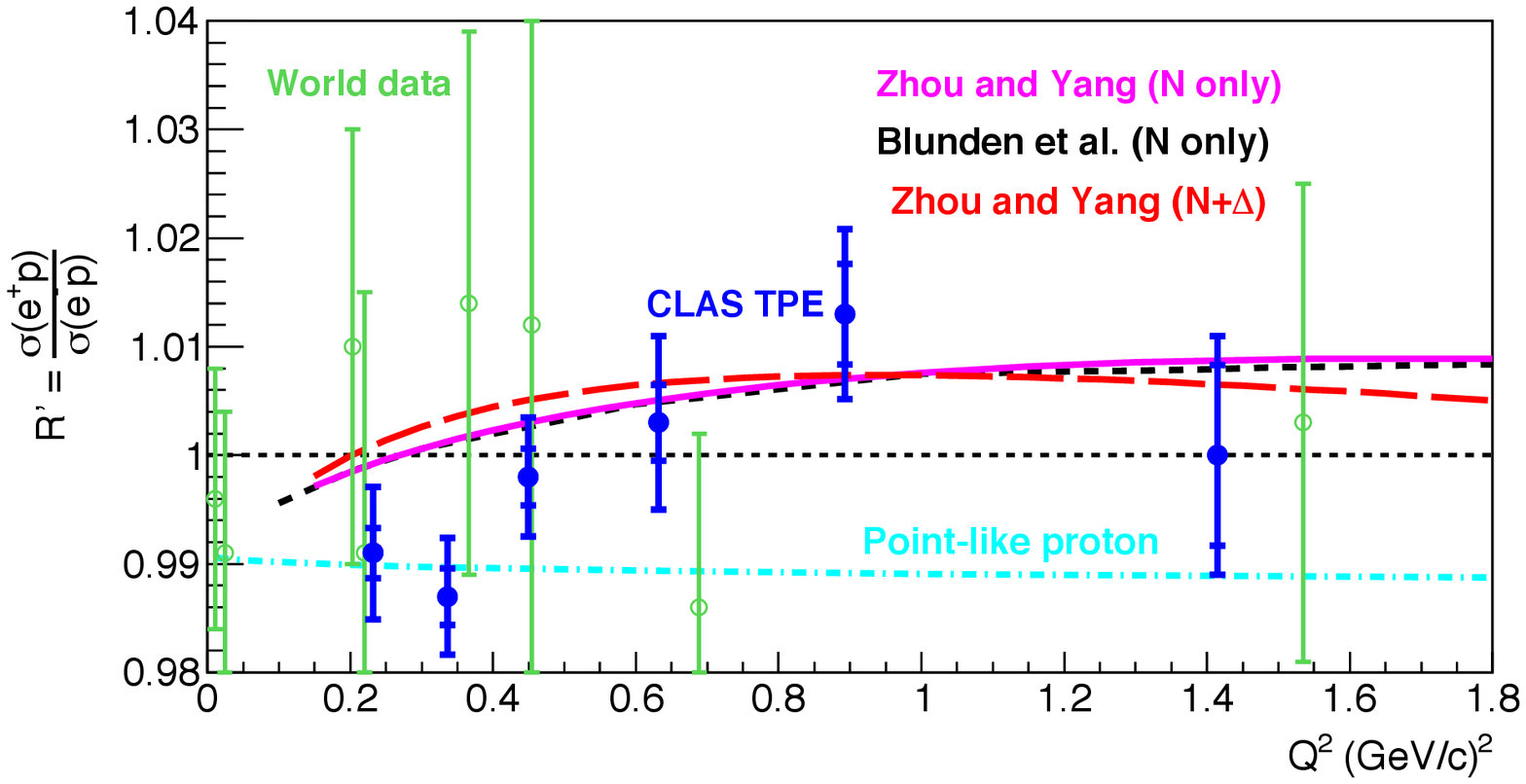}
\caption{The ratio of $e^+p$/$e^-p$ cross sections corrected for $\delta_{brem}$ as a function of $\varepsilon$ at $Q^2 = 1.45$ GeV$^2$ (top) and as a function of $Q^2$ at $\varepsilon = 0.88$ (bottom). The filled circles (blue) are from \cite{Adikaram:2014ykv}. The inner error bars are the statistical uncertainties and the outer error bars are the statistical, systematic and radiative-correction uncertainties added in quadrature. The line at $R'=1$ is the limit of no two-photon-exchange contribution.  The theory curves are:  dotted line (black) - \cite{blundenB}, solid (magenta) and dashed (red) - \cite{Zhou2014} including $N$ only and $N+\Delta$ intermediate states, respectively, dot-dashed line (cyan) - \cite{arrington11b}.  The open circles (green) show the previous world data (at $Q^2 > 1$ GeV$^2$ for the top plot)~\cite{arring04}.}
\label{hallb2gammafig}
\end{center}    
\end{figure}

The results for $R_{2\gamma}$ from Runs I and II at VEPP-3 \cite{Rachek:2014fam} are shown in Fig.~\ref{vepp3}, together with several theoretical predictions.  At JLab, in Hall B, the CLAS collaboration obtained two-photon exchange data for $Q^2$ between 0.5 and 3 GeV$^2$, with 0.15 $< \epsilon <$ 0.95 \cite{Adikaram:2014ykv}.  Results from this experiment for the ratio $R^\prime$ are shown in Fig.~\ref{hallb2gammafig} as a function of both $Q^2$ and $\epsilon$.
The data from each of these experiments indicate that the hard two-photon-exchange effect is significant, and they are in moderate agreement with several two-photon-exchange predictions which also explain the form factor ratio discrepancy at higher $Q^2$ values, thus pointing to two-photon-exchange as a likely source of at least part of the discrepancy. 

\subsection{The Proton Charge Radius}
\label{subsec:protonradius}

Non-relativistically, the elastic {\it {ep}}  cross section is related to the product of the Mott cross section for a point-like spin-$\frac{1}{2}$ electron
and the Fourier transform of the charge and/or magnetization density of the target nucleon, as follows:
\begin{equation}
\sigma(\theta_e)=\sigma_{Mott}\times\left|\int_{volume}\rho(\vec r)e^{i\vec{q}.\vec r}d^3\vec r\right|^2,
\end{equation}    
\noindent where $\rho(\vec r)$ is either the electric or the magnetic spatial distribution function. For the particular case of 
the electric form factor G$_{Ep}(Q^2)$, it follows that for short distances it can be expanded in terms of even moments of the distance $<r_{Ep}^{2n}>$ as:
\begin{equation}
G_{Ep}=1 - \frac{1}{6}{Q^2<r_{Ep}{^2}>} + \frac{1}{120}{Q^4<r_{Ep}{^4}>} ... 
\label{eq:rms}
\end{equation}
\noindent Hence, for very small distances within the nucleon, the mean-square radius of the proton can be obtained from the derivative of
 Eq.~(\ref{eq:rms}):
\begin{equation}
\frac{dG_{Ep}}{dQ^2}=-\frac{1}{6}\left|{r_{Ep}{^2}}\right|_{at Q^2=0}
\label{eq:sloperms}
\end{equation}
\noindent from which it follows that 
\begin{equation}
<r_{Ep}{^2}> = - 6\left|\frac{dG_{Ep}}{dQ^2}\right|_{at Q^2=0}; 
\end{equation}
Similar relations hold for the magnetic form factor, G$_{Mp}(Q^2)$, and the magnetic radius, $<r_{Mp}{^2}>$.
The cumulative cross section data from electron scattering experiments at low $Q^2$ have been used to obtain values of $<r_{Ep}{^2}>$ \cite{sick:2003,sick:2014,Hill:2010yb,Lorenz:2012tm,Lorenz:2014vha}. 
The extraction of $<r_{Mp}{^2}>$ is more difficult to obtain as its contribution to the cross section is suppressed by the factor $\tau$ (see Eq.~(\ref{eq:csgegm})). 

In a completely complementary fashion, the proton radius can also be obtained from precise measurements of the Lamb shift energies either in the hydrogen atom \cite{melnikov:2000} or in muonic hydrogen \cite{antognini:2013}; indeed, a (different) fraction of the Lamb shift is related to the finite size of the proton nucleus in each case.
In both cases, the Lamb shift under consideration depends upon the overlap of the appropriate S-, P-, or D-state wave functions and the proton nucleus; it is in this way that the finite proton size contributes to the Lamb shift.  Recent measurements of the muonic Lamb shift energies at PSI have produced
values of $<r_{Ep}{^2}>$ which are smaller than the mean value of all electron scattering experiments (the so-called CODATA value - see Refs. \cite{Mohr:2005,Mohr:2008,Mohr:2012}) by about 4\% (a 7 $\sigma$ difference). This discrepancy has become known as the proton radius puzzle, and its resolution has become a topic of great current interest, and the aim of several new and novel experimental efforts.   We focus here on the experimental determinations of the proton radius; indeed, there has been a plethora of works concerned with the theoretical aspects of the problem.  We refer the reader to \cite{pohl:2013} and \cite{carlson:2015}, and references therein, for further details.

\subsubsection{Previous Results}
\label{subsubsec:radiuspast}

\begin{figure}[t]
\begin{center}
\includegraphics[width = 1.0\columnwidth]{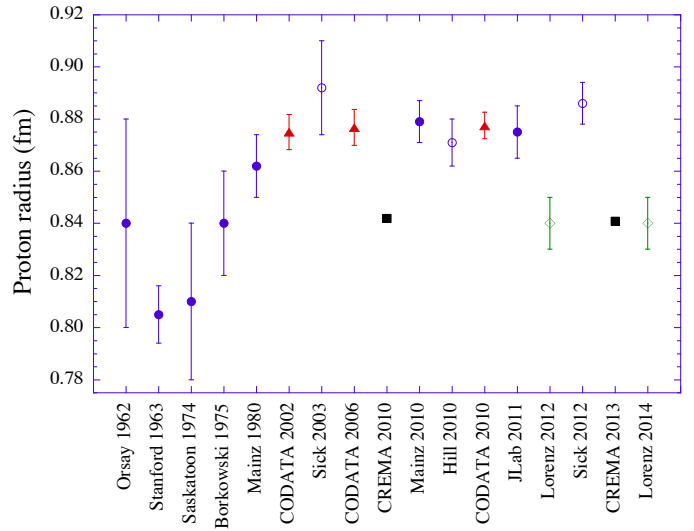}
\caption{Proton radius determinations over time. Electronic measurements
  seem to settle around $r_p$=0.88\,fm, whereas the muonic hydrogen
  value is at 0.84\,fm.  The solid circles (blue) are all
  electron scattering measurements;
  Orsay~1962~\cite{Lehmann:1962dr},
  Stanford~1963~\cite{Hand:1963},
  Saskatoon~1975~\cite{Murphy:1974zz},
  Borkowski~1975~\cite{bork}, and
  Mainz~1980~\cite{Simon:1990}
  are early measurements, whereas
  the more recent measurements are 
  Mainz~2010~\cite{Bernauer:2010wm} and 
  JLab~2011~\cite{Zhan:2011ji}.
  The open circles (blue)~\cite{sick:2003,sick:2014,Hill:2010yb} and open diamonds (green) ~\cite{Lorenz:2012tm,Lorenz:2014vha} denote reanalyses of the world electron 
  scattering data using various fitting functions (See Sec.~\ref{subsubsec:cf}).
  The solid triangles (red), (CODATA) originate from laser spectroscopy of 
  atomic hydrogen and advances in hydrogen QED theory~\cite{Mohr:2005,Mohr:2008,Mohr:2012}.
  The solid squares (black) are the values extracted from muonic hydrogen~\cite{Pohl:2010zza,Antognini:1900ns}.
  %
}
\label{Rp_vs_t}
\end{center}
\end{figure}

In Fig.~\ref{Rp_vs_t}, we show several determinations of the RMS proton charge radius, $r_p=\sqrt{<r_{Ep}{^2}>}$ over the last several decades.  Earlier extractions were based on data from elastic $\it{ep}$ scattering experiments at Orsay~\cite{Lehmann:1962dr}, Stanford~\cite{Hand:1963},
Saskatoon~\cite{Murphy:1974zz} and
Mainz~\cite{bork,Simon:1990}, together with the various re-analyses of these world
data~\cite{sick:2003,sick:2014,Hill:2010yb,Lorenz:2012tm,Lorenz:2014vha}.  In general, the extraction of the proton radius from these data involves fitting the form factors with either simple mathematical or in some cases theoretically inspired parametrizations in order to determine the radius.    As can be seen clearly in Fig.~\ref{Rp_vs_t}, the consistency between various approaches has improved over the years.  However, several issues related to the extraction procedure remain.  There is, inherently, a model dependence uncertainty in the extraction (as evidenced for example in the discrepancy between the open circles and open diamonds in Fig.~\ref{Rp_vs_t}) but this is typically not included in the quoted uncertainty for a given extraction.  Moreover, it appears at this point that the treatment of systematic uncertainties in many experiments was overly optimistic. This can be inferred from the fact that in many global fits, the $\chi^2$ per degree of freedom is larger than unity.  In some analyses, the absolute normalization of the cross section data was allowed to float, but this is done at the expense of sensitivity to the radius.  In addition, there are issues connected with the range of $Q^2$ data that is included in the global fits.  It can be seen from Eq.~(\ref{eq:rms}) that the coefficients of successive $<r_{Ep}^{2n}>$ terms increase with order.  Therefore, it is not possible to define a value of $Q^2$ where any one term in the expansion sufficiently dominates such that its value could be fixed and then used in fitting to lower $Q^2$ data.

In order to address some of the issues related to systematic uncertainties in the cross section data, the Mainz A1 collaboration at MAMI~\cite{Bernauer:2010wm} measured 1422 precise relative $\it{ep}$ cross sections in the low $Q^2$ region (0.0038 GeV$^2$ to 0.98 GeV$^2$) and a wide range of beam energies and scattering angles.  Experimental systematic uncertainties were controlled by using one spectrometer as a luminosity monitor, and then moving the other two spectrometers through multiple, overlapping angle settings.  The cross section data were subsequently fit with a variety of functional forms, in order to assess model dependent uncertainties.  Interestingly, it was found that satisfactory goodness of fit could only be obtained through the use of more flexible mathematical fitting functions (polynomials or splines) as opposed the more traditional physically motivated forms, such as dipoles.  The final extracted value for the proton radius was 
$r_p$ = 0.879 $\pm$ 0.005$_{stat}$ $\pm$ 0.004$_{syst}$ $\pm$ 0.002$_{model}$ $\pm$ 0.004$_{group}$,
where the final uncertainty comes from the polynomial vs.\ spline difference.

In parallel with the cross section measurements, the Jefferson Lab LEDEX collaboration~\cite{Zhan:2011ji,Ron:2007vr,Ron:2011rd} measured the proton form factor ratio $\mu_p{G_{Ep}}/{G_{Mp}}$ using polarization transfer in the $Q^2$ $\approx$ 0.1 $\rightarrow$ 0.4 GeV$^2$ region.  It is particularly interesting that while measurements of the form factor ratio do not give the proton radius directly, accurate and precise knowledge of the ratio helps to constrain normalizations of cross section data during fits, which in turn leads to an improved value of the extracted radius.  The analysis of Ref.~\cite{Zhan:2011ji} gives $r_p$ = 0.875 $\pm$ 0.008$_{exp}$ $\pm$ 0.006$_{fit}$, in agreement with the value extracted in the Mainz analysis.  It is interesting to note that this analysis uses an entirely independent data set from the Mainz analysis.

\begin{figure}[t]
\begin{center}
\includegraphics[width = 1.0\columnwidth]{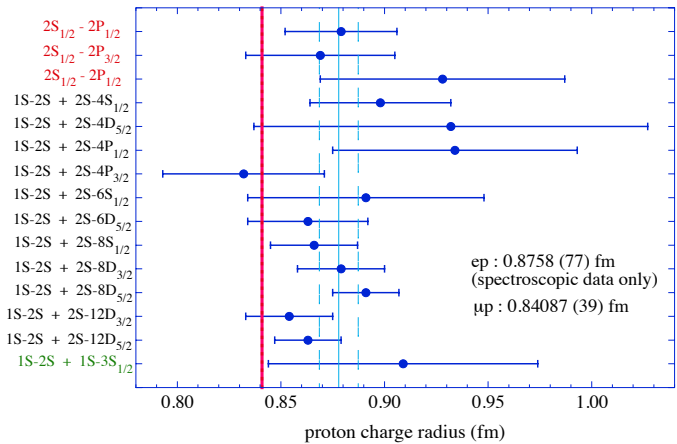}
\caption{Proton charge radii $r_p$ obtained from hydrogen
  spectroscopy. The proton charge radius can best be
  extracted from a combination of the
  1S-2S transition frequency~\cite{Parthey:2011} and one of
  the indicated 2S-NL$_J$
  transitions~\cite{Beauvoir:1997,Schwob:1999}. The thin vertical line (cyan) is the weighted average of the data points shown; the error bar on this average is represented by the vertical dashed lines.  
  The value from muonic
  hydrogen~\cite{Pohl:2010zza,Antognini:1900ns} is represented by the thick vertical line (red); the thickness of this line represents the error bar.}
\label{Rp_from_H}
\end{center}
\end{figure}

The Lamb shifts in both normal electronic as well as muonic hydrogen are sensitive to the non-point-like structure of the proton nucleus.  While this effect is indeed minuscule, the ability of quantum electrodynamics (QED) to predict the energy levels of hydrogen with remarkable accuracy, together with experimental measurements of the relevant transition frequencies to around one part in 10$^{11}$, results in a measurable sensitivity to the proton radius.  The energies of the S-states in hydrogen are given by:
\begin{equation}
\label{eq:E_simple}
E(nS) \simeq - \frac{R_{\infty}}{n^2} + \frac{L_{1S}}{n^3}
\end{equation}
where $n$ is the principal quantum number, and $L_{1S}$ denotes the Lamb shift
of the 1S ground state which is given by QED and contains the effect of the
proton charge radius, $r_p$. Numerically, $L_{1S} \simeq ( 8172 + 1.56\,
r_p(fm)^2 )$\,MHz, so the finite size effect on the 1S
level in hydrogen is about 1.2\,MHz.  The different $n$-dependence of the two terms in Eq.~(\ref{eq:E_simple}).
permits the determination of both $R_{\infty}$ and $r_p$ from at least two
transition frequencies in hydrogen.  The approach taken to date has been to use the 1S-2S transition~\cite{Parthey:2011}, which has been measured to a stunning accuracy of one part in 10$^{15}$, together with one of
  the 2S-NL$_J$
  transitions~\cite{Beauvoir:1997,Schwob:1999}.  The former transition is maximally sensitive to $r_p$, whereas the latter contain much smaller Lamb shift contributions, due to the $1/n^3$ scaling in Eq.~(\ref{eq:E_simple}).

In Fig.~\ref{Rp_from_H}, we show the values of $r_p$ extracted from the various transition combinations.  The $r_p$ values extracted from measurements in normal electronic hydrogen
favor a $r_p$ value of $\approx$ 0.88\,fm, consistent with the world-averaged electron scattering results.
The discrepancy between the combined value from just electronic hydrogen alone, as obtained in the
elaborate CODATA adjustment of the fundamental
constants~\cite{Mohr:2012}, and the muonic hydrogen value, is about
$4.4\sigma$.

The difference between the muonic hydrogen determination of the proton radius and the results obtained from either electron scattering or from electronic hydrogen transitions is highly enigmatic.  The theoretical QED calculations of the portion of the Lamb shift that is not due to the finite proton size have been checked and re-checked by multiple independent groups.  Also, the many measurements of the transition frequencies in electronic hydrogen are in agreement with one another.  Thus, from an experimental point of view, the solution to this puzzle is not obvious at this time.  On the theoretical side, there has been a significant amount of interest in this problem as well; this is discussed in more detail in Section~\ref{sec:theory}.    

\subsubsection{Future Experiments}

Experimental approaches to unraveling the proton radius puzzle break down into several distinct categories:  improving the precision of the proton radius extraction from electron scattering experiments, extending the atomic spectroscopy measurements to other ions or exotic atoms, and/or determining the proton radius in elastic muon-proton scattering.

At Jefferson Laboratory, in Hall B, there is an approved experiment which aims
to improve the electron scattering radius determination
by extending the $Q^2$ range from the 0.0038 GeV$^2$ of the
Mainz experiment down to 1-2 $\times$ 10$^{-4}$ GeV$^2$ \cite{gasparian2011}. 
This is indeed a challenging experiment. If the proton radius is to be determined in this experiment with similar accuracy compared to the current
world-averaged result from electron scattering,
then relative cross sections need to be determined at the 0.2\% level.
Fortunately, the large cross sections and corresponding event rates for low $Q^2$ scattering make statistical uncertainties of 0.1\% achievable.
However, systematic uncertainties must also be controlled at a similar level.  One of the largest sources of systematic error
is the determination of the electron scattering angle; at the lowest $Q^2$ value of the experiment, the scattering angle
is only 10~mrad, and due to the severe angular dependence of the Mott cross-section at small angles, a 10 $\mu$rad knowledge of the scattering angle is needed
to limit shifts in the relative cross sections to 0.2\%.

At MAMI, an experiment is underway that aims to measure the proton electromagnetic form factors in $ep$ scattering at very low momentum transfers by using a technique based on initial state radiation~\cite{ISR:2013}. The basic premise is that initial state radiation degrades the energy of the incoming electron so that the momentum transfer to the proton can be quite low. The outgoing electron angle and energy are measured as usual, and together with theoretical input, an accurate form factor can in principle be obtained at $Q^2$ values as low as 10$^{-4}$ GeV$^2$. The full experiment ran in 2013, and the analysis is continuing.

As can be seen from Eq.~(\ref{eq:E_simple}), the determination of $r_p$ using transitions in hydrogen relies heavily on a precise determination of $R_{\infty}$.   Currently, $R_{\infty}$ is known from measurements of transition frequencies in hydrogen and deuterium.  However, since the correlation between $r_p$ and $R_{\infty}$ is significant, new precise measurements of $R_{\infty}$ (at the level of a few parts in 10$^{12}$) that are independent of $r_p$ could potentially impact the extracted value of $r_p$ from electronic hydrogen measurements.   There are several such efforts currently underway.  These include single-~\cite{Beyer:2013jla} and two-photon~\cite{Flowers:2007} measurements in hydrogen, as well as experiments which aim to use laser spectroscopy of neutral and ionic helium~\cite{Herrmann:2009,Kandula:2011,Rooij:2011}.

Noticeably, a crucial missing piece of the proton radius puzzle is a measurement of the proton radius using muon scattering; indeed, there is a proposal at PSI, the MUSE Experiment, to make just such a measurement~\cite{MUSE2012}.    The MUSE experiment will measure elastic $\mu p$ scattering to a minimum $Q^2$ of 0.002~GeV$^2$ - about half the lower limit of the Mainz electron scattering experiment - using both positively and negative charged incident muons, so that any possible two-photon effects can be taken into account directly from the data, rather than relying on theoretical calculations.  In addition, the experiment will simultaneously collect $ep$ scattering data, so that the extracted proton radius from muon and electron scattering can be compared directly within a single experimental apparatus.  Preliminary estimates are that the proton radius that can be extracted from muon scattering will be similar in precision to that extracted from the Mainz experiment.  

\label{subsubsec:radiusfuture}
\subsection{Flavor Separation of Nucleon Form Factors}
\label{flavor}
Charge symmetry implies that the proton and neutron wave functions are identical under the interchange of the up and down quark contributions.  Measurements of asymmetries in parity non-conserving electron scattering on the proton have found that the strange quark form factors are small (see the review article by Ref.~\cite{Armstrong}). Ignoring the contributions
of higher mass quarks, the proton and neutron form factors can be written in terms of the contributions from the up and down dressed quark form factors as:
\begin{eqnarray}
G_{(E,M)p} &=& \frac{2}{3}G_{(E,M)u} - \frac{1}{3}G_{(E,M)d} \nonumber \\
G_{(E,M)n} &=& \frac{2}{3}G_{(E,M)d} - \frac{1}{3}G_{(E,M)u}.
\end{eqnarray}
The up and down form factors, $G_{(E,M)u}$ and $G_{(E,M)d}$ are defined by convention to represent the up and down dressed quark form factors in the proton. The anomalous magnetic moments of the up and down quarks can be expressed as $\kappa_{u} = 2\kappa_{p} +\kappa_{n}$ 
and $\kappa_{d} = \kappa_{p} + 2\kappa_{n}$, respectively. 
The   Dirac and Pauli form factors for 
the up and down quarks can be written as:
\begin{eqnarray}
F_{(1,2)u} &=& 2F_{(1,2)p} + F_{(1,2)n} \nonumber \\
F_{(1,2)d} &=& F_{(1,2)p} + 2F_{(1,2)n}.
\end{eqnarray}
The recent precision data on $G_{Mn}$ in the region of $Q^2$ between 1.5 to 4.8~GeV$^2$ \cite{lachniet:2008} and data on $G_{En}$/$G_{Mn}$ to $Q^2 = 3.4$~GeV$^2$ by Ref.~\cite{Riordan:2010} have enabled precise phenomenological fits to the proton and neutron form factors  and detailed comparison to theory predictions. This allows one  to extract information about the underlying contributions of the up and down quarks to the nucleon form factors. 

\begin{figure}[tb]
	\begin{center}
		\resizebox{\columnwidth}{!}{%
			\includegraphics[]{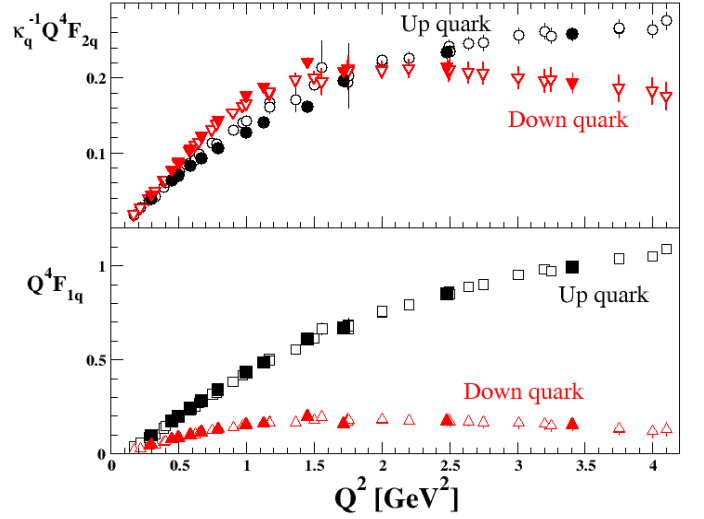}
		}
		\caption{(Top panel) Up quark's $\kappa_{u}^{-1}Q^4$$F_{2u}$ filled circle(black) and down quark's $\kappa_{d}^{-1}Q^4$$F_{2d}$ 
filled triangle down (red) from Ref.~\cite{cates:2011}.  Up quark $\kappa_{u}^{-1}Q^4$$F_{2u}$ empty circle (black) and down quark's 
$\kappa_{d}^{-1}Q^4$$F_{2d}$  empty triangle down (red) from Ref.~\cite{qattan:2013}. 
(Bottom panel) Up quark's $Q^4$$F_{1u}$ filled square (black) and down quark's  $Q^4$$F_{1d}$ filled triangle (red) 
are from Ref.~\cite{cates:2011}. Up quark's $Q^4$$F_{1u}$ empty square (black) and down quark's  $Q^4$$F_{1d}$ empty triangle (red) are from Ref.~\cite{qattan:2013}.
			.   
			 }
		\label{fig:up-down-cates-f1f2}
	\end{center}
\end{figure}
	A calculation of the up and down quark form factors from the available proton and neutron data was done by Ref.~\cite{cates:2011}. The quark form factors were calculated to $Q^2$~=~3.4~GeV$^2$ by combining the measurements of $G_{En}$/$G_{Mn}$ by Ref.~\cite{plaster,zhu,Riordan:2010,glazier:2004,warren,bermuth} 
	with the Kelly fit \cite{kelly04} to $G_{Mn}$, $G_{Mp}$ and $G_{Ep}$. Fig.~\ref{fig:up-down-cates-f1f2} is a plot of $Q^4$$F_1$ and $Q^4$$F_2$/$\kappa$$Q^2$ versus $Q^2$ for the up and down quarks. The data is plotted at the $Q^2$ of the  $G_{En}$/$G_{Mn}$ measurements and the error on the quark form factors is determined by the error on the $G_{En}$/$G_{Mn}$ measurements. 
	 For $Q^2 > 1.0$~GeV$^2$, the $Q^2$ dependence of both the $F_1$ and $F_2$ changes for the up and down quarks. For the up quark, $Q^4$$F_1$ and $Q^4$$F_2$/$\kappa$ continue to rise, while, for the down quark $Q^4$$F_1$ and $Q^4$$F_2$/$\kappa$ are plateauing or slightly dropping.

Another separation of the up and down form factors was done by Ref.~\cite{qattan:2013}. They used $G_{Ep}$ and  $G_{Mp}$ from a
extraction using cross section and polarization data which included two-photon exchange contributions \cite{Qattan:2011ke}. In addition, they added the data of Ref.~\cite{Zhan:2011} for $G_{Ep}/G_{Mp}$ at low $Q^2$.   For the neutron form factors, they used the fit of Ref.~\cite{Riordan:2010} to $G_{En}/G_{Mn}$
and an updated parametrization of $G_{Mn}$ using the data of Ref.~\cite{lung,anklin,Anklin:1998ae,kubon,lachniet:2008,anderson}. The up and down form factors are calculated at the $Q^2$ of the proton data and are plotted in Fig.~\ref{fig:up-down-cates-f1f2}.
In the region of $Q^2$ between 0.5 to 1.5 GeV$^2$, $F_{2u}$ from Ref.~\cite{qattan:2013} is slightly larger than that in Ref.~\cite{cates:2011} and, correlated with that, $F_{2d}$ from Ref.~\cite{qattan:2013} is slightly smaller.  
In general, comparisons between the two different flavor separations of the form factors  give a sense that the size of the uncertainty due to two-photon exchange contributions and tensions in the data sets is relatively small and does not obscure the general trends  in the $Q^2$ dependence of the up and down quarks form factors.

\begin{figure}[b]
	\begin{center}
		\resizebox{\columnwidth}{!}{%
			\includegraphics[angle=0,width = 3.5 in]{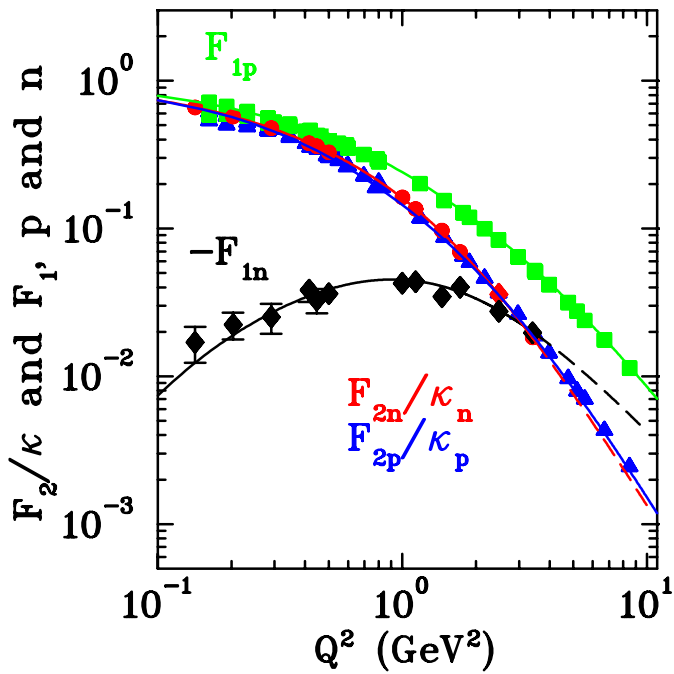}
		}
		\caption{Neutron and proton $F_1$ and $F_2$/$\kappa$ versus $Q^2$. The data points are from the polarization experiments. The neutron $F_{1n}$ and $F_{2n}$/$\kappa_n$ data are plotted as filled diamond (black) and filled circle (red).
			The proton $F_{1p}$ and $F_{2p}$/$\kappa_p$ are plotted as filled square (green) and filled  triangle (blue).
			The solid lines are calculations of $F_1$ and $F_2$ based on the Kelly fit \cite{kelly04} for  $G_{Mn}$ and $G_{Mp}$ and new fits to $\mu_{n}G_{En}/G_{Mn}$ and $\mu_{p}$$G_{Ep}/$$G_{Mp}$ using Eqs.~(\ref{eq:genfit}) and (\ref{eq:gepfit}).}
		\label{fig:f1f2-cfp-fit}
		
	\end{center}
\end{figure}

 Great interest exists in extending the  separation of quark flavors to higher $Q^2$. In the spirit of Ref.~\cite{cates:2011} and Ref.~\cite{qattan:2013}, we use the Kelly fit \cite{kelly04} for  $G_{Mn}$ and $G_{Mp}$ while refitting $\mu_{n}$$G_{En}/$$G_{Mn}$ and $\mu_{p}$$G_{Ep}/$$G_{Mp}$ to include the data since the Kelly fit was done. The fit form for $\mu_{n}$$G_{En}/$$G_{Mn}$  is:
\begin{equation}
	\label{eq:genfit}
\frac{\mu_{n}G_{En}}{G_{Mn}}=\frac{A_1\tau}{1+A_2\sqrt{\tau}+A_3*\tau},
\end{equation}
with $A_1 = 2.6316 $, $A_2 = 4.118  $ and $A_3 = 0.29516  $. 
The fit form for $\mu_{p}$$G_{Ep}/$$G_{Mp}$  is:
\begin{equation}
	\label{eq:gepfit}
	\frac{\mu_{p}G_{Ep}}{G_{Mp}}=\frac{1+B_0\tau+B_1\tau^2+B_2\tau^3}{1+B_3\tau+B_4\tau^2+B_5\tau^3+B_6\tau^4},
\end{equation}
with $B_0=-5.7891$,$B_1=14.493$, $B_2=-3.5032$, $B_3 =-5.5839$, $B_4=12.909$, $B_5=0.88996$ and $B_6=1.5420$.
In Fig.~\ref{fig:f1f2-cfp-fit}, the fits are compared to the world
data for 
the proton and neutron $F_1$ and $F_2/{\kappa}$. The shapes for the proton and neutron $F_{2}/{\kappa}$ are nearly identical with the data on top on each other. A small shape difference in the proton and neutron $F_2/{\kappa}$ dependence on $Q^2$  accounts for the difference between the up and down quark's $F_2/\kappa$ which is seen in Fig.~\ref{fig:up-down-cates-f1f2}.

\begin{figure}[th]
 	\begin{center}
 		\resizebox{\columnwidth}{!}{%
 			\includegraphics[angle=90]{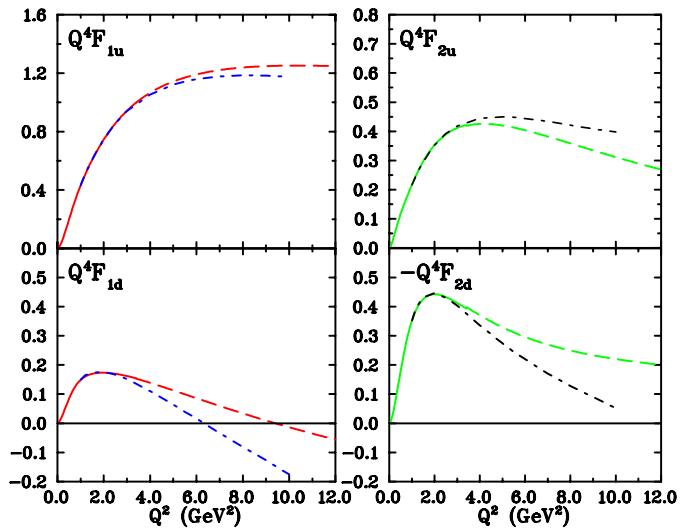}
 		}
 		\caption{Calculations of $F_1$ and $F_2$ for the up and down quarks based on the Kelly fit \cite{kelly04} for  $G_{Mn}$ and $G_{Mp}$ and new fits to $\mu_{n}G_{En}/G_{Mn}$ and $\mu_{p}$$G_{Ep}/$$G_{Mp}$ using Eqs.~(\ref{eq:genfit}) and (\ref{eq:gepfit}). The results from the fit are shown as a solid 
line in the region of $Q^2 < 3.4$~GeV$^2$ where $G_{En}$ data exists, then extended as a dashed line when the fit is extrapolated to $Q^2 = 12$~GeV$^2$. 
Calculations, which replace $G_{En}/G_{Mn}$ from fit using Eq.~(\ref{eq:genfit}) with $G_{En}/G_{Mn}$ from the DSE model of Ref.~\cite{cloet:2008}
 			while keeping the same $G_{Mn}$, $G_{Mp}$ and $\mu_{p}$$G_{Ep}/$$G_{Mp}$ fits, are plotted as a dash-dotted line.}
 		\label{fig:up-down-f1f2}
 	\end{center}
 \end{figure}

Using the fits the nucleon form factors, the flavor separation can be extrapolated to higher $Q^2$. To investigate the sensitivity of the extrapolation of the quark form factors, a different shape for $G_{En}/G_{Mn}$ can be used in calculating the quark form factors. As an example, the $G_{En}/G_{Mn}$ prediction  from  the Dyson Schwinger equation  (DSE) model of Ref.~\cite{cloet:2008} is plotted as a dash dotted line in Fig.~\ref{fig:gen_pol}. The DSE models are discussed in the theory Sec.~\ref{subsec:dse}. The up and down quark's $F_1$ and $F_2$  from the fit using Eqs.~(\ref{eq:gepfit}) and (\ref{eq:genfit}) are plotted in Fig.~\ref{fig:up-down-f1f2} as a solid line in the region of $Q^2 < 3.4$~GeV$^2$ where $G_{En}$ data exists, then extended as a dashed line when the fit is extrapolated to $Q^2 = 12$~GeV$^2$. The fit shows that $Q^4 F_{1d}$ will have a zero crossing at $Q^2 \approx 11.5$~GeV$^2$ while $Q^4 F_{1u}$ starts to plateau above $Q^2$ of 7~GeV$^2$.   Both $Q^4 F_{2u}$ and $Q^4 F_{2d}$ slowly drop-off above $Q^2$ of 3~GeV$^2$ with $Q^4 F_{2d}$ falling slightly faster. When the $G_{En}/G_{Mn}$ prediction from the DSE model is combined with the other form factors from the fit to calculate the quark form factor, then mainly the down quark's form factors are modified. This is shown in Fig.~\ref{fig:up-down-f1f2} where the up and down quark's $F_1$ and $F_2$ using the $G_{En}/G_{Mn}$ prediction from the DSE model are plotted as a dash-dotted line.  The zero crossing in  $Q^4 F_{1d}$ moves to lower $Q^2$. This demonstrates the need for precision measurements of all nucleon form factors to large $Q^2$. Future experiments at JLab to extend the $Q^2$ range of the nucleon form factors measurements are discussed in Sec.~\ref{sec:conclusion}.  
 
{\tiny }\section{Theoretical Interpretations of Nucleon Form Factors}
\label{sec:theory}


We give an overview of theoretical work 
on nucleon electromagnetic form factors. 
These form factors encode the information on the structure of a  
strongly interacting many-body system of quarks and gluons, 
such as the nucleon. 
This field has a long history and many theoretical 
attempts have been made to understand the nucleon form factors. 
This reflects the fact that a direct calculation of nucleon form factors from the 
underlying theory, Quantum Chromodynamics (QCD), is complicated as it 
requires, in the few GeV momentum transfer region,  
nonperturbative methods. Hence, in practice it involves approximations 
which often have a limited range of applicability. 
Despite their approximations and limitations, some of these nonperturbative 
methods do reveal interesting insight into the nucleon's structure.  

Section~\ref{subsubsec:cf} describes work simply fitting the form factors using an expansion in a transformed variable that allows convergence of a polynomial expansion for all values of $Q^2$.  These techniques have long been used in parameterizing form factors measured in Weak decays, and part of the motivation here is to ensure a better extrapolation to obtain the slope at zero $Q^2$, \textit{i.e.,} in determining the charge radius.

Section~\ref{subsubsec:vmd} describes vector meson dominance of the photon-nucleon coupling, which in addition to being physically motivated, provides good forms for fitting the nucleon form factors.  Notable here is the fact that the falling $G_E^p/G_M^p$ ratio was obtained from early on (although fits to the neutron form factors and form factors in the timelike region required tuning after data became available).

Section~\ref{subsubsec:da} uses the ideas of dispersion relations to fit the form factors.  In principle, if one knew the couplings and locations of all the poles and cuts in the $N \bar N$ channels, one could calculate without further approximation the form factors at all $Q^2$.  In practice, the information is incomplete, and what one has are good parameterizations of the form factors that obey all the necessary analyticity properties and will converge everywhere.  One then uses the available data to fit parameters in these functions, and hence obtain an accurate analytic representation of the data.  

Section~\ref{subsubsec:cqm} reviews the extensive work that has been done using constituent quark models to calculate the form factors.  Relativity is crucial here, and many of the works use one of the Hamiltonian dynamical approaches enumerated by Dirac long ago~\cite{dirac}.  The constituent quarks are often thought to be representatives not only for elementary quarks but also to represent not explicitly included contributions from gluons and higher Fock states, and as such may themselves have form factors that need to be parameterized.  The outcome is usually a physically motivated form with parameters that need to be fit to data, and the fits in modern times are quite good for all the form factors. 

QCD has chiral symmetry, and chiral symmetry in our world leads to the existence of light mesons.  These light mesons can then be part of the long range structure of the nucleons, and the pion cloud models that take these degrees freedom into account are described in Sec.~\ref{subsubsec:pion}.
 
To simply Fourier transform the electric and magnetic form factors to obtain the charge and magnetic densities is not valid relativistically, so this door to obtaining structure information about the nucleons is closed.  However, it can be shown that projections of the densities onto the transverse plane for a fast moving nucleon can be validly obtained  from two-dimensional Fourier transforms of the Pauli and Dirac form factors.  Work on transverse densities is described in Sec.~\ref{subsubsec:td}.

A new approach to obtain approximate predictions for nucleon form factors is to use correspondences that have been discovered between gravitational theories in five dimension and approximately conformal field theories, like QCD when considering only the light quarks.  Results from this approach are described in Sec.~\ref{subsubsec:ads}.

The Dyson-Schwinger equations (DSE) are an infinite set of equations for the vertices, propagators, and other quantities related to calculating observables from a field theory.  They can be truncated with some success, and they have been well studied in QCD, with DSE results for nucleon form factors presented in~\ref{subsec:dse}.

Perturbative QCD results for the form factors are on the face of it vitiated by the $G_E^p/G_M^p$ measurements.   The pQCD basics, and improvements to and extensions of the pQCD techniques, are reviewed in~\ref{subsubsec:pqcd}.

Form factors may also be obtained as integrals of generalized parton distributions (GPDs), which are the amplitudes for removing a quark from a nucleon and substituting another quark with a different momentum and possibly differing also in other quantum numbers.   GPDs are important metrics of nucleon structure, and are measured in exclusive processes like $\gamma^* N \to \gamma N$ or $\gamma^* N \to \rho N$.  Their consequence for nucleon form factors is highlighted in Sec.~\ref{subsubsec:gpd}.

Finally, nucleon form factor results from lattice gauge theory are reviewed in Sec.~\ref{subsec:lattice}.  The results so far are only for the isovector form factors, but the uncertainty limits are decreasing as further work is done, and lattice gauge theory has the advantages of being really QCD and not a model of QCD, and of obtaining results that are systematically improvable.

\subsection{Models of Nucleon Form Factors}
\label{subsec:models}

%
\subsubsection{Conformal Fits to Form Factors}
\label{subsubsec:cf}

There has been a lively discussion of what sorts of functions to use in fitting the form factors.  Simple polynomial fits, for example,  will not converge for moderate or large momentum transfers.   The reason flows from the fact that the form factors are, from a mathematical viewpoint, analytic functions of their argument $Q^2 = -q^2$, except for cuts at known locations.  The cuts are on the timelike side, and begin where one can have a photon to two pion transition at $q^2 = 4 m_\pi^2$.  A cut can be viewed as a weighted continuum of poles, so that there is a contribution to the form factor containing a factor $1/(q^2 - 4 m_\pi^2)$.  The weighting of this pole may be weak, but in principle its existence means that a polynomial expansion of the form factor will not converge for $Q^2 \ge 4 m_\pi^2$.  It is like the expansion of the geometric series $1/(1-x)$, which does not converge for $|x| \ge 1$.

However, it is possible to make a mapping of $Q^2$ to another variable, denote it $z$, where a polynomial expansion in $z$ is allowed.  The trick is to find a transformation where spacelike momentum transfers all map onto the real line $|z| < 1$ and timelike momentum transfers map onto the circle $z =1$ (in the complex $z$-plane).  Then since all poles of the form factors lie on the unit circle in $z$, a polynomial expansion in $z$ is convergent everywhere inside the unit circle, \textit{i.e.,} for all spacelike momentum transfers.

This trick has been applied in the context of Weak interaction form factors, as for semileptonic meson decay, for some time~\cite{Boyd:1995sq,Bourrely:2008za}.  It has now also been applied to fitting electromagnetic form factors by Hill and Paz~\cite{Hill:2010yb} and by Lorenz \textit{et al.}~\cite{Lorenz:2014vha}.

The variable $z$ is given by the conformal mapping~\cite{Hill:2010yb,Lorenz:2014vha}, with $t = q^2$,
\begin{equation}
z(t,t_{\rm cut}) = \frac{ \sqrt{t_{\rm cut} - t} - \sqrt{t_{\rm cut}}}
					{ \sqrt{t_{\rm cut} - t} + \sqrt{t_{\rm cut}}}	\,,
\end{equation}
where $t_{\rm cut} = 4 m_\pi^2$ and one can easily enough verify that the mapping has the properties stated above.

Fitting with a nonconvergent expansion can give good analytic fits to the data in any region where there is data to be fit.  The danger lies in extending them outside the region where there is data.  Such extrapolations can go awry,  sometimes diverging wildly from physical expectation and sometimes, depending on how far one extrapolates, there may be problems that are less visible.  Here enters also the proton radius question, whose evaluation from form factors requires an extrapolation from the lowest $Q^2$ where there is data, down to $Q^2 = 0$.  Extrapolating a fit made with an intrinsically convergent expansion is arguable safer.

The two fits made to the electromagnetic form factors using the conformal variable $z$, however, differ in their conclusions regarding the proton charge radius.  The earlier fit~\cite{Hill:2010yb}  used electron-proton scattering data available before the Mainz experiment~\cite{Bernauer:2010wm} published in 2010.   They found a proton radius $r_p =  0.870\pm 0.023 \pm 0.012$,  so their central value is closer to the CODATA value than to the muonic Lamb shift value.  The other fit~\cite{Lorenz:2014vha} used the 1422 data points from the Mainz experiment, and obtained $r_p = 0.840 \pm 0.015$ fm (see also~\cite{Lorenz:2014yda}).

Also in this section we may mention two recent reanalyses of world $e$-$p$ scattering data that obtain larger proton radii, one by Sick and Trautmann~\cite{sick:2014,Sick:2012zz} and another by Graczyk and Juszczak~\cite{Graczyk:2014lba}.  The former was particularly mindful of effects of the charge density at large distances upon the charge radius, and obtained $r_p = 0.886(8)$ fm, including all known data; the latter used a Bayesian framework and obtained $r_p = 0.879(7)$ fm, albeit without including the Mainz 2010 data~\cite{Bernauer:2010wm}.


\subsubsection{Vector Meson Dominance (VMD)}
\label{subsubsec:vmd}

\begin{figure}[h]
\centerline{
\includegraphics[width = 3 cm]{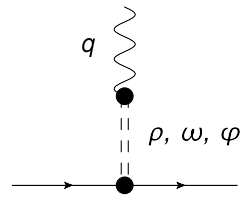}
}
\caption{\small Vector meson dominance picture for the coupling of the  
photon (with four-momentum $q$) to a nucleon. }
\label{vmd}
\end{figure}

The photon has the same $J^{PC}$ quantum numbers as the lowest lying vector mesons $\rho(770)$, $\omega(782)$, and $\phi(1020)$.  These mesons are prominent in the process $e^+e^- \to hadrons$ at the relevant timelike values of the CM energy squared $q^2 > 0$.  One could hence expect that in elastic electron nucleon scattering at low spacelike momentum transfers $q^2 < 0$, some or much of the behavior of the coupling could be explained by the emitted photon converting to the strongly interacting meson which then attaches to the nucleon, as illustrated in Fig.~\ref{vmd}.

That vector meson dominance (VMD) can explain much of the low $Q^2$ behavior of the form factors is confirmed by history.  Before the $\rho$~\cite{Erwin:1961ny}, $\omega$~\cite{Maglic:1961nz}, and $\phi$~\cite{Schlein:1963zz} were explicitly discovered in the early 1960's, in reactions such as $\pi N \to \pi \pi N$ or $e^+ e^- \to {\rm pions}$, there were hints or predictions of the existence gleaned from the behavior of the proton form factor.  Nambu~\cite{Nambu:1997vw} in 1957 suggested that the observed form factor was consistent with the existence of a vector meson intermediary, and Frazer and Fulco~\cite{Frazer:1959gy} in 1959 in a famous dispersion analysis were more explicit, even suggesting a later confirmed mass range for the $\rho$ meson.

A single vector meson exchange with simple couplings gives an $m_V^2/(m_V^2 - q^2)$ factor, from its propagator, for the falloff of the form factor.  One can obtain a $Q^{-4}$ high momentum falloff, in accord with observation or with pQCD, by having cancellations among two or more vector meson exchanges with different masses, or more commonly in practice by giving the vector mesons themselves a form factor in their coupling to nucleons.

\begin{figure}
	\begin{center}
		\resizebox{0.45\textwidth}{!}{%
			\includegraphics[angle=0]{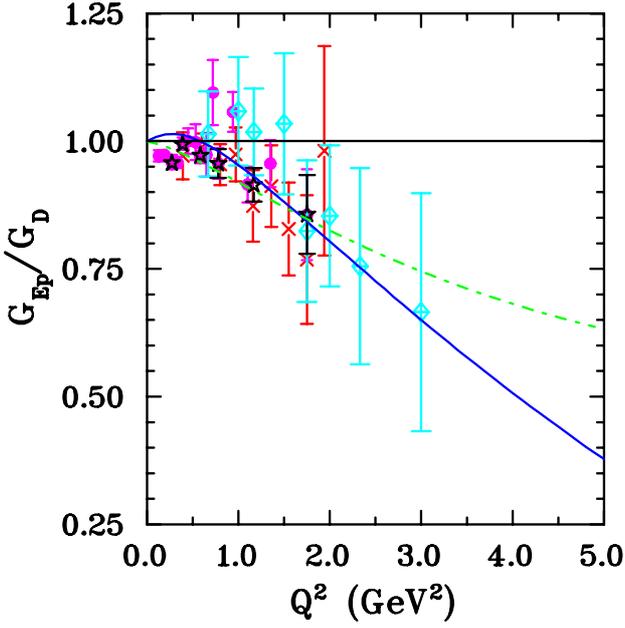}}
		\caption{Two early VMD fits are shown for nucleon form factor data from Iachello {\it et al.} \cite{iachello} 
solid curve (blue),
and Gari and Kr\"umpelmann \cite{gariA,gariB,gariC} short dot-dashed curve (green). The data 
are from Hand {\it et al.} \cite{hand63} open star (black), 
Price {\it et al.} \cite{price} filled circle (magenta), Berger {\it et al.} \cite{berger} multiplication sign (red), 
and Bartel {\it et al.} \cite{bartel} crossed diamond (cyan).} 
		\label{fig:earlygep}
	\end{center}
\end{figure}

An early example of a VMD fit to form factor data was given by Iachello, Jackson, and Lande~\cite{iachello} or IJL.  They had several fits, but the one most cited is a 5 parameter fit with a more complicated $\rho$ propagator that the form noted above, to account for the large decay width of the $\rho$ meson.  (The $\omega$ and $\phi$ are narrow enough that modifying their propagators gives no numerical advantage.)  They in 1973 predicted the falloff of $G_{Ep}/G_{Ep}$ later seen experimentally, as illustrated along with some early data in Fig.~\ref{fig:earlygep}.

The IJL work was improved by Gari and Kr\"umpelmann \cite{gariA,gariB,gariC} to better match the power law pQCD expectations at high $Q^2$, that $F_1 \sim Q^{-4}$ and $F_2 \sim Q^{-6}$, but also including some $\log Q^2$ corrections to the falloffs based on the running behavior of the coupling $\alpha_s(Q^2)$.

Further improvement in VMD fits was made by Lomon \cite{lomon,Lomon:2006xb}, who included  a second $\rho$ as the $\rho'(1450)$, and later also a second $\omega$ as the $\omega'(1419)$, and obtained a good parameterization for all the nucleon form factors.  The first of the polarization transfer $G_{Ep}/G_{Mp}$ measurements~\cite{jones} was available in time for Lomon's 2001 work~\cite{lomon}.  Lomon further tuned his fits~\cite{Lomon:2006xb} when the second set of polarization transfer data became available~\cite{gayou:2002}.

Viewing the form factors as analytic functions of $q^2$, the VMD forms are straightforward to analytically continue to the timelike region (see, for example, Brodsky \textit{et al.}~\cite{Brodsky:2003gs} or Dubnickova \textit{et al.}~\cite{Dubnickova:1992ii}), and compare to data that is now available.  Workers in the field have done so, and have modified the VMD fits to give good accounts of data in both the timelike and spacelike region.

In addition, the original IJL fits~\cite{iachello} were not as good for the neutron as for the proton.  Both the spacelike neutron form factors and timelike nucleon form factors were addressed in what may be termed IJL updates, by Iachello and Wan~\cite{Iachello:2004aq} and Bijker and Iachello~\cite{bijker}, both in 2004.   Further, Lomon and Pacetti~\cite{Lomon:2012pn} have updated and analytically continued the earlier Lomon fits in order to also give a good account of data in both timelike and spacelike regions. 

\begin{figure}[h]
\begin{center}
\includegraphics[angle=0,width = 3.4 in]{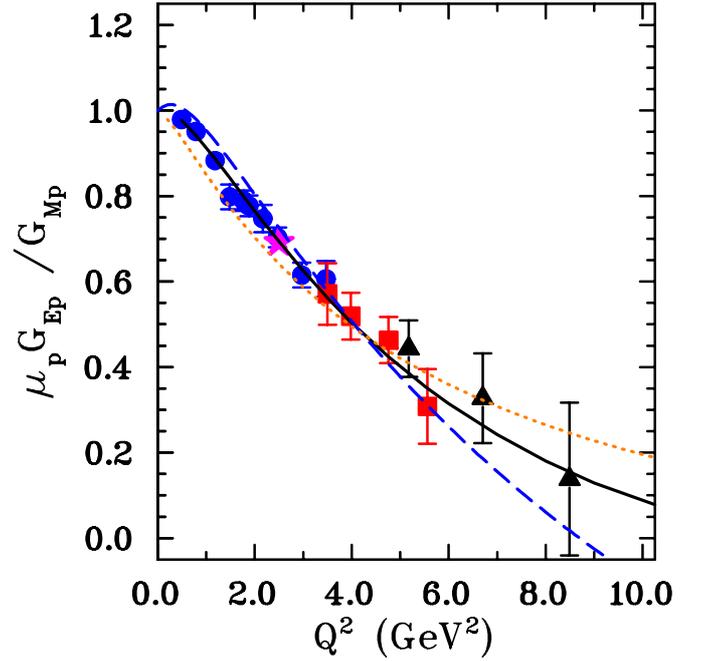}
\caption{Several VMD fits compared to the JLab $G_{Ep}/G_{Mp}$ data.  The solid curve (black) is the fit of Lomon~\cite{Lomon:2006xb}, the 
dashed curve (blue) is that of Iachello, Jackson, and Lande~\cite{iachello}, and the dotted curve (red) is that of Bijker 
and Iachello~\cite{bijker}.}
\label{fig:f2andf1}
\end{center}
\end{figure}

The VMD fits of course are fits to existing data, and they have been regularly updated as new data appeared.  It will be interesting to check the ``predictions'' for the neutron form factors as newer data appears.  A plot of the existing situation for protons is given in Fig.~\ref{fig:f2andf1}.

\subsubsection{Dispersion Analyses}
\label{subsubsec:da}

Newer works here include the dispersive analyses of the nucleon form factors by workers in Bonn~\cite{Lorenz:2014vha,Lorenz:2014yda,Lorenz:2012tm}.  There is also work that is only slightly older by Beluskin \textit{et al.} and Baldini \textit{et al.}~\cite{Belushkin:2006qa,Baldini:2005xx,BaldiniFerroli:2012pr}.  The works include general analyses and fits to the form factors, as well as aspects directly aimed at the resolution of the proton radius puzzle~\cite{Lorenz:2014vha,Lorenz:2014yda,Lorenz:2012tm}.

Dispersion relations relate the form factors in the spacelike and timelike regions.  More generally, the form factors are complex functions of $q^2$ that are analytic except for known cuts, and the form factors anywhere can be calculated if one knows just their imaginary parts at the cuts.  The cuts are all on the real axis for timelike $q^2$.  The cuts run from $q^2 \approx 4 m_\pi^2$ to $q^2 = \infty$.  In practice, one cannot know the imaginary part of the form factors over this whole range, and uncertainty in knowing the form factors in the timelike region builds nonlinearly to larger uncertainty in predicting the form factors in the spacelike region, especially at higher $Q^2$.

At lower $Q^2$, one specific boon of the dispersive treatment is that the connection between the timelike and spacelike regions puts an extra constraint on the form factors and their slope at spacelike threshold.  This means that the determination of the charge and magnetic radii are not purely extrapolations of the scattering data, but is an interpolation in this procedure and hence arguably more reliable.

A technique involving dispersion relations, used by both \cite{Lorenz:2012tm} and by~\cite{Baldini:2005xx}, is to parameterize the imaginary part of the timelike form factors, and determine the parameters by making a least squares fit to the known spacelike and timelike data.  One thereby obtains a representation of the form factors that one can use in regions where there is not yet data.

Ref.~\cite{Baldini:2005xx}, published in 2006, applies dispersion relations to the ratio $G_{Ep}/G_{Mp}$, with the assumption of no zeros in $G_{Mp}$.  They used some large-uncertainty-limit $G_{Ep}/G_{Mp}$ data in the timelike region, obtained from angular distributions in $e^+ e^- \to p \bar p$ or the reverse, to supplement the polarization data in the spacelike region.  One of the main goals was to compare to models that fit the spacelike data, especially to the continuations of those models to the timelike regions~\cite{Brodsky:2003gs}.  

They found that there was a zero in $G_{Ep}(q^2)$ at about $11$ GeV$^2$ spacelike momentum transfer squared, and found that the Phragm\'en-Lindel\"of theorem, which leads to the statement that $|G_{Ep}(q^2)/G_{Mp}(q^2)|$ should be the same at very large momentum transfers, whether spacelike or timelike, was satisfied, albeit with opposite signs.  Many of the more purely phenomenological models differed on the latter point.  This work~\cite{Baldini:2005xx} preceded the completion of the polarization experiments at the highest current $Q^2$~\cite{Puckett:2011,Puckett:2010}.  The dispersive aspects have not been updated in more recent works by some of the same authors, \textit{e.g.}~\cite{BaldiniFerroli:2012pr}, but one can see that the results would not be materially changed by the newer data.

Lorenz \textit{et al.}~\cite{Lorenz:2012tm}, in their update of~\cite{Belushkin:2006qa}, apply the dispersion analysis to $G_{Ep}$ and $G_{Mp}$ separately.  An important improvement in the newer work~\cite{Lorenz:2012tm} is that it includes the recent Mainz data~\cite{Bernauer:2010wm} in its fit.  A salient outcome of this analysis is that the proton charge radius comes out at a value $r_p = 0.84\, (1)$ fm, in agreement with the value found in the muon hydrogen Lamb shift measurement.

\subsubsection{Constituent Quark Models}
\label{subsubsec:cqm}


Constituent quark models (CQMs) have been used to understand the structure of baryons, beginning when quarks were first conjectured and predating the establishment of QCD as the theory of the strong interactions.  Indeed, the observed spectroscopy, particularly the existence of the $\Delta^{++}$, played a crucial role in bringing to light the quantum number of color.  In the CQM, the nucleon is a quantum mechanical ground state of three quarks in a confining potential.  More generally, ground state baryons are composed of three quarks, selected from up ($u$), down ($d$) and strange ($s$) flavors, and are described using $SU(6)$ spin-flavor wave functions and a completely antisymmetric color wave function.

Early CQMs concentrated on explaining static properties, including magnetic moments and transition amplitudes.  Examples are the models of De R\'ujula, Georgi, and Glashow~\cite{derujula} and of Isgur and Karl~\cite{isgurA}.  In the latter, the quarks were in a harmonic oscillator potential, and at least at first the wave functions were nonrelativistic product wave functions, and the ground state baryons appeared as a 56-plet of $SU(6)$.  QCD by this point having been discovered, the hyperfine splittings, \textit{e.g.,} between the nucleon and the $\Delta(1232)$, were given by a one-gluon exchange potential added to the confining potential.  The one-gluon exchange also generates a small tensor interaction that leads to some $D$-state admixture into the ground state baryons.  This in turn allows some non-zero electric quadrupole ($E2$) and Coulomb quadrupole ($C2$) nucleon to $\Delta(1232)$ transitions, in accord with observation.

However, form factors require a relativistic treatment.  At high $Q^2$, nonrelativistic treatments lead to form factors that are far too small compared to data.  At low $Q^2$, the charge radius defined from the slope of the form factor has contributions corresponding to the RMS charge radius known from nonrelativistic treatments, but also has recoil terms proportional to the Compton wavelength or inverse mass, squared,  of the target.  The latter are absent in any nonrelativistic model, and the nucleon is light enough for this to be a problem.  

A crucial question for a form factor calculation, since the nucleon must be moving after or before the interaction or both, is how the wave function in the rest frame transforms to a moving frame.  This is not a trivial question, and the answer can, depending on the formalism, be dependent on the interactions binding the quarks.  Formally, one needs to know how the eigenfunctions of the mass and spin operators can be viewed as unitary representations of the Poincar\'e group, whence it will be known how they change under Poincar\'e, which includes Lorentz, transformations.  The generic ways this can be done were laid out by Dirac~\cite{dirac}.  There are three forms of dynamics, which are the instant, point, and light-front forms.  These differ in which generators form the kinematic subgroup of the Poincar\'e group.  This is the subgroup whose commutators do not involve the interactions among the constituents.  The Poincar\'e group has ten generators, four space-time translations (momentum operators), three spatial rotations, and three boosts.  In a given representation each of these may be kinematical or interaction dependent, or dynamical, meaning dependent on the specifics of the interaction potential.  The latter cannot usually be dealt with in an exact way, but must be dealt with approximately or numerically in a practical calculation.  

The \textit{point} form has all boosts and rotations kinematical,  meaning that as operators in a field theory they can be written out without having to know the interaction Lagrangian or interaction Hamiltonian.  Straightforwardly, the angular momenta and Lorentz boost are the same as in the free case.  However, all four components of the momentum operators are interaction dependent in this case.

The \textit{instant} form has the rotation operators and space components of the momenta kinematical.  Eigenstates of the angular momentum are then easy to construct.  However, the time component of the momentum, or Hamiltonian, and the boosts are dynamical.  Boosts, then, require knowing and including effects of the interaction in order to ascertain important infomation, \textit{e.g.,} the momentum space wave function, of the boosted state.

The \textit{light-front} form has seven kinematical generators.  This is the maximum possible.  The three dynamical generators are one component of the four-momentum operator (for which the mass operator obtained from $\mathcal M^2 = p_\mu^2$ may be substituted) and two light-front transverse rotations (or light-front transverse boosts, meaning here two particular linear combinations of the two ordinary transverse rotations and the two ordinary transverse boosts).  Light-front calculations have the advantage that states can be easily and exactly transformed from one frame to another the using the (kinematic) longitudinal boost, and the two kinematic light-front transverse rotations (the linear combinations orthogonal to the one previously mentioned).  However, light-front calculations have the disadvantage that states of definite angular momentum are difficult to construct because the rotation operators are interaction dependent.

A calculation of the form factors also requires knowing the electromagnetic current operator.  It is usually assumed that the photon only interacts with one quark in the nucleon.

The relative ease of exactly transforming states from the frame where the wave functions are calculated or otherwise given, to any other frame, makes the light-front form attractive for form factor calculations.  The light-front form in this context was introduced by Berestetsky and Terentev~\cite{BerestetskyA,BerestetskyB}, and later developed by Chung and Coester~\cite{chung}.  In these calculations one begins with some wave function that has been developed in CQMs designed to study the baryon spectrum.  The light-front form of the wave function is obtained by a Melosh or Wigner rotation of the Dirac spinors for each quark.  These come about because the usual CQM models use spinors that in momentum space are obtained from rest spinors by a direct boost, while the light-front spinors are obtained by a longitudinal boost followed by a kinematical light-front transverse boost.  Undoing one and then applying the others leads to the same momentum, but leaves a residual rotation that does not affect the momentum, but does affect other quantities such as the spin.  While the Melosh rotation is not difficult conceptually, the expressions it leads to are tedious to write out.  

If, in addition, one calculates in the so-called light-front (or Drell-Yan) frame characterized by $q^+ =0$, then momentum conservation ensures that the current matrix elements connect only states whose Fock components have the same number of constituents.  There are, for example, no matrix elements connecting $qqq$ to $q^4 \bar q$ states, so that a consistent calculation can be done using only three-quark states.

Chung and Coester~\cite{chung} used Gaussian wave functions.  They did obtain a falling $G_{Ep}/G_{Mp}$ ratio.  This apparently~\cite{miller02b} is a feature shared by many relativistic calculations and occasioned by the Melosh transformation.  However, the form factors fell far too fast at large $Q^2$.  Schlumpf~\cite{schlumpfA,schlumpfB} used instead a wave function with a power law falloff, fitting parameters in his wave function to static baryon properties.  The high $Q^2$ falloff was now in line with data, including some at that time new neutron data~\cite{lung}, and also showed a $F_{2p}/F_{1p}$ ratio that fell more slowly than $1/Q^2$, in qualitative agreement with later data.

French, Jennings, and Miller~\cite{gamiller,miller02} focused on the effects of the nuclear medium upon the form factors, but also calculated the form factors of single free nucleons.  They followed the work of Schumpf~\cite{schlumpfA,schlumpfB} and similarly found a decreasing $G_{Ep}/G_{Mp}$ ratio, obtaining a zero between 5 and 6 GeV$^2$ for $Q^2$.

Improvements in the detailed quality of the fit can come by introducing Dirac and Pauli form factors for the quarks.  This can be justified by arguing that nucleons are not just bound states of three quarks, but have further constituents in the form of gluons and quark-antiquark pairs.  Modeling the nucleon with three quarks means the constituent quarks are also representing the additional components present in a complete Fock space expansion of the nucleon, and this gives them an effective structure represented by the quark form factors.  This viewpoint was taken by Carderelli \textit{et al.}~\cite{rome,pace} (plus further references contained in the latter) to produce good fits to both the proton and neutron form factor data then available.  They used the light-front formalism and quark wave functions obtained from a potential of Capstick and Isgur~\cite{Capstick:1986bm}, and made the point that the one-gluon exchange is crucial to obtaining sufficient high momentum components in the wave function to explain the form factor data.

A different starting wave function, now in the context of the point form formalism, appears in the hypercentral constituent quark model of De Sanctis, Santopinto, and others (\cite{Sanctis:2007zz,DeSanctis:2011zz} and references therein).  The feature here is that the confining potential is treated as a function of an average separation defined by the RMS sum of the quark positions, relative to the CM, and there is also a term to give the hyperfine splitting.  Parameters of the potential are fit to the baryon mass spectrum.  With the inclusion here also of form factors for the constituent quarks, good fits are obtained for the nucleon form factors, with updates~\cite{DeSanctis:2011zz} succeeding the latest polarization transfer $G_E^p$ results~\cite{Puckett:2011}.

\begin{figure}[htbp]
\begin{center}
\includegraphics[angle=0,width = 3.37 in]{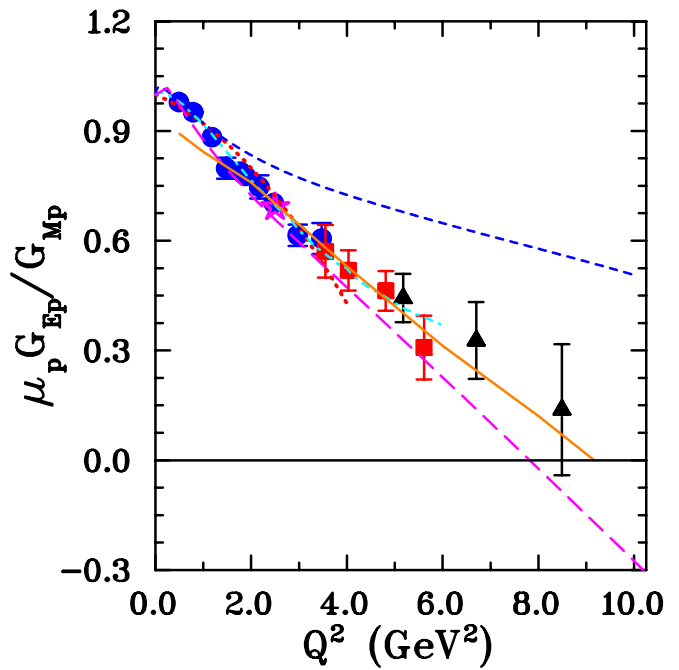}
\caption{The JLab $G_{Ep}/G_{Mp}$ data compared to the results of a selection of constituent quark models.   The short dashed curve (blue) 
is from Boffi \textit{et al.}~\cite{boffi}, the solid (orange) from de Melo \textit{et al.}~\cite{deMelo:2008rj}, the long dash (magenta) from Gross \textit{et al.}~\cite{Gross:2006fg}, the dotted (red) from Chung and Coester~\cite{chung}, and the dash-dot (cyan) from Cardarelli \textit{et al.}~\cite{rome}.}
\label{fig:cqm}
\end{center}
\end{figure}

A comparable amount of high-momentum components in the nucleon wave function was obtained in the Goldstone-boson-exchange (GBE) quark model \cite{Glozman:1997fs,Glozman:1997ag}. This model relies on constituent quarks and Goldstone bosons, which arise as effective degrees of freedom of low-energy QCD from the spontaneous breaking of the chiral symmetry. The resulting CQM assumes a linear confinement potential supplemented by a quarkÐquark interaction based on the exchange of pseudoscalar Goldstone bosons, which is the source of the hyperfine interaction. It was shown \cite{Glozman:1997fs,Glozman:1997ag} that the GBE CQM yields a unified description of light- and strange-baryon spectra. The GBE CQM was used in \cite{Wagenbrunn:2005wk,boffi} to calculate the nucleon e.m. form factors in the point-form. The neutron charge radius is well described in this model and is driven by the mixed-symmetry component in the neutron wave function. In contrast to the light-front calculation \cite{pace,cardarelli}, it was found that when performing a point-form calculation of the nucleon e.m. form factors at larger $Q^2$ within the impulse approximation, i.e. considering only single-quark currents, a surprisingly good overall description of the nucleon e.m. form factors can be obtained, using point-like constituent quarks only. When looking at details of Refs.~\cite{Wagenbrunn:2005wk,boffi}, the agreement is worse though for $G_{Mp}$ which is underpredicted at larger $Q^2$, and the ratio of $G_{Ep}/G_{Mp}$ is overpredicted at larger $Q^2$, see Fig.~\ref{fig:cqm}. Similar findings have also been obtained in the point-form calculation of \cite{wagenbrunn} for the one gluon exchange CQM. The overall success of the point-form result using point-like constituent quarks was attributed in \cite{Wagenbrunn:2005wk,boffi,wagenbrunn} to the major role played by relativity. Such a finding is remarkable in view of the expected finite size of the constituent quarks, as discussed above.

An explanation for the above finding for the nucleon e.m. form factors in the point form, using the single-quark current approximation, has been suggested by Coester and Riska \cite{Coester:2003rw}. When the spatial extent of the three-quark wave function is scaled (unitarily) to zero, both instant and front forms yield form factors independent of the momentum transfer. Therefore, to reproduce the experimental fall-off of the nucleon e.m. form factors at large momentum transfers requires the introduction of constituent quark form factors. In contrast, when the wave function in point form is scaled unitarily to zero (so-called point limit), a non-trivial scaling limit is obtained for the form factors, depending on the shape of the wave function. At high values of momentum transfer, the scaled form factors decrease with an inverse power of the momentum transfer. The power is determined by the current operator and is independent of the wave function. An explicit comparative calculation of the baryon e.m. form factors between the three different forms was performed in \cite{Julia-Diaz:2003gq} using a simple algebraic form for the three-quark wave function, depending on two parameters. It was verified that a qualitative description of the nucleon form factors data demands a spatially extended wave function in the instant- and front-form descriptions, in contrast to the point-form description which demands a much more compact wave function.

A manifestly covariant CQM calculation within the Bethe Salpeter formalism and using an instanton-induced interaction between quarks has been performed by Merten et al. \cite{Merten2002}. Although this model reproduces the baryon spectrum, it can only qualitatively account for the $Q^2$ dependence of the nucleon e.m. form factors.

Another type of covariant CQM calculation was done by Gross, Ramalho, and Pe\~na~\cite{Gross:2006fg}, partly based on earlier work of Gross and Agbakpe~\cite{gross}, avoiding questions of dynamical forms by staying in momentum space.  They use a covariant spectator model, where the photon interacts with one quark and the other two quarks are treated as an on-shell diquark with a definite mass.  They too take the view that the constituent quark includes, at least at lower $Q^2$, effects from higher Fock states and so the quark itself should have a form factor, including the possibility of nonzero quark anomalous magnetic moments.  They note, as others have, that obtaining a good fit to the neutron electric form factor $G_{En}$ requires isospin breaking.  Since they use VMD forms for the quark form factors, the difference between the $\rho$ (isospin-1) and the $\omega$ (isospin-0) couplings is one sufficient way to obtain this.  They obtain forms with parameters that they fit to the data.  Their fit from the 9-parameter ``model IV'' is quite precise.  

Of interest, especially noted in~\cite{Gross:2006fg} although perhaps somewhat relaxed in~\cite{Gross:2012si}, also by Gross \textit{et al.}, is that only $S$-state quark wave functions are needed to obtain good fits to the form factors.  Other work (the article of Brodsky and Drell~\cite{Brodsky:1980zm} is an early example) argues, on the other hand, that nonzero angular momenta are necessary for fitting the form factor data.  This can be a difference in organization of the calculation.  Looking in particular at the magnetic moments, if there are only $S$-states, the anomalous magnetic moments of the nucleons can only come from intrinsic anomalous magnetic moments of the quarks, which can be nonzero if quark form factors are allowed;  whereas if the quarks are all elementary particles their anomalous magnetic moments are zero, and higher Fock states or nonzero orbital angular momenta are needed to produce the observed magnetic moments of the nucleons.

\subsubsection{Pion Cloud Models}
\label{subsubsec:pion}


In nature, the up and down quarks are nearly massless.  In the exact massless limit the QCD Lagrangian is separately invariant under rotations in flavor space of the left handed and right handed quarks.  \textit{I.e.,} there is invariance under a \textit{chiral symmetry} group $SU(2)_L \times SU(2)_R$.  Parity is also a symmetry of the Lagrangian, and fully compatible with the chiral symmetry, and so it would seem that for any state of QCD, there would be an equal mass state of opposite parity.  However, parity doubling is not seen in nature.  Instead we find the alternative, which is that there exist massless negative parity scalars, the \textit{Goldstone bosons}.  For two quark flavors, there are three Goldstone bosons, the pions, which acquire a mass in nature because of the symmetry breaking due to the quark masses.  Since the pions are light, they dominate the long distance behavior of the nucleon wave functions, and have a potentially important effect on the low-momentum transfer behavior of hadronic form factors.  Hence one can improve the constituent quark models by including pionic degrees of freedom.

One early improvement to constituent quark models was in the context of the bag model of hadrons, where a number of workers, in or about 1980, including Brown and Rho~\cite{Brown:1979ui}, Jaffe \cite{Jaffe:1979df}, and Thomas, Th\'eberge, and Miller~\cite{Miller:1979kg} secured the chiral symmetry of the model by including coupling to pions.  The vision was then of quarks within the boundary of the bag and pions without, and the model was called the Cloudy Bag Model.  However, states of the bag model are expressed in coordinate space as set of independent particle wave functions for each quark.  Turning these states into momentum eigenstates is not a trivial problem in general, because there are center-of-mass fluctuations that must be removed, and momentum eigenstates are needed to discuss the form factors.  Additionally, the bag has a simple spherical boundary in the rest frame, and it must be understood how the states alter under Lorentz transformation in order to make accurate form factor calculations for nucleons.  Hence some time passed before Lu \textit{et al.} in 1998~\cite{lu} calculated nucleon electromagnetic form factors in the cloudy bag model. They used the Pierels-Thouless formalism and a plausible hypothesis for the effects of Lorentz transformations  (details in~\cite{lu}), and obtained a good description of the nucleon electromagnetic form factors for $Q^2 < 1$~GeV$^2$.

Miller extended the calculation to larger $Q^2$ using a 
light-front version of the cloudy bag model calculation \cite{miller02b}. 
Starting from constituent quarks \cite{miller02},  using the Schlumpf wave function instead of bag wave functions for the quark core, Miller calculated 
the effects of the pion cloud through one-loop diagrams. The model gives a relatively good overall 
account of the form factor data at both lower $Q^2$ and higher $Q^2$.  

The cloudy bag model is one chiral quark model which 
treats the effect of pions perturbatively. Other quark models 
which calculated nucleon electromagnetic form factors using perturbative pions can be found 
e.g. in the early works of~\cite{Oset:1984tvA,Oset:1984tvB,Jena:1992qx}, as well as in 
the already discussed works of~\cite{Glozman:1997fs,Glozman:1997ag}.  
Recently, the above chiral quark models where pions are included 
perturbatively have been improved in~\cite{faessler}. 
This work extends a previous work of ~\cite{Lyubovitskij:2001nm} by 
dynamically dressing bare constituent quarks by mesons to fourth order 
within a manifestly Lorentz covariant formalism. 
Once the nucleon and $\Lambda$ hyperon magnetic moments are fitted, 
other e.m. properties, such as the
nucleon e.m. form factors at low momentum transfers, follow as a prediction. 
It was found in~\cite{faessler} 
that the meson cloud is able to nicely describe 
the form factor data in the momentum transfer region up to about 0.5 GeV$^2$.  
To extend the calculations to larger $Q^2$, a phenomenological approach 
has been adopted in \cite{faessler} by introducing 
bare constituent quark form factors which were parameterized 
in terms of 10 parameters. Such parameterization makes it plausible to 
simultaneously explain the underlying dipole structure in the nucleon e.m. form factors
as well as the meson cloud contribution at low $Q^2$ which results from 
the underlying chiral dynamics. In a later paper~\cite{Faessler:2006ky}, a 
model calculation for the bare constituent quark form factors has been performed and 
applied to the electromagnetic properties of the $N \to \Delta$ transition, and also of the $N\to$ Roper transition~\cite{2014PhRvD..89a4032O}.   The latter paper also includes updates of that group's fits to the nucleon form factors.

When pion effects dominate nucleon structure, their effects have to be 
treated non-perturbatively. A nonperturbative 
approach which has both quark and pion degrees of freedom 
and interpolates between a CQM and 
the Skyrme model (where the nucleon appears as a soliton solution of an 
effective nonlinear pion field theory) is 
the chiral quark soliton model ($\chi$QSM). 
As for the Skyrme model, 
the $\chi$QSM is based on a $1/N_c$ expansion  
(with $N_c$ the number of colors in QCD).  
Its effective chiral action has been
derived from the instanton model of the QCD vacuum \cite{Dia86}, which
provides a natural mechanism of chiral symmetry breaking 
and enables one to generate dynamically the constituent 
quark mass.  
Although in reality the number of colors $N_c$ is equal to three, 
the extreme limit of large $N_c$ is 
known to yield useful insights. At
large $N_c$ the nucleon is heavy and can 
be viewed as $N_c$ ``valence" quarks bound by a self-consistent pion
field (the ``soliton")~\cite{Dia88}.
A successful description of static properties of baryons, 
such as mass splittings, axial constants, magnetic moments, 
form factors, has been achieved (typically at the 30 \% level or better, 
see~\cite{Chr96} for a review of early results). 
After reproducing masses and decay constants in the mesonic sector, 
the only free parameter left to be fixed in the baryonic sector
is the constituent quark mass. 
When taking rotational ($1/N_c$) corrections into account, 
this model achieved a qualitative good description of the nucleon  electromagnetic 
form factors in the range $Q^2 < 1$~GeV$^2$, using a constituent 
quark mass around $420$~MeV~\cite{Christov:1995hr}.
The chiral soliton model obtained a decrease of the 
$G_{Ep} / G_{Mp}$ ratio with increasing $Q^2$ already in the late 1990's.


Holzwarth~\cite{HolzwarthB} extended the chiral soliton model by including the 
$\rho$ and $\omega$ meson propagators for the isovector and isoscalar channels, 
respectively. Furthermore, to extend the range in $Q^2$ of the predictions, 
he uses a relativistic prescription to boost the soliton rest frame densities 
to the Breit frame. Such prescription is also used to extract 
radial charge and magnetization rest frame densities 
from experimental form factors, as will be discussed in Sect.~\ref{subsubsec:td}. 
Using 4 fit parameters (one effective boost mass and three free parameters 
to fix the couplings of $\rho$ and $\omega$ mesons), 
the model was found to provide a good account of 
the detailed structure of the nucleon e.m. form factors in the low $Q^2$ region. 
In particular, for $G_{Ep}/G_{Mp}$ it 
predicts a decreasing ratio in good agreement with the data. 
At larger $Q^2$, the boost prescription gives a reasonably good account of the 
data (except for $G_{Mn}$) and predicts a zero in $G_{Ep}$ around 
10~GeV$^2$. Due to the uncertainty introduced from 
the particular choice for the boost prescription, the 
high $Q^2$ behavior (for $Q^2$ larger than about $4 m_p^2$) 
of the e.m. form factors is however not a profound prediction of the 
low-energy effective model.  

Clo\"et and Miller~\cite{Cloet:2012cy}, in addition to fitting the electromagnetic form factors with a quark plus pion cloud model, have the further goal to accurately reproduce the spin fraction of the proton that comes from quark spin.  It will be remembered that the EMC collaboration found that little of the proton spin came from quark spin, and a more modern analysis~\cite{deFlorian:2008mr} gives the quark spin fraction of the proton spin as $36.6^{+1.2}_{-1.6}$\% (for $x_{\rm min}=0.001$ and $1 \sigma$).  The Clo\"et-Miller model uses a light-front formalism with a quark-diquark system accounting for 71\% of the nucleon state, by probability, plus a quark-diquark core with a pion accounting for the rest.  Gluons are not explicitly included.  They have 10 parameters, which they fit to the electromagnetic form factor data, obtaining a good representation of the data and/or the Kelly fit thereto.  They have a zero of $G_{Ep}$ at $Q^2 = 12.3$ GeV$^2$.  The total quark plus diquark spin they obtain is 36.5\% of the proton's $\hbar/2$, in fine accord with expectation.

\subsubsection{Transverse Densities}
\label{subsubsec:td}


Nonrelativistically, form factors and charge distributions are Fourier transforms of each other.  Relativistically, this is no longer true, because of quandaries in transforming the nucleon wave function from one reference frame to another.  In terms of the charge or magnetic radius this leads to additive terms that may be called recoil terms, proportional to $Q^2/m_N^2$ with coefficients that are not trivial to calculate.  For heavy targets, these terms may be ignored, but the nucleon is light enough that precision work cannot proceed nonrelativistically.  One cannot obtain accurate spatial images of the nucleon charge or magnetic densities by just Fourier transforming the charge or magnetic form factors.

However, one can obtain kinematically correct, and accurate to the same level that the input data is accurate, charge distributions if one is willing to adopt a new viewpoint.  The viewpoint is that of a light front moving towards a nucleon, or equivalently of an observer viewing a nucleon approaching at nearly the speed of light.  The charge density that will be seen is two dimensional, or the 3D charge density of the nucleon projected onto a plane transverse to its direction of approach.  Additionally, the charge density that will be seen is not the charge density obtained from the electromagnetic current component $J^0$, but rather densities coming from $J^+ = J^0 + J^3$.

Typically one chooses the z-direction as special, choosing it along the direction $P = (p+p')/2$, where $p$ and $p'$ ate the incoming and outgoing nucleon momenta, and  further arranges the frame so the photon momentum has $q^+ = 0$, and its transverse (lying in the $xy$ plane) momentum is denoted $\vec q_\perp$.  The charge density projected onto the transverse plane is shown in works of Burkardt and of Miller and of others~\cite{Burkardt:2000za,Miller:2007uy,Carlson:2007xd,Miller:2010nz} to be
\begin{eqnarray}
\rho_{\lambda N}(\vec b) &=& \int \frac{d^2 \vec q_\perp}{(2 \pi)^2 \, 2P^+} \,  
e^{- i \, \vec q_\perp \cdot \vec b}
										\nonumber \\
&&	\ \ 	\times  
\langle P^+, \frac{\vec q_\perp}{2}, \lambda \,|\, J^+(0) \,|\, 
P^+, -\frac{\vec q_\perp}{2}, \lambda  \rangle,
\end{eqnarray}
where $\vec b$ is the position in the $xy$ plane relative to the nucleon's CM, $\lambda$ is the (light-front) helicity, and the longitudinal and transverse components  of the incoming and outgoing nucleon's momenta are indicated separately.  For the density $\rho_{0N}$ of an unpolarized nucleon one finds, following Miller~\cite{Miller:2007uy}, that
\begin{eqnarray}
\rho_{0N}(b) = \int_0^\infty \frac{d Q}{2 \pi} \, Q \, J_0(b \, Q) F_1(Q^2), 
\end{eqnarray}
where $Q^2 = \vec q_\perp^2$,  $J_0$ is the Bessel function, and $F_1$ is the Dirac form factor.  Further, if one polarizes the nucleon transversely in the $x$-direction, one obtains~\cite{Carlson:2007xd},
\begin{eqnarray}
\rho_{TN}(\vec b) &=& \rho_{0N}(b) 		\nonumber	\\
	&-& \sin \phi_b  \, 
		\int_0^\infty \frac{d Q}{2 \pi} \frac{Q^2}{2 M_N} \, J_1(b \, Q)  F_2(Q^2), 
\end{eqnarray}
where $\phi_b$ is the azimuthal angle of $\vec b$, and $F_2$ is the Pauli form factor.

\begin{figure}[htbp]
\begin{center}
\includegraphics[width = 83 mm]{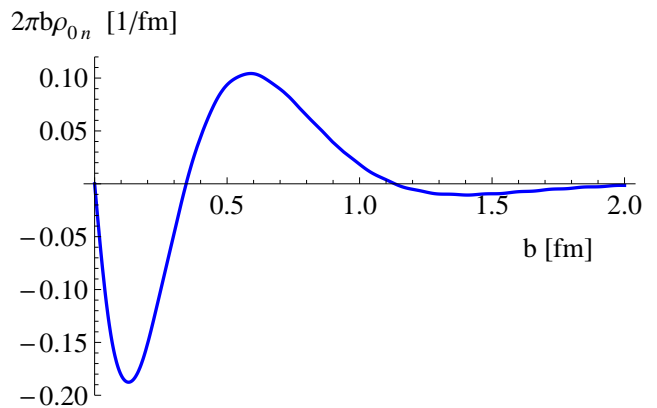}
\caption{The 2D projected charge density of a neutron moving rapidly towards an observer, shown as a function of the distance from the CM of the neutron.}
\label{fig:transdensn}
\end{center}
\end{figure}

The result for the unpolarized density distribution of the proton is not startling and is not shown here, but the result for the neutron is quite striking~\cite{Miller:2007uy}.  The neutron charge density is found to be negative near its center as shown in Fig.~\ref{fig:transdensn};  it had long been known to be negative far from the center.

\begin{figure}[htbp]
\begin{center}
\includegraphics[width = 73 mm]{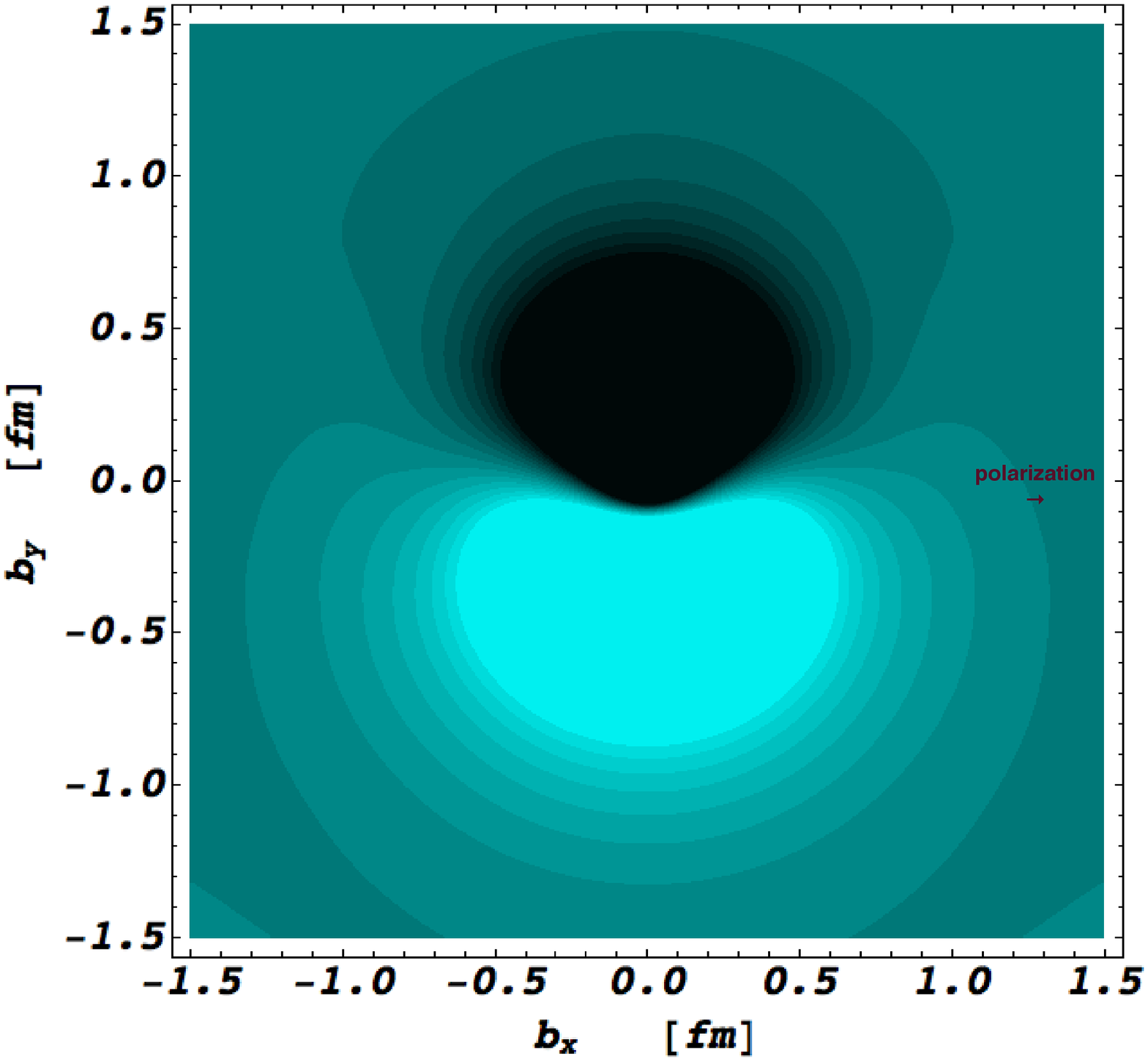}

\vskip 5 mm
\includegraphics[width = 78 mm]{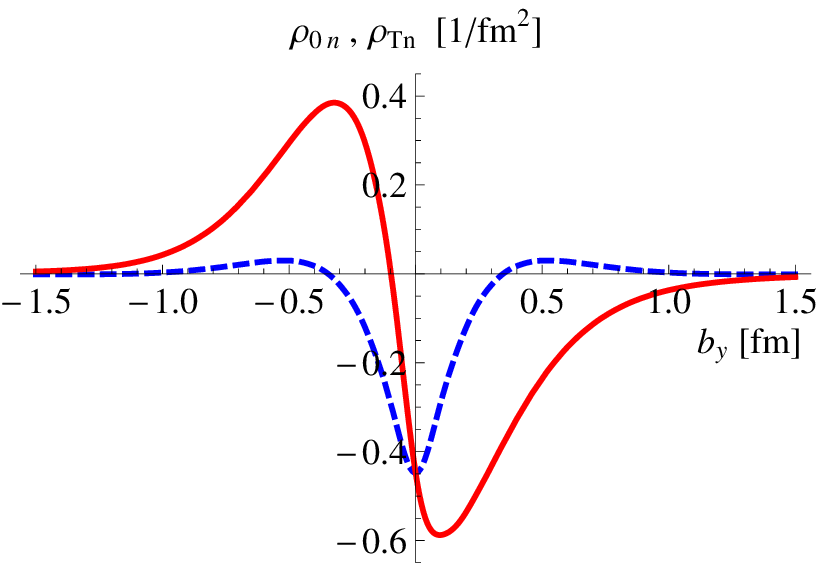}
\caption{Upper panel: the 2D projected charge density $\rho_T^n$ for a neutron polarized in the $x$-direction (to the right, in the figure).  Dark areas represent negative charge, light areas, positive charge.  Lower panel: the charge densities along the $y$-axis for a neutron polarized along the $x$-direction;  $\rho_{0n}$ is given by the dashed curve (blue), and $\rho_{Tn}$ is given by the solid curve (red).}
\label{fig:transspinn}
\end{center}
\end{figure}

For a transversely polarized neutron, one finds a charge separation as shown in Fig.~\ref{fig:transspinn}, based on figures in~\cite{Carlson:2007xd}.   The upper panel shows the difference between the charge densities of the polarized and unpolarized neutron.  One sees negative charge above and positive charge below.  The lower panel gives similar information by comparing the polarized and unpolarized charge distributions along a single line, the $y$-axis.  (One should know that time reversal invariance forbids an electric dipole moment for a stationary elementary particle, but for a moving particle the electric dipole moment $\vec d$ is given by $\vec d = \vec v \times \vec \mu$, where $\vec v$ is the velocity of the particle and $\vec \mu$ is its magnetic moment.)

\subsubsection{Correspondences with Higher Dimensional Theories}
\label{subsubsec:ads}


A recent, exotic, and interesting way to obtain approximate QCD results is to use the anti-de Sitter space/conformal field theory (AdS/CFT) correspondence, initiated by Maldacena~\cite{Maldacena:1997re}, where the most relevant approximately conformal field theory is QCD and we talk of the AdS/QCD correspondence, where early applications were made by Erlich \textit{et al.}~\cite{Erlich:2005qh}.  The idea is that some string theory in ten dimensional space has a symmetry that is divided so that for five dimensions we are and remain in the ground state in those dimensions, and the theory in the other five dimensions mimics a gravitational theory in a five dimensional anti-de Sitter space, or AdS$_5$.  The AdS$_5$ possesses a $SO(2,4)$ symmetry, which is important because it is the same symmetry possessed by a conformal (in practice, a theory with all masses zero) field theory in four dimensions.   One can exploit the sameness of the symmetry group to find numerically viable relations between the solutions to the gravitational theory in five-dimensions and the conformal field theory in four-dimensions.    For an extensive review, see~\cite{Brodsky:2014yha}.

Since QCD is not a conformal theory, maintaining a correspondence requires also breaking the symmetry of the AdS space, which in practice is done either with the hard wall model, where the AdS space is cut off at long distances in the fifth dimension, or by the soft wall model, where an extra potential is introduced that suppresses long distance propagation in the fifth dimension. 
 
Much of the work on individual particles has focused on the bosonic sector, studying both quarkic hadrons and glueballs of various spins, and obtaining masses, decay constants, and charge radii.  The actual correspondence is between operators in the four-dimensional space, such as the electromagnetic current or the energy momentum tensor $T_{\mu\nu}$, and fields in the five-dimensional space with corresponding quantum numbers, such as a vector field or the metric $g_{\mu\nu}$.   Results can be compared to experiment at the few times 10\% level~\cite{Brodsky:2014yha}.

 Obtaining results for fermions is more involved than for bosons.  One approach is to build from the bosonic sector, and treat the fermions as Skyrmions within that model~\cite{Hata:2007mb,Pomarol:2008aa}.  Another approach is to begin in AdS with fundamental fermions that interact with the AdS gravitational background~\cite{Henningson:1998cd,Mueck:1998iz,Contino:2004vy,Hong:2006ta,Brodsky:2008pg,Abidin:2009hr,Gao:2009ze,Lyubovitskij:2014lja}.  In this version the five-dimensional Lagrangian is still relatively simple.  It has terms for gravity (the scalar curvature and the cosmological constant terms) and for vector fields, and the terms for the fermions are simple given the context of interactions with gravity and the vector field, 
\begin{equation}
\mathcal L_F = 	
\sqrt{g}\,  \Big( \frac{i}{2} \bar{\Psi} e^N_A \Gamma^A D_N \Psi	 
 	    -\frac{i}{2}(D_N \Psi)^\dagger \Gamma^0 e^N_A \Gamma^A \Psi
	    - M \bar{\Psi}\Psi  \Big),
\end{equation}
for the hard wall version of the model, and the covariant derivative is $D_N=\partial_N +\frac{1}{8}\omega_{NAB}[\Gamma^A,\Gamma^B]-iV_N$, where $V_N$ is the vector field which will be dual to the electromagnetic current when we make the correspondence.  The indices $N$, $A$, $B$, \textit{etc.,} each run over five values as appropriate to the five-dimensional space.  In the above Lagrangian, gravity enters via the metric $g_{MN}$ and is seen in its determinant $g$ and also in the spin connection $\omega_{NAB}$, which will not be given in detail here.  

\begin{figure}[htbp]
\begin{center}
\includegraphics[width = 73 mm]{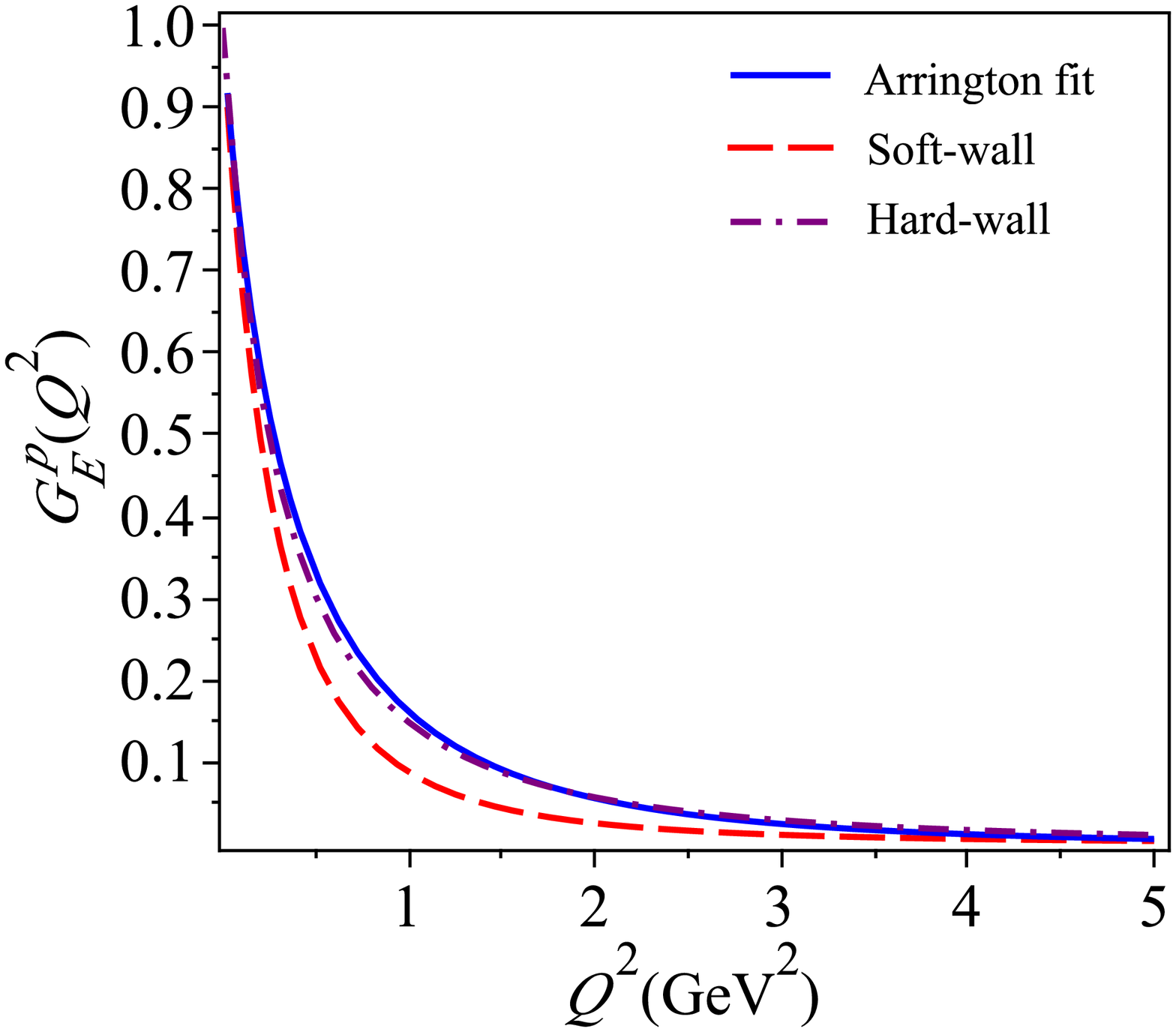}
\includegraphics[width = 78 mm]{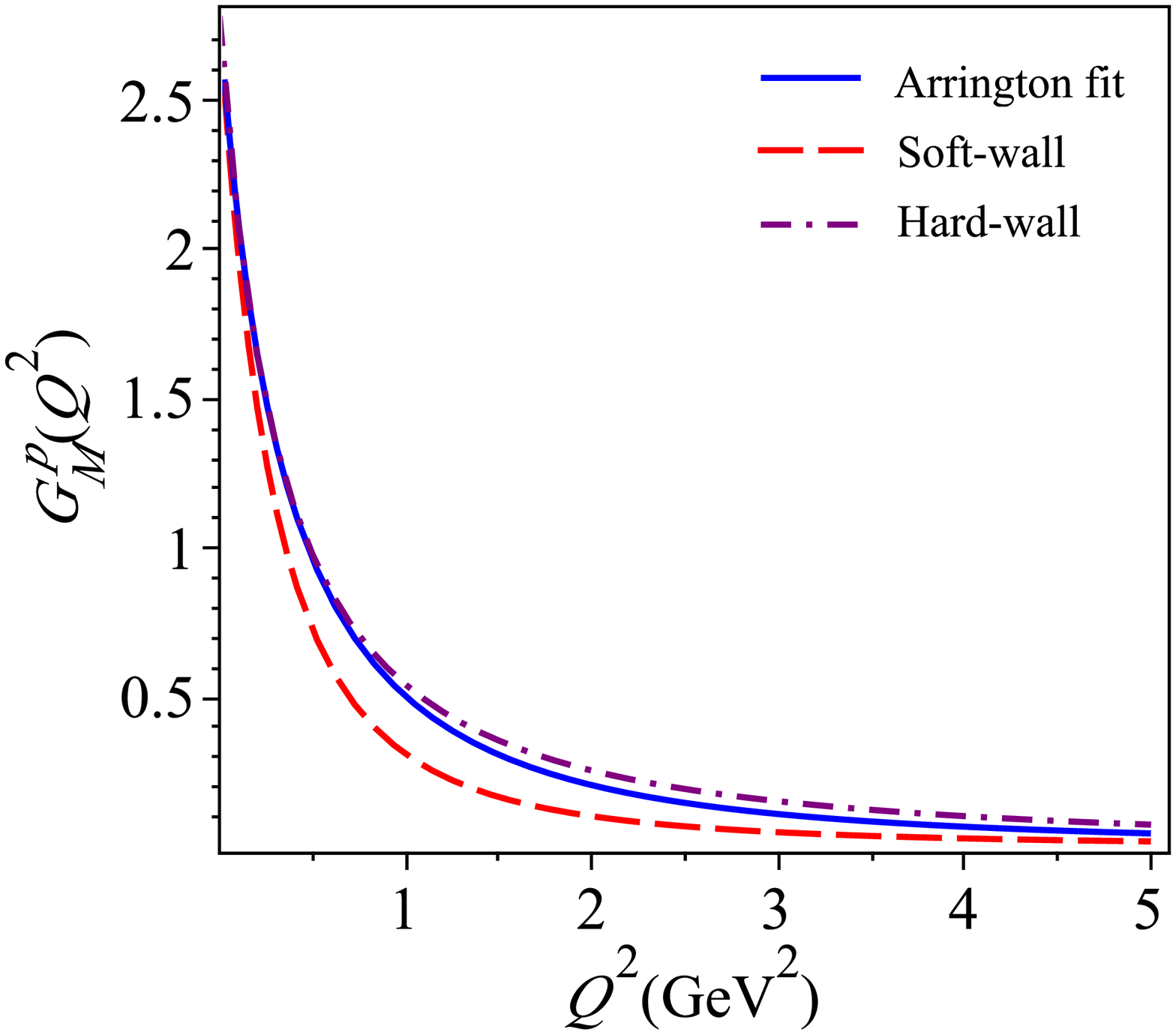}
\caption{Results for the proton electromagnetic form factors obtained using the AdS/CFT correspondence~\cite{Abidin:2009hr}, with the soft wall model in dashes (orange) and the hard wall version in dot-dash (violet), compared to the fits of Arrington \textit{et al.}~\cite{Arrington:2007ux} shown as solid lines (blue).}
\label{fig:ads}
\end{center}
\end{figure}

As an example, results obtained using the AdS/CFT correspondence for the proton electromagnetic form factors by Abidin \textit{et al.}~\cite{Abidin:2009hr} are shown in Fig. \ref{fig:ads} for both the hard wall and soft wall models.

\subsection{Dyson-Schwinger Equations and Diquark Models}
\label{subsec:dse}


The Dyson Schwinger equations (DSE) are generically a non-perturbative approximation for obtaining results for a field theory, in the present case QCD.  The equations are, in principle, an infinite set of coupled integral equations.  In practice, they must be truncated, in a way that preserves all symmetries of QCD, in order to proceed with any calculation. For a general DSE review, see~\cite{Bashir:2012fs}.   

One accomplishment of the DSE follows the solution for the full quark propagator, represented in momentum space as
\begin{equation}
S_F(p^2) = \frac{ i F(p^2) }{ \not\! p + M(p^2) }	\,,
\end{equation}
where the normalization $F(p^2)$ and the mass $M(p^2)$ become functions of momentum because of interactions.  With relatively simple truncations and modeling of the QCD interactions, the DSE obtain a mass function in good agreement with lattice calculations.  The mass $M(p^2)$ is several hundred MeV at small $p^2$ and falls smoothly to the small values at large $p^2$ that one might expect in perturbation theory.

Also early in the DSE program is building a model of the quark-quark and quark-antiquark interactions that will reproduce data on, among other quantities, the pion mass and decay constant.

One then uses the same quark-quark interactions developed in meson studies to obtain a three quark wave function model for the nucleon by solving the three-cody Fadeev equations.  There arise significant diquark contributions, \textit{i.e.,} significant quark-quark correlations, which have a strong effect on the form factors one obtains.  Since the quarks in this model are dressed,  many of their features are different from expectations for pointlike fermions.  One finds in particular large quark anomalous chromomagnetic moments, which affect the quark-gluon interactions, which lead to large quark anomalous magnetics moments in the quark-photon interactions, which in turn are needed to obtain good fits to the nucleon electromagnetic form factor data.  

\begin{figure}[htbp]
\begin{center}
\includegraphics[width = 3.37 in]{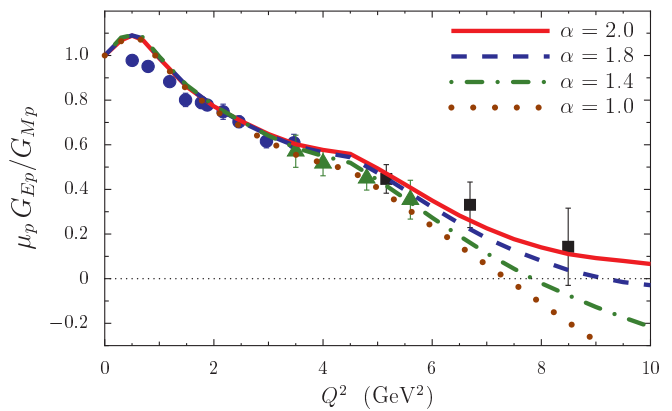}
\caption{An illustration of what may be learned about bound state substructure, in the context of a DSE model~\cite{Cloet:2013gva}, from the measured $G_{Ep}/G_{Mp}$ ratio.  Parameter $\alpha$ measures how quickly the dressed or constituent quark mass approaches its asymptotic or perturbative value; $\alpha =1$ is a benchmark favored by fits to meson masses, and larger values of $\alpha$ accelerate the rate of approach to the asymptotic quark masses.  The data is from~\cite{gayou:2002,jones,punjabi05B,Puckett:2010}.}
\label{fig:crt}
\end{center}
\end{figure}

The theoretical DSE results, from the work of Clo\"et, Roberts, and others~\cite{Cloet:2014rja,Cloet:2013gva,Cloet:2013jya}, show a falloff of the $G_E^p$/$G_M^p$ ratio similar to what is seen in the data;  see Fig.~\ref{fig:crt}.

Qualitatively, the behavior of the form factors in the DSE approach is related to the behavior of the mass function $M(p^2)$.  At lower momenta, where the mass function is far from its perturbative or current quark value, the Pauli form factor is also large compared to its perturbative value and is falling more slowly than perturbation theory predicts.  (For reference, perturbative QCD predicts a $Q^{-4}$ power law falloff for $F_1(Q^2)$ at large $Q^2$, and a $Q^{-6}$ falloff for $F_2(Q^2)$. )  Hence one can get a zero in $G_E(Q^2)$,
\begin{equation}
G_E(Q^2) = F_1(Q^2) - \frac{Q^2}{4 M_p^2} F_2(Q^2)	\,,
\end{equation}
and hence a falloff in the ratio $G_E(Q^2)/G_M(Q^2)$.

For the newer DSE calculations reported by Segovia \textit{et al.}~\cite{Segovia:2014aza}, the zero in $G_E$ is at $Q^2 = 9.5$ GeV$^2$.  If the mass function fell to its low perturbative value more quickly than it does, the quarks would behave more like free quarks, and the value of the Pauli or anomalous magnetic moment, form factor would be small as well as quickly falling.  In such a case, the zero of $G_E(Q^2)$ would be pushed to higher values of $Q^2$ or possibly not occur at all~\cite{Segovia:2014aza}.

We may mention that models based on the Dyson-Schwinger equations, as well as many of the other models discussed, do extend to  form factors for other hadronic reactions, such as the electromagnetic $N \to \Delta$ transition~\cite{Segovia:2014aza,Segovia:2013uga}.  The result for the ratio of the electric and magnetic transition form factors for this process, $R_{EM}$, turns out to be small in the DSE approach,  even at momentum transfers above 5 GeV$^2$,  in accord with experimental data.   The perturbative QCD result, that $R_{EM} \to 1$, may well ensue, but only at momentum transfers well above those now experimentally accessible.

\subsection{Links between Deep-Inelastic Scattering and Nucleon Form Factors}
\label{subsec:dis}

%

\subsubsection{Perturbative QCD Inspired Models}
\label{subsubsec:pqcd}

\begin{figure}[h]
\centerline{  
  \includegraphics[width = 6.5 cm]{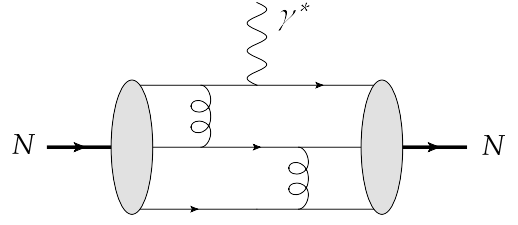} 
}
\caption{\small Perturbative QCD picture for the nucleon 
electromagnetic form factors. The highly virtual photon resolves 
the leading three-quark Fock states of the nucleon, described by 
a distribution amplitude. The large momentum is transferred between the quarks 
through two successive gluon exchanges (only one of several possible 
lowest-order diagrams is shown). }
\label{ff_pqcd}
\end{figure}

Perturbative QCD (pQCD) predicts the scaling behavior of the nucleon electromagnetic form factors are high $Q^2$.  The predictions were given by Brodsky and Farrar~\cite{brodsky,Brodsky:1973kr} and by Matveev, Muradyan, and Tavkheledze \cite{Matveev:1973}, and pQCD not only predicts the scaling behavior of each amplitude, but also predicts that the leading amplitude is the one where the hadron helicity is maintained in the interaction.  A photon of high $Q^2$ sees the nucleon as a set of three parallel moving quarks, one of which absorbs the high momentum photon, whose momentum is brought in from a sideways direction. In order to reconstitute the proton, the momentum must be shared out by two hard gluon exchanges, so that one also has three parallel moving quarks in the final state, as illustrated in the Fig.~\ref{ff_pqcd}.  

The overall hard amplitude can be factored~\cite{Chernyak:1977asA,Chernyak:1977asB,Chernyak:1977fkA,Chernyak:1977fkB,Efremov:1979qk,brodlep} as a product of a hard scattering amplitude that takes three parallel moving quarks into three parallel moving quarks, and two distribution amplitudes (DA) that specify how the longitudinal momentum of the nucleon is divided among the quarks.  Each gluon carries a virtuality proportional to $Q^2$ (and there are also factors $1/Q$ from each of the internal quark propagators, factors of $Q$ from each of the thoroughgoing quark lines, and a $1/Q$ involved in the definition of $F_1$), leading to a pQCD prediction that the helicity conserving Dirac form factor $F_1$ will fall like $1/Q^4$, with possible $\log Q^2$ factors, at high $Q^2$.   The Pauli form factor $F_2$ requires a helicity flip between the final and initial nucleon, which in turn requires, thinking of the quarks as collinear, a helicity flip at the quark level, which is suppressed at high $Q^2$.  The result is a prediction that $F_2$ will fall like $1/Q^6$ at high $Q^2$.  Hence, $G_E$ and $G_M$ will both fall like $1/Q^4$ asymptotically.

\begin{figure}[h]
\centerline{  \includegraphics[angle=90,width = 3.37 in ]{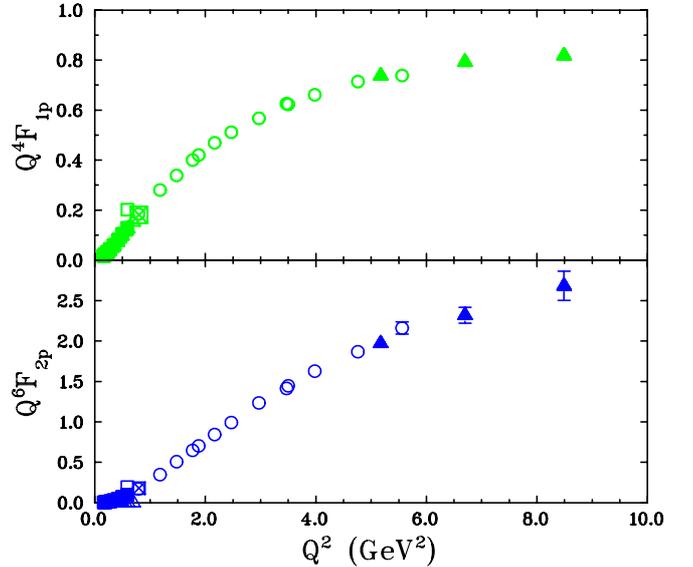}  }
\caption{\small Test of the scaling behavior of the proton form factors.  
Upper panel: proton Dirac form factor multiplied by $Q^4$.   
Lower panel: proton Pauli form factor multiplied by $Q^6$. In both panels $G_{Ep}$ was obtained from the JLab $\mu_pG_{Ep}/G_{Mp}$ data using
the Kelly parametrization for $G_{Mp}$ \cite{kelly04}.  
}
\label{fig:scaling}
\end{figure}

We can see how well pQCD predicts current electromagnetic proton form factor data by examining Fig.~\ref{fig:scaling}.  The figure shows data up to 10 GeV$^2$ for $Q^4 F_1$ (upper panel) and $Q^6 F_2$ (lower panel).  For $F_1$, it appears that the curve is flattening out, as pQCD would predict, and indeed there is data for $F_1$ up to 31 GeV$^2$ to corroborate this.  However, for $Q^6 F_2$, where there is no further data currently, the existing data does not match the simple pQCD expectation.

\begin{figure}[htbp]
\begin{center}
\includegraphics[angle=90,width = 83 mm]{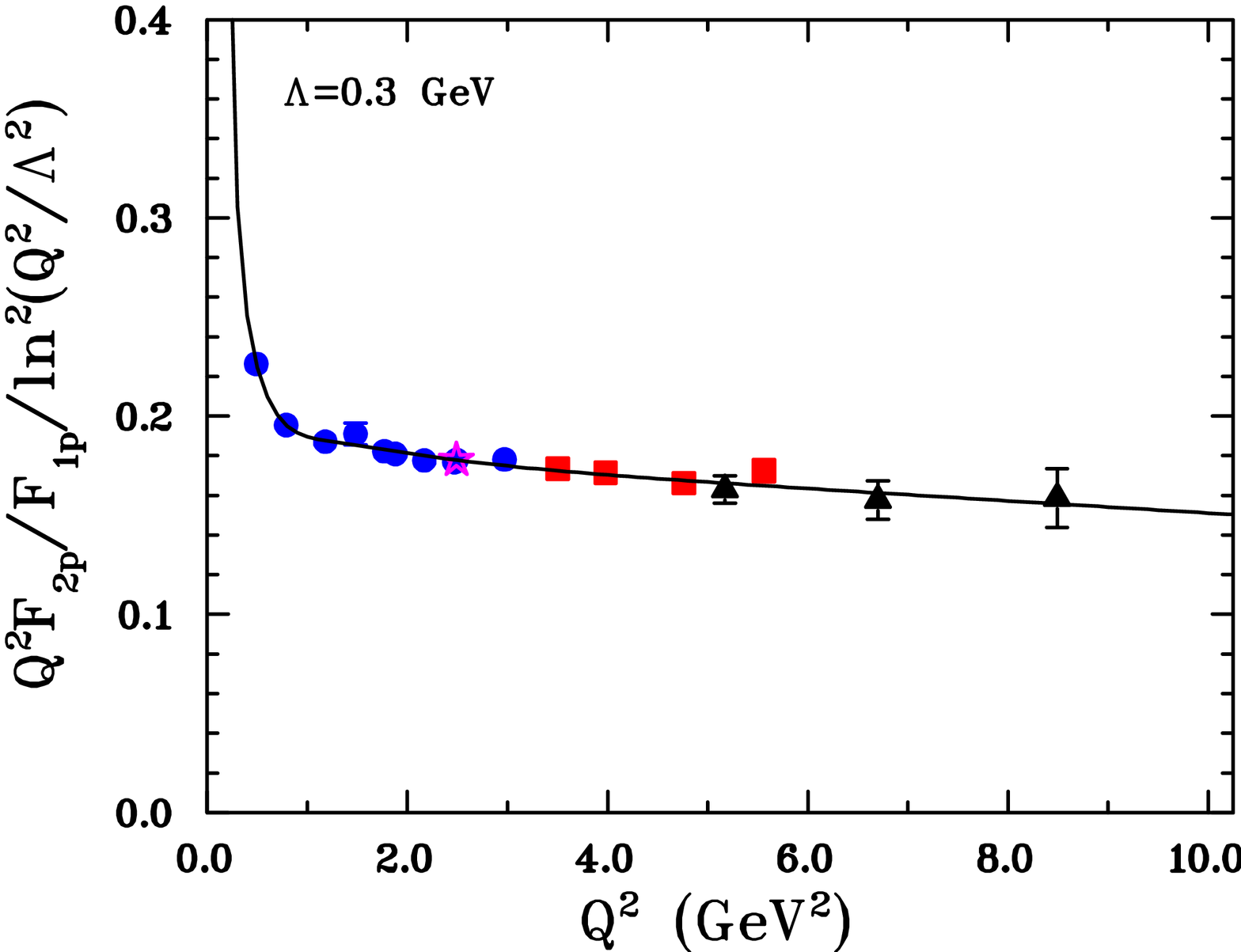}
\vskip 5 mm
\includegraphics[angle=90,width = 83 mm]{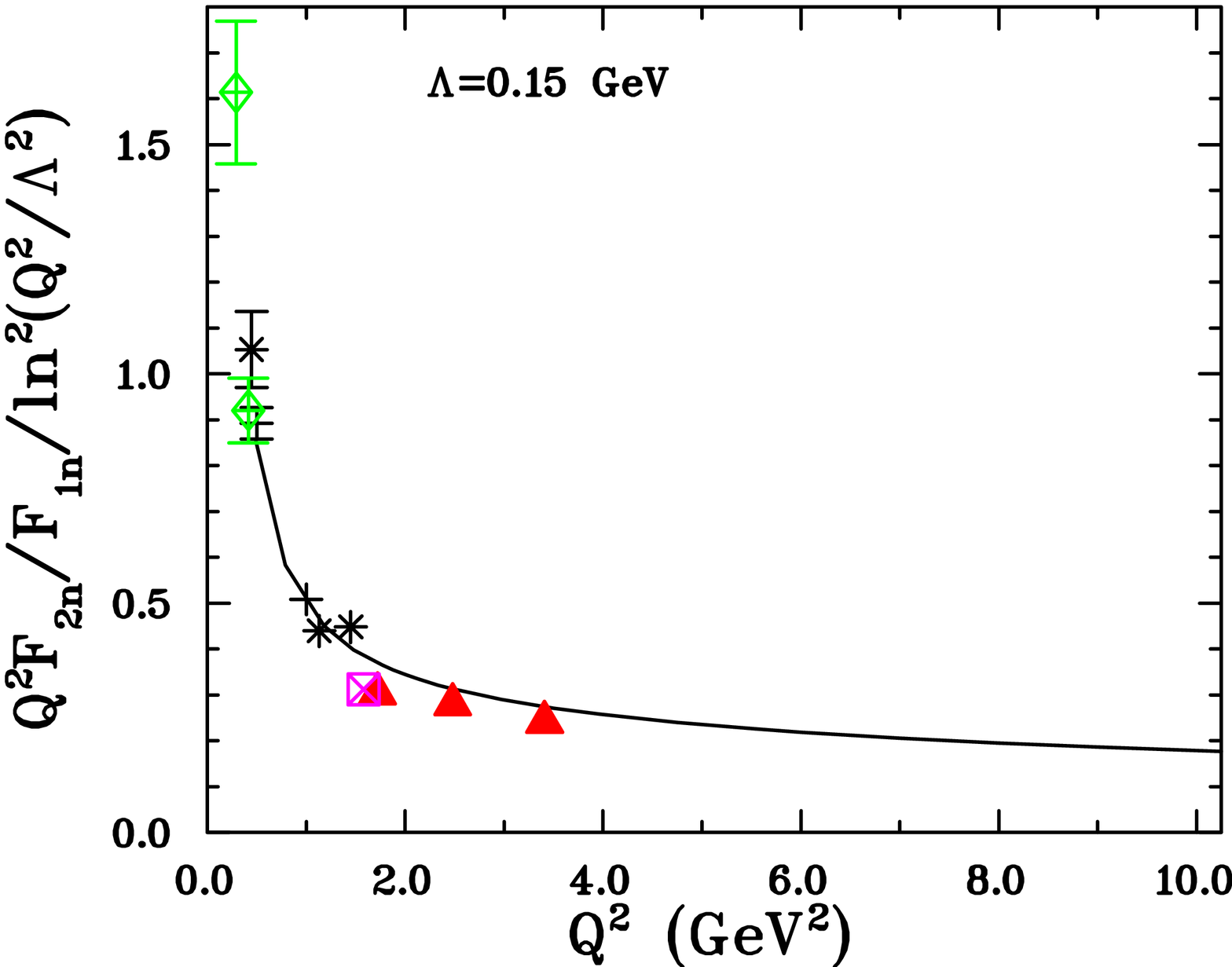}
\caption{Upper panel: Data from the JLab plotted as $Q^2 F_{2p}/F_{1p}/ln^2(Q^2/\Lambda^2)$ as proposed by Belitsky \textit{et al.}~\cite{Belitsky:2002kj}.  The solid curve is the fit to the data using Eqn. 44. Lower panel: the same but for the neutron, data from \cite{warren,Riordan:2010,madey,geis:2008,schlimme:2013}. The solid curve is calculated from Kelly fit \cite{kelly04}.}
\label{fig:bjy}
\end{center}
\end{figure}

The data show that $F_{2 p}/F_{1 p}$ falls slower than $1/Q^2$ with increasing $Q^2$.
Belitsky, Ji, and Yuan \cite{Belitsky:2002kj} investigated the assumption of quarks moving collinearly with the proton, which underlies the pQCD prediction.  They have shown~\cite{Belitsky:2002kj} that by including components in the nucleon light-cone wave functions with nonzero quark 
orbital angular momentum projection, they obtain a behavior $F_2/F_1 \to \ln^2 (Q^2 / \Lambda^2)/ Q^2$ at large $Q^2$,  with $\Lambda$ a non-perturbative mass 
scale.   (Refs.~\cite{jain,Brodsky:2003pw} also discuss using quark orbital angular momentum to get a ratio $F_{2p}/F_{1p}$ which drops slower than $1/Q^2$ with increasing $Q^2$.)
With $\Lambda$ around $0.3$~GeV~\cite{Belitsky:2002kj}, the 
data for $F_{2 p}/F_{1 p}$ agree with such double-logarithmic enhancement, as seen in Fig.~\ref{fig:bjy}, where it may be noted that the higher $Q^2$ data was obtained after~\cite{Belitsky:2002kj} was published.    The arguments of~\cite{Belitsky:2002kj} do still rely on pQCD, and it remains to be seen if still higher $Q^2$ data will continue to support this amended prediction. 

Hard scattering is calculated as if all three quarks are moving fast.   An alternative is that one quark carries nearly all the nucleon's momentum, and the other two quarks are soft.  It is not necessary to transfer momentum to the soft quarks before reconstituting the proton.  Nesterenko and Radyushkin~\cite{Nesterenko:1983ef} point out that the hard scattering mechanism requires the exchange of two gluons, each of which brings in a suppression factor $\alpha_s / \pi \sim 0.1$. One therefore see that the hard scattering mechanism for $F_1^p$  
could be numerically suppressed relative to the soft term, also called the Feynman mechanism;  see also~\cite{bolz,kroll}. 

\begin{figure}[b]
\begin{center}
\includegraphics[width = 86 mm]{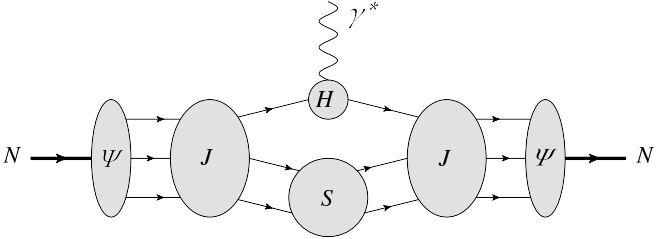}
\caption{Soft contributions to the nucleon form factor.  The kernels $J$ connect three quarks with finite momentum to a configuration with two soft quarks and a hard quark interacting with the photon.  In the simplest case, each kernel $J$ would contain two gluon exchange.}
\label{fig:kv}
\end{center}
\end{figure}

Early on, Duncan and Mueller~\cite{Duncan:1979hi} showed that the soft or Feynman process also gave a $1/Q^4$ falloff, with logarithmic corrections, for the Dirac form factor $F_1$.  This has been taken up more recently by Kivel and Vanderhaeghen~\cite{Kivel:2010ns,Kivel:2012zz}, who were able to show that also for the Feynman process, a type of factorization was possible, where the second step is given by a process independent kernel that transfers momentum among the initially all finite momentum fraction quarks to make two of them soft, as in Fig.~\ref{fig:kv}.  They also considered the Pauli form factor $F_2$, not with the same success in proposing a factorization theorem, but showing $F_2/F_1 \sim 1/Q^2$ at high enough $Q^2$ also for the soft process.

We shall also mention work where the soft contribution  
was evaluated within the light-cone sum rule (LCSR) approach of Braun \textit{et al.}~\cite{braun06}. 
Using asymptotic distribution amplitudes for the nucleon, 
the LCSR approach yields values of $G_{Mp}$ and 
$G_{Mn}$ which are within 20\% compatible with the data in the range 
$Q^2 \sim 1$--$10$~GeV$^2$. The electric form factors however were found to be much
more difficult to describe, with $G_{En}$ overestimated, and 
$G_{Ep}/G_{Mp}$ near constant when using an asymptotic nucleon distribution amplitude. 
Only when including twist-3 and twist-4 nucleon distribution amplitudes within a simple model, 
is a qualitative description of the electric
proton and neutron form factors obtained. Such higher twist components hint at the
importance of quark angular momentum components in the nucleon wave function.

\subsubsection{Generalized Parton Distributions}
\label{subsubsec:gpd}

\begin{figure}[htbp]
\centerline{  
  \includegraphics[width = 8.0 cm]{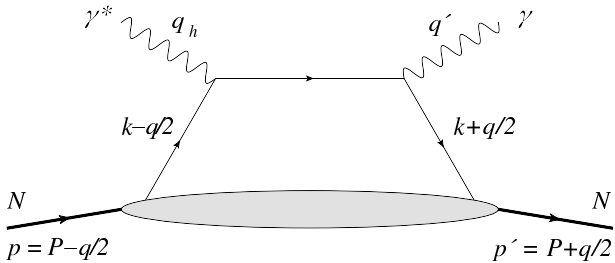} 
}
\caption{\small 
The ``handbag'' diagram for the nucleon DVCS process. 
Provided the virtuality of the initial photon (with momentum $q_h$) 
is sufficiently large, the 
QCD factorization theorem allows to express the 
total amplitude as the convolution  
of a Compton process at the quark level and a non-perturbative 
amplitude parameterized in terms of generalized parton distributions 
(lower blob). The diagram with the photon lines crossed is also understood.    
}
\label{fig:dvcs}
\end{figure}

Generalized parton distributions (GPDs) represent an amplitude for removing a quark from a nucleon and replacing it with another quark with a different momentum, and possibly also with different spin projection and flavor.  These amplitudes can be measured in virtual Compton scattering, $\gamma^*(q_h) + N(p) \to \gamma(q') + N(p')$, or in meson electroproduction, \textit{e.g.,} $\gamma^* + N \to \rho +N$.  The momenta are indicated, with ${q'}^2 = 0$, the virtuality $Q_h^2 = -q_h^2 >0$, and $q$ will be the momentum transfer to the nucleon.  A diagram for the virtual Compton process is shown in Fig.~\ref{fig:dvcs}, in a diagram where both photons interact with the same quark.  The upper part of the diagram is to be calculated perturbatively,  and the lower part of the diagram is given by the GPD.  For ``deep'' virtual Compton scattering, where $Q_h^2 \gg Q^2, m_N^2$, it has been shown that the dominant contributions are given by diagrams like the one shown, and that one can separate or factor the perturbative process specific stage of the interaction from the non-perturbative process independent part (see~\cite{Ji:1998pc,Goeke:2001tz,Diehl:2003ny,Belitsky:2005qn,Ji:2004gf,Diehl:2013xca} 
for reviews and references).

Further notation is that $P = (p+p')/2$ is the average nucleon momentum, $k$ is the average momentum of the quarks entering and leaving the nucleon, $x$ is the light-front momentum fraction defined from $x = k^+/P^+$, and the asymmetry between the quark momenta is given by the skewedness $\xi = -q^+/(2P^+)$.  In the high $Q_h^2$ limit, one can show that $\xi$ is related to the Bjorken variable $x_B$ by $2\xi=x_B/(1-x_B/2)$, where $x_B = Q_h^2/(2 p \cdot q_h)$.

GPDs were introduced by Ji~\cite{ji} and by Radyushikin \cite{Radyushkin:1996nd}. Formally, in the notation of Ji~\cite{ji}\footnote{A gauge link P$\exp(ig\int dx^\mu A_\mu)$, ensuring color gauge invariance, is tacit.} and in a frame where $P$ and $q_h$ are collinear with $\vec P$ in the positive $z$-direction, one obtains the GPDs from,
\begin{eqnarray}
&& \frac{1}{2\pi} \, \int dy^{-}e^{ix  P^{+}y^{-}}
		\nonumber\\
&& \qquad		\times
\left. \langle N(p^\prime)| \bar{\psi_q } (-y/2) \; \gamma^+ \; \psi_q (y/2)
| N(p) \rangle \right|_{y^{+}=\vec{y}_{\perp }=0} 
						\nonumber \\
&&=\; H^{q}(x,\xi ,Q^2)\; \bar{N}(p^{'}) \; \gamma^+ \; N(p)
						\nonumber\\
&& \qquad
+ \  E^{q}(x,\xi ,Q^2)\; \bar{N}(p^{'}) \; \frac{i}{2m_N} \sigma^{+ \nu} 
 \, q_\nu \; N(p) ,
\label{eq:qsplitting}
\end{eqnarray}
where $\psi_q$ is the quark field for flavor $q$ and $N$ is the nucleon Dirac spinor.   The matrix element is non-perturbative and is given in terms of two functions $H^q$ and $E^q$ for each flavor $q$.  There is a similar matrix element with operator $\bar{\psi_q}  \gamma^+ \gamma_5 \psi_q$ and two further, polarized, GPDs $\tilde H^q$ and $\tilde E^q$.

The notation is such that positive momentum fractions correspond to quarks and negative ones to antiquarks.  Hence in $x>\xi$, the fermions leaving and entering the nucleon are both quarks.  Also possible is that both are antiquarks ($x< -\xi$), or that there is a quark-antiquark pair.

GPDs enter this review because of their relation to form factors.  An integral over $x$ on the LHS of Eq.~(\ref{eq:qsplitting}) forces both quark fields to be at the same point, as in the matrix elements of the electromagnetic current.  One can show that the matrix elements have support for $-1 < x <1$, and that
\begin{eqnarray}
\int_{-1}^{+1}dx\, H^{q}(x,\xi ,Q^2)&=& F_{1}^{q}(Q^2)\, ,
				\nonumber\\
\int _{-1}^{+1}dx\, E^{q}(x,\xi ,Q^2)&=& F_{2}^{q}(Q^2)\, ,
\label{eq:ffsumrulehe}
\end{eqnarray}
where the nucleon form factor are given in terms of the quark flavor form factors $F_i^Q$ in the expected ways,
\begin{eqnarray}
F_{ip} &=& \frac{2}{3} F_i^u - \frac{1}{3} F_i^d - \frac{1}{3} F_i^s  ,	\nonumber\\
F_{in} &=& - \frac{1}{3} F_i^u + \frac{2}{3} F_i^d - \frac{1}{3} F_i^s	,
\end{eqnarray}
where $i=1,2$ and $F_1^{u,d}$ are specifically for the proton.

These relations allow us, if we have complete measurements or good models (the latter is more the case at present) for the GPDs, to obtain the electromagnetic form factors from them.   Alternatively, the measured form factors can be used as constraints upon GPD models.

An example of model GPDs is the modified Regge parameterization for $H$ and $E$ that was proposed by Guidal \textit{et al.}~\cite{guidal},
\begin{eqnarray}
H^q (x,0,Q^2) &=& q_v (x)\,  x^{\alpha^\prime \, (1 - x) \, Q^2}, 
			\nonumber\\
E^q (x, 0, Q^2) &=& \frac{\kappa^q}{N^q} 
\, (1 - x)^{\eta^q} \, q_v(x) \, 
{{x^{\alpha' \, (1 - x) \, Q^2}}} \, , 
\label{eq:gpd_r2}
\end{eqnarray}
depending on 3 parameters.  
The Regge slope $\alpha^\prime$ is determined from the Dirac radius, 
and two parameters $\eta^u$ and $\eta^d$, entering the GPD $E$, ensure that 
the $x\sim 1 $ limit of $E^q$ 
has extra powers of $1-x$ compared to that  of $H^q$. 
This results in a proton helicity flip form factors $F_2$ which has a faster 
power fall-off at large $Q^2$ than $F_1$, as observed experimentally.   


\begin{figure}
\begin{center}

	\includegraphics[width = 84 mm,angle=0]{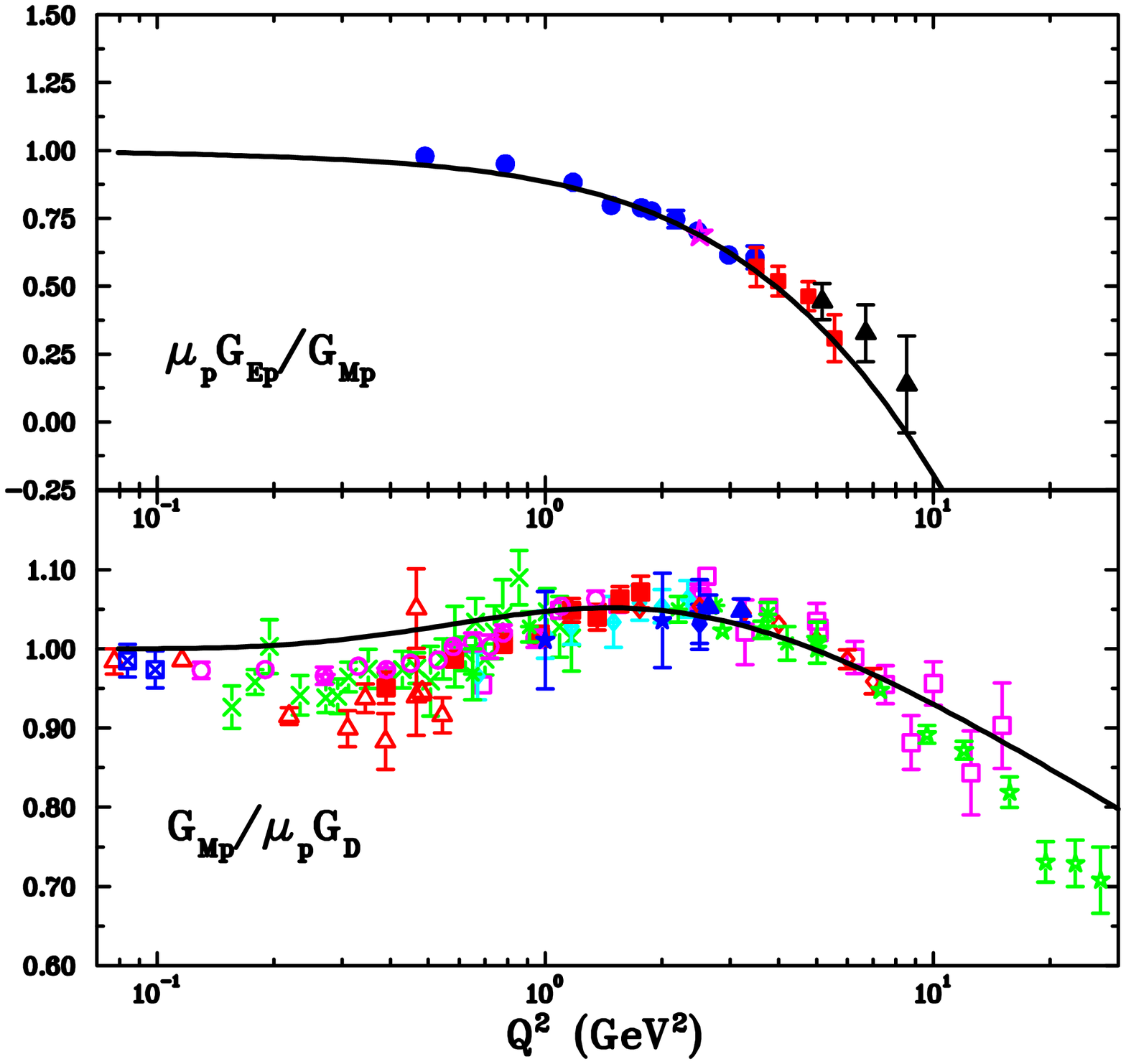}
	
	\includegraphics[width = 84 mm,angle=0]{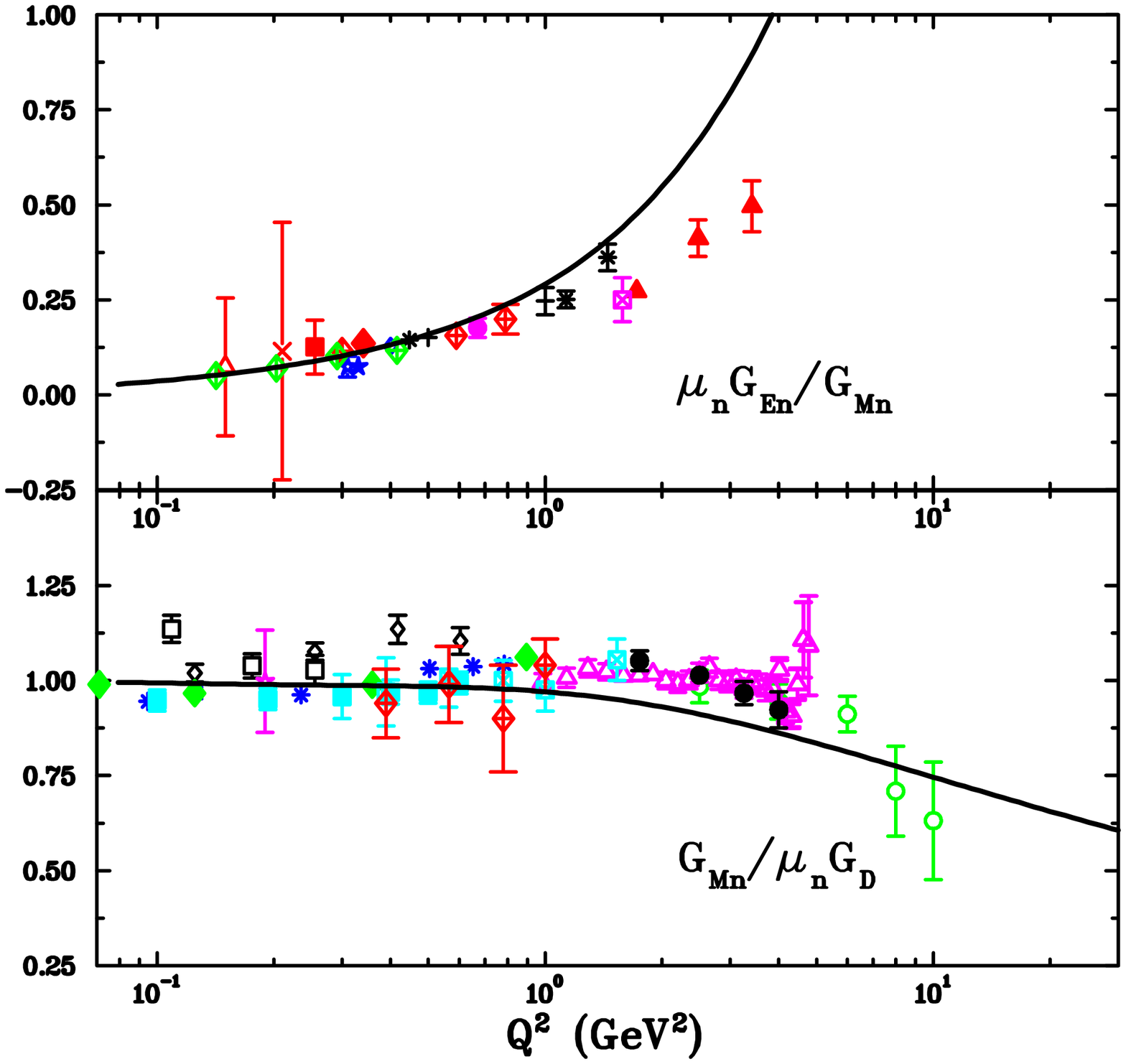} 

\caption{\small GPD calculation of 
$G_{Mp}$ relative to the dipole $G_D$ and of  $G_{Ep}/G_{Mp}$ in the upper panels, and the corresponding plots for the neutron in the lower panels, using GPD's from Ref.~\cite{guidal}. 
Data for $G_{M p}$ are from   
\cite{janssens} (open squares), \cite{litt} (open circles),
\cite{berger} (blue solid stars), \cite{bartel} (green open stars), 
\cite{andivahisA} (red solid circles), \cite{sill} 
(red solid squares), 
according to the recent re-analysis of Ref.~\cite{brash}. 
Data for the ratio $G_{E p} / G_{M p}$ are from 
\cite{gayou:2001} (blue open triangles), 
\cite{gayou:2002} (red solid squares),
\cite{punjabi05B} (blue solid circles), 
and \cite{crawford} (green solid triangles). 
The data for $G_{M n}$ are from 
\cite{xu00} (red solid circles), 
\cite{xu02} (red solid squares),    
\cite{Anklin:1998ae} (open triangles), \cite{kubon} (green open stars), 
\cite{lung} (open squares), \cite{rockA} (solid triangles), 
and \cite{brooksA} (blue solid stars).
The data for $G_{E n}$ are from 
double polarization experiments at 
MAMI~\cite{herberg,ostrick,glazier:2004,becker,rohe} (red solid circles), 
NIKHEF~\cite{passchier} (green solid triangle),  
and JLab~\cite{zhu,madey,warren} (blue solid squares). 
}
\label{fig:gegmpn}

\end{center}
\end{figure}

In Fig.~\ref{fig:gegmpn}, the proton and neutron Sachs electric  and magnetic form factors are shown.  One observes that the 3-parameter  modified Regge model gives a rather good overall 
description of the available form factors data for both proton and neutron in the whole $Q^2$ range,   using as value for the Regge trajectory $\alpha^\prime $ = 1.105 \,GeV$^{-2}$, and the following values for the coefficients governing the $x \to 1$ behavior of the $E$-type GPDs:  
$\eta^u$ = 1.713 and $\eta^d$ = 0.566.  Note that a value $\eta^q = 2$ corresponds to a $ 1/Q^2$ asymptotic behavior of the ratio $F_2^q / F_1^q$ at large $Q^2$. The modified Regge GPD parameterization allows one to accurately describe the decreasing ratio of $G_{E p} / G_{M p}$ 
with increasing $Q^2$, and also leads to a zero for $G_{E p}$ at a momentum transfer of $Q^2 \simeq 8$~GeV$^2$.

\subsection{Lattice QCD Calculations of Nucleon Form Factors}
\label{subsec:lattice}

Strictly speaking, lattice calculations of nucleon form factors are currently available only for the isovector form factors.  

Isoscalar form factors require calculations of disconnected diagrams, which are diagrams with quark loops not connected to the quark lines emanating from or ending on the lattice nucleon source or sink.  There are gluons that attach the quark loops to the valence quarks, but these are not indicated in lattice diagrams, hence the phrase ``disconnected.''  Contributions from the disconnected loops require computer time intensive calculations, and remain undone.  However, the disconnected diagrams contribute equally to proton and neutron, so the isovector case can be considered without them.  

A review including lattice form factor results up to 2010 is available in~\cite{Hagler:2009ni}, and newer lattice form factor results are reported in~\cite{Alexandrou:2013joa,Bhattacharya:2013ehc,Green:2014xba}.

The new calculations reported in Green \textit{et al.}~\cite{Green:2014xba} have pion masses from 373 MeV down to close to physical 149 MeV.  The latter also strove to reduce contamination from excited nucleons.  They analyze their lattice data using three methods which they call the standard ratio method, the summation method, and the generalized pencil-of-function method (GPoF), with varying outcomes.  The best results, judged by comparison to data as represented by one of the standard fits~\cite{Alberico:2008sz}, come from the summation method.  Here agreement with experimental data is good for both $G_{Ev}$ and $G_{Mv}$ in the region considered, which is $Q^2$ from scattering threshold up to about 0.5 GeV$^2$, with uncertainty limits about $20\%$ at $Q^2$ of $0.4$ GeV$^2$.

The works of Alexandrou \textit{et al.}~\cite{Alexandrou:2013joa} and Bhattacharya \textit{et al.}~\cite{Bhattacharya:2013ehc} have pion masses in the 213--373 MeV range, and quote results for somewhat higher $Q^2$.  For $Q^2$ above about $0.6$ GeV$^2$, their isovector form factors results tend to be 50\% or so above the data for $G_{Ev}$ (or $F_{1v}$), with uncertainties indicated at about 10\%.  For $G_{Mv}$ (or $F_{2v}$) the results are closer to data .  The authors of these works do point out that the lattice treatments with these pion masses are all consistent with each other.

One may specifically focus on nucleon radii calculated from lattice gauge theory.  In the future, it may be possible and desirable to calculate using a dedicated correlator which gives directly the slope of the form factor at zero momentum transfer.  Finding such correlators by taking derivatives of known correlators is suggested and studied~\cite{deDivitiis:2012vs} for lattice calculations of form factors at points where the Lorentz factors they multiply go to zero.  Applications in~\cite{deDivitiis:2012vs} are to form factors for semi-leptonic scalar meson decay, and to hadronic vacuum polarization corrections to the muon $(g-2)$.

At present, lattice calculations of nucleon radii proceed by calculating the form factor at several non-zero $Q^2$, fitting to a suitable form, typically a dipole form, and finding the radius by extrapolating to zero $Q^2$.  Truly complete results are available only for the isovector nucleon.  Ref.~\cite{Green:2014xba} presents a plot of radius results for lattice calculations at various pion masses.  They use the Dirac radius, obtained from the slope of $F_{1v}$, rather than the charge radius obtained from $G_{Ev}$, but these are related by, using the proton as an example,
\begin{equation}
\langle r_{1p}^2 \rangle = \langle r_{p}^2 \rangle - \frac{3}{2} \frac{\kappa_v}{m_p^2}	\,,
\end{equation}
where $\kappa_v$ is the isovector anomalous nucleon magnetic moment.  Hence, given the great accuracy of the magnetic moment measurements, one knows the Dirac radii to the same accuracy as the charge radii.  

The great interest is to obtain sufficient accuracy from the lattice results to be able to adjudicate between the electron and muon measured values of the isovector charge or Dirac radii.  The electron measured isovector radius is straightforward to look up, the muon measured value of the Dirac or charge radius is for now a defined quantity obtained by using the electron value for the neutron radius-squared.  Using the summation method, Ref.~\cite{Green:2014xba} obtains, by extrapolation to the physical pion mass, a value of the isovector Dirac radius between the muonic and electronic results, with uncertainties that accommodate both at about the one standard deviation level.  However, using the GPoF or ratio method gives a smaller $\langle r_{1}^2 \rangle_v$, on the order of $2/3$ the value from the summation method.

\begin{figure}
	\begin{center}
		\resizebox{0.45\textwidth}{!}{%
			\includegraphics[angle=90]{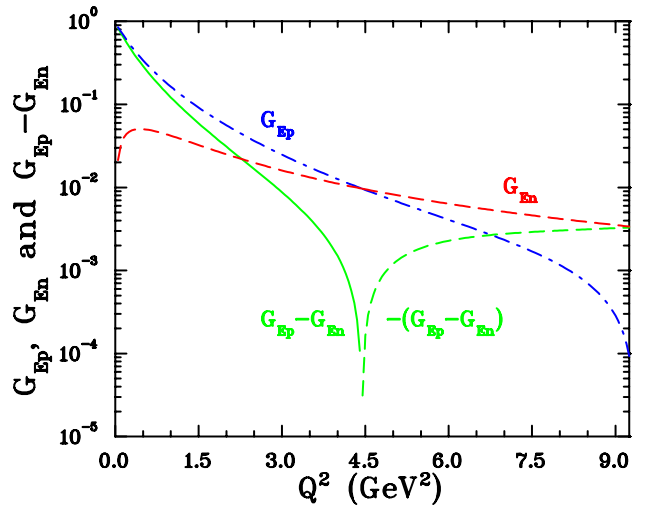}}
		\caption{The isovector electric form factor $G_{Ep}^v=G_{Ep}-G_{En}$ obtained from fits to the experimental form factor data shown in Figs.~\ref{fig:gepgmp_large_qsqr_pol} and~\ref{fig:gen_pol}.}
\label{fig:isovector}
\end{center}
\end{figure}	

One may say there is opportunity for further work.  An uncertainty of 1\% or less for the proton alone is needed for a lattice calculation to impact the proton radius puzzle.  Two extrapolations are needed to obtain the charge radius for a physical nucleon.  On is in the lattice pion mass, commented upon above.  The other is in $Q^2$.  Currently on the experimental side, the lowest $Q^2$ from scattering data is about $0.004$ GeV$^2$, and there are experiments planned or running to reduce this number.  Further there are discussions, alluded to in Sec.~\ref{subsubsec:cf}, regarding the best fit forms to use for the extrapolation.  Currently on the lattice, represented by Ref.~\cite{Green:2014xba}, the lowest $Q^2$ is about $0.04$ GeV$^2$, and the fit to the lattice data is only done using a simple dipole form.

And of course one wants the isoscalar as well as the isovector form factors.   However, a challenge involving just the isovector form factors, albeit at a higher $Q^2$ than lattice form factor results currently display,  is to obtain the zero in the isovector $G_{Ev}$ that is visible in the fits to  the experimental data shown in Fig.~\ref{fig:isovector}.

\section{Outlook}
\label{sec:conclusion}

The experimental and theoretical status of the nucleon form factors were reviewed extensively in the 15 years following
publication of the results of the first recoil polarization experiment at Jefferson Lab \cite{jones} for the proton. Chronologically
these reviews include Gao \cite{gaoA}, Hyde-Wright and de Jager \cite{charleskees}, Perdrisat {\it et al.}\cite{perdrisat:2006}, 
Arrington {\it et al.} \cite{arrreview},  Clo\"et  {\it et al.} \cite{cloet:2008}, Arrington {\it et al.} \cite{arrington:2011}, Perdrisat and Punjabi \cite{scholar} and S. Pacetti {\it et al.} \cite{pacetti:2015}. The completion of the GEn(1) and GEp(3)  experiments, which reached a maximum $Q^2$ of 3.4 and 8.4 GeV$^2$ respectively, has brought the field into previously unexplored regions of four-momentum transfer squared, and correspondingly, generated a burst of theoretical investigations along old and new paths.   

The proton form factors were originally introduced in the approximation of non-relativistic scattering, as the three-dimensional Fourier transform of the charge density \cite{hofs53,wilsonrr}. However the proton recoil implies that the electron is interacting with a moving charge distribution. Already for $Q^2$=0.25 GeV$^2$, the recoil proton relativistic boost factor $\gamma$ is 1.1, corresponding to $v/c=0.42$. The argument that form factors are Fourier transforms of nucleon density in the Breit frame had to be abandoned, as this frame's velocity in the Lab frame is significantly different for every $Q^2$.

\begin{figure}[htbp]
	\begin{center}
		\includegraphics[width = 83 mm]{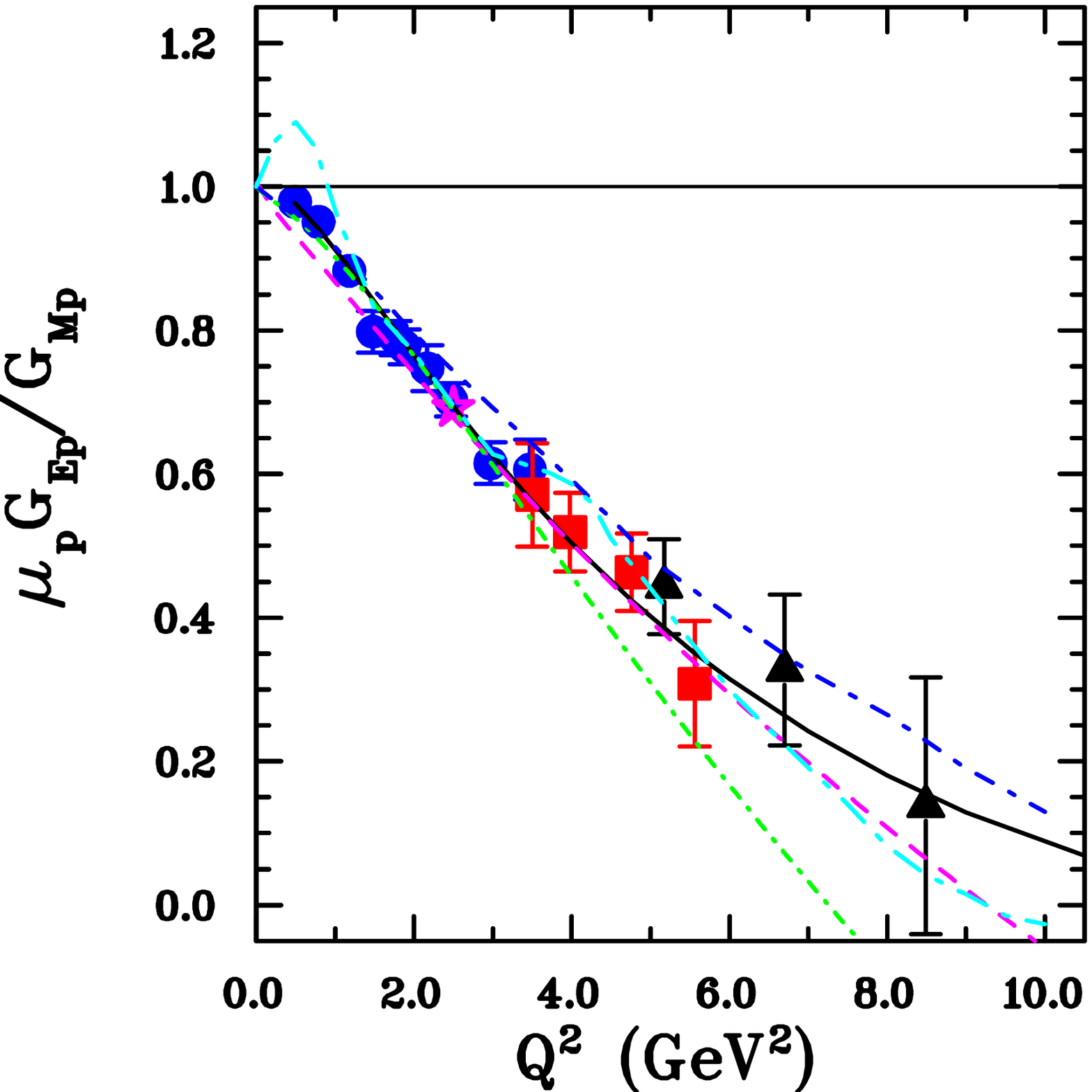}
		\vskip 5 mm
		\includegraphics[width = 83 mm]{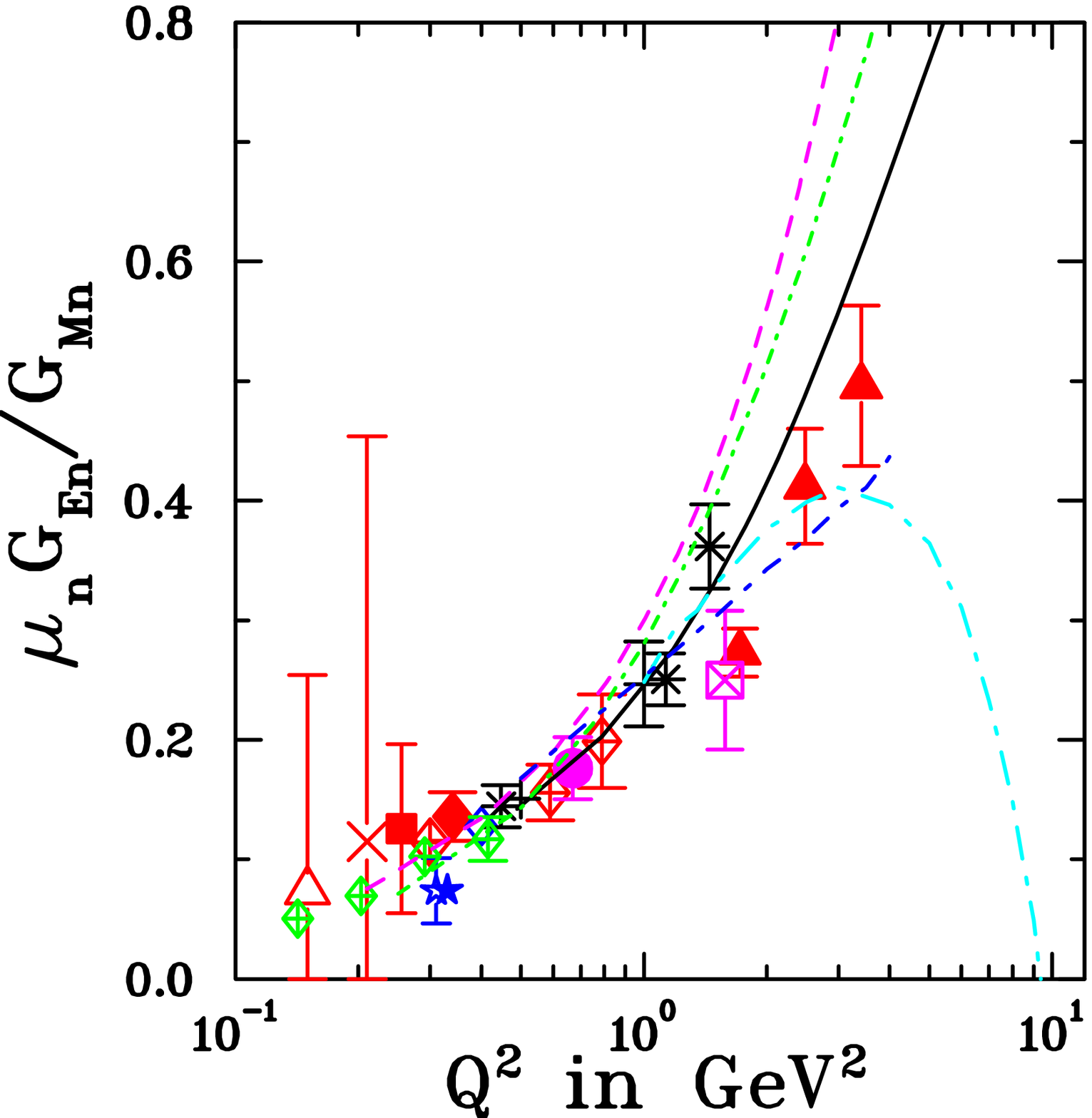}

		\caption{Upper panel: Five theoretical model predictions of the proton form factor
                         ratio $\mu_pG_{Ep}/G_{Mp}$ shown together with the data of JLab recoil polarization experiments, symbols 
                         are same as in Figure 10. Lower panel: The same five theoretical model predictions for the neutron form factor
                          ratio $\mu_nG_{Ep}/G_{Mn}$, data are same as in Figure 11. The theoretical curves are from the VMD model of Lomon 
                           \cite{Lomon:2006xb},
                           solid (black) curve; from the GPD's of Guidal {\it et al} \cite{guidal}, medium dashes (magenta), the
                           covariant spectator model of Gross {\it et al.} \cite{Gross:2006fg} as short dash-dot (green)
                           Dyson-Schwinger equation solutions of Clo\"{e}t {\it et al.} \cite{Cloet:2013gva}, long  
                         dash-dots (cyan), and the quark-diquark model of Clo{\"e}t and and Miller \cite{Cloet:2012cy}, 
                          medium dash-dots (blue). The symbols for the data are explained in Fig.~\ref{fig:gen_pol}.} 
		\label{fig:cncd}
	\end{center}
\end{figure}
Changes in our view of the structure of the proton are many. For example, the proton in its ground state is not necessarily spherically symmetric, but can show a typical multipole shape, 
when referred to the spin direction of one of its quarks (constituents) relative to the nucleon spin orientation
 \cite{miller:2003}. 
Also, the wave front or infinite momentum frame charge and magnetization densities are invariant, two-dimensional transverse distributions which are drastically different
from the non-relativistic ones \cite{miller:2003,carlson:2007}. 

A selection of model predictions for the form factor ratios $\mu_nG_{En}/G_{Mn}$ and $\mu_pG_{Ep}/G_{Mp}$  is shown together 
with the data obtained in double polarization  experiments at JLab and in other laboratories in Fig. \ref{fig:cncd}. These 
two figures emphasize the importance of future experiments which will establish whether either ratio does, or does not cross 
zero near $Q^2=$10 GeV$^2$. Whereas for $\mu_pG_{Ep}/G_{Mp}$, all model predictions discussed in this review anticipate 
a zero crossing somewhere above 9 GeV$^2$, for $\mu_nG_{En}/G_{Mn}$ only the calculation based on the Dyson-Schwinger 
equations predict such a zero crossing in the 10 GeV$^2$ region of $Q^2$.

A recent development has been the calculation of the flavor separated form factors of the "dressed" quarks from simple linear relations between the nucleon form factors, assuming charge symmetry applied to the data available.
The dressed up and down quarks have significantly different form factors 
\cite{cloet:2008,cates:2011,rohrmoser:2011,wilson:2011,Cloet:2013gva,qattan:2013}. 
Nucleon form factors determine the parameters of the valence quark GPDs;
these can be used to obtain corresponding valence quark densities \cite{Diehl:2013xca}. They can be compared with the GPDs obtained from real and virtual Compton scattering.

\begin{figure*}
	\begin{center}
		\resizebox{0.9\textwidth}{!}{%
			\includegraphics[angle=90]{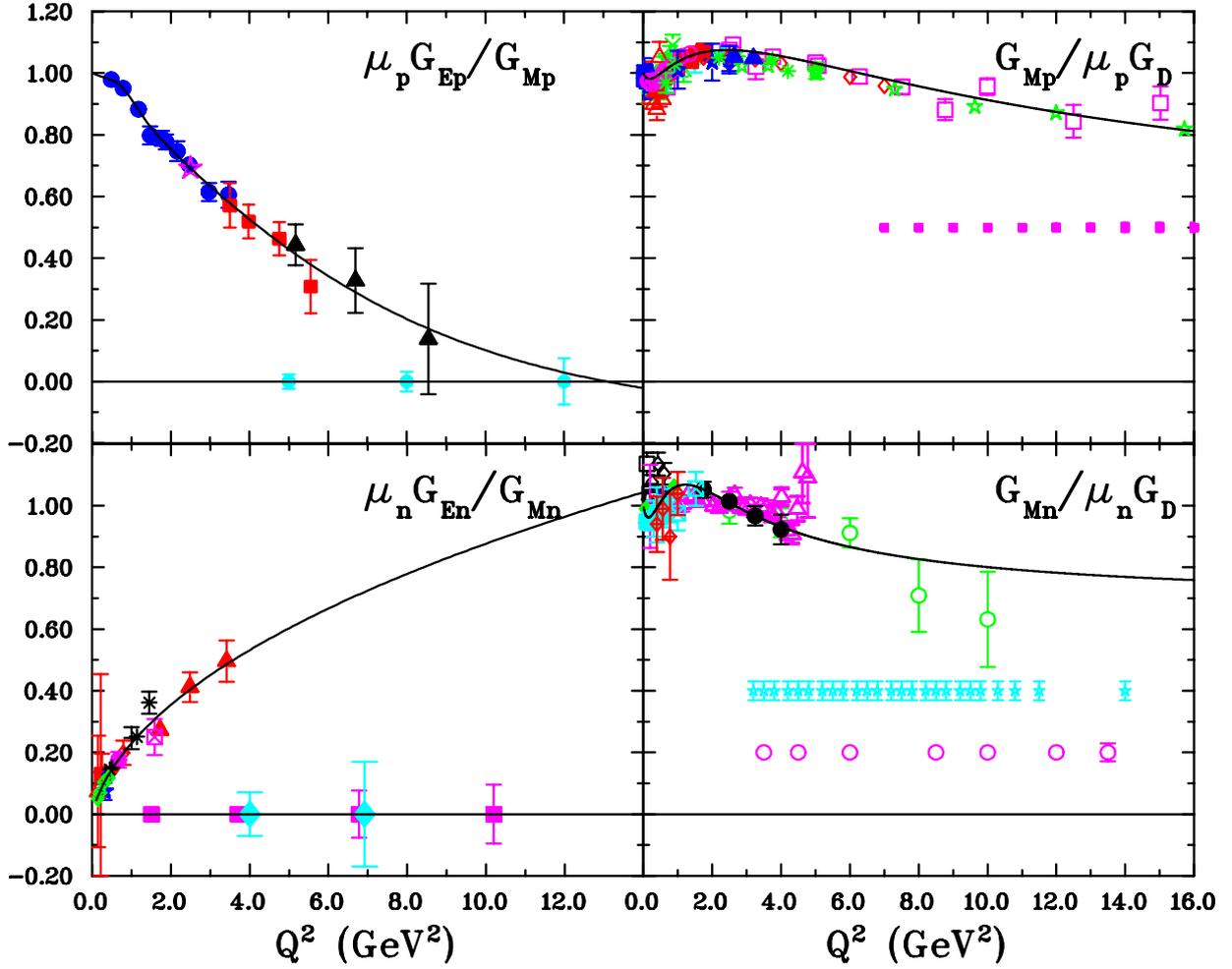}
		}
		\caption{The projected error bars for the approved nucleon form factors experiments at Jefferson Lab in the 12 GeV era. 
			For the Hall~A SBS experiment E12-07-108 \cite{gep12GeV}, the anticipated error bars on the ratio $\mu_p G_{Ep}/G_{Mp}$ are shown as filled circles (cyan). The anticipated error bars on the ratio $G_{En}/G_{Mn}$ are shown as filled squares (magenta) for the Hall A SBS experiment E12-09-016 \cite{gen12GeV} and as filled diamonds (cyan) for the Hall C experiment E12-11-009 \cite{genhallc12GeV}. The anticipated error bars for  $G_{Mp}/\mu_p G_D$ from the Hall A experiment E12-09-019 \cite{gmp12GeV} are shown with square 
symbols (magenta). Finally the ratio $G_{Mn}/\mu_n G_D$ will be measured in two experiments: E12-09-019 in Hall~A \cite{gmn12GeV} and E12-07-104 in Hall B \cite{gmnclas12GeV}. The expected error bars are shown as empty circles (magenta) and filled stars (cyan), respectively.
		\label{fig:formfactors}}
	\end{center}
\end{figure*}

The doubling of the energy of the Jefferson Lab accelerator to 12~GeV will lead to a much enhanced program of experiments investigating the structure of the nucleon. 
An experiment will use the existing High Resolution Spectrometers in Hall A at Jefferson Lab to measure $G_{Mp}$ with greatly improved error bars up to 14~GeV$^2$ \cite{gmp12GeV}. An new, versatile Super BigBite Spectrometer (SBS), consisting of a simple dipole magnet and associated detectors, is being built for three form factor experiments in Hall~A. One SBS experiment will measure $G_{Ep}/G_{Mp}$ up to $Q^2$~=~12 GeV$^2$ using the recoil polarization technique \cite{gep12GeV}. Sitting behind the SBS dipole magnet will be a recoil polarimeter which will have two analyzers with multiple GEM chambers used for incoming and scattered track determination. Another SBS experiment will extract $G_{En}/G_{Mn}$ up to $Q^2$~=~10 GeV$^2$ from  beam-target asymmetry measurements using an upgraded polarized $^3$He target \cite{gen12GeV}. The experiment will detect the scattered electrons in the BigBite spectrometer and the scattered neutron in a large solid angle hadron calorimeter sitting behind the SBS magnet.  In Hall C, measurements of the neutron recoil polarization in quasi-free electron deuteron scattering will be done to extract  $G_{En}/G_{Mn}$ up to $Q^2$~=~6.9~GeV$^2$ \cite{genhallc12GeV}.
With the new CLAS12 spectrometer in Hall B at Jefferson Lab , the measurement of $G_{Mn}$ will be done to $Q^2$~=~14~GeV$^2$ \cite{gmnclas12GeV}. A third SBS experiment will also measure $G_{Mn}$ to $Q^2$~=~14~GeV$^2$ \cite{gmn12GeV}.
The proposed error bars for all these experiments are shown in Fig.~\ref{fig:formfactors}.

One of the most stringent constraints that nucleon elastic form
factor data at large $Q^2$ can provide, relates to the issue of 
the various contributions from quarks, gluons, and orbital angular momentum
to the total angular momentum of the nucleon. 
The elastic form factors also provide a powerful check of lattice QCD.  
The lattice calculations of form factors are making impressive progress, and  the comparison
of these results with experimental measurements will be extremely important.
There is an indication from the results of GEp(3) experiment that we may 
be entering the range of momentum transfers where the pQCD prediction 
is vindicated. Yet a continuation of the fast 
decrease of the ratio toward negative values cannot be excluded.
Great progress in the theoretical description of the structure of the nucleons can be expected.

\begin{acknowledgement}

We would like to thank Dr.~C.~Ayerbe~Gayoso for useful discussions and critical reading of the 
manuscript. This work was supported by U.S. Department of Energy grant DE-FG02-89ER40525 (VP) 
and by DOE contract DE-AC05-06OR23177, under which Jefferson Science Associates, LLC, operates
the Thomas Jefferson National Accelerator Facility (MKJ), and by National Science Foundation (USA) 
grants PHY-1208056 (EJB), PHY-1205905 (CEC), and PHY-1066374 (CFP).   

\end{acknowledgement}

%
\bibliographystyle{epj}
\bibliography{EPJA_bibli_feb-18-2015}
\end{document}